\pdfoutput=1
\RequirePackage{ifpdf}
\ifpdf
\documentclass[pdftex]{sigma}
\else
\documentclass{sigma}
\fi

\numberwithin{equation}{section}

\newcommand{\Nint}{\mathbb{N}}
\def\tr{\,{\rm Tr}\, }
\def\det{\,{\rm det}\, }

\def\tr{\,{\rm tr}\, }

\def\rangl{\right\rangle   }
\def\langl{\left\langle  }
\def\({\left(}
\def\){\right)}
\def\[{\left[}
\def\]{\right]}
\def\p{\partial}
\newcommand{\de}{\mathrm{d}}
\newcommand{\I}{\mathrm{i}}

\def\11{1\!\! 1}

\def\hf{\frac{1}{2}}

\def\eps{\varepsilon}

\def\l{\lambda}

\def\t{\tau}

   \def\CD {{\cal D}}

   \def\CG {{\cal G}}
   \def\CH {{\cal H}}
   \def\CI {{\cal I}}
   
   \def\CK {{\cal K}}
   
   \def\CM {{\cal M}}

   \def\CP {{\cal P}}

   \def\CS {{\cal S}}

   \def\CV {{\cal V}}

   \def\CZ {{\cal Z}}

\newcommand{\unit}{{\mathbb{I}}}

\newcommand{\Cmat}{{\mathbb C}}

\newcommand{\Rmat}{{\mathbb R}}
\newcommand{\Zmat}{{\mathbb Z}}
\newcommand{\gmat}{\mathfrak{g}}
\newcommand{\hmat}{\mathfrak{h}}

\newcommand{\hT}{\hat T}

\newcommand{\im}{\gamma}

\newcommand{\tE}{\lefteqn{\smash{\mathop{\vphantom{<}}\limits^{\;\sim}}}E}
\newcommand{\tP}{\lefteqn{\smash{\mathop{\vphantom{<}}\limits^{\;\sim}}}P}

\newcommand{\Pt}{\lefteqn{\smash{\mathop{\vphantom{\Bigl(}}\limits_{\sim}
\atop \ }}P}

\newcommand{\tNn}{\lefteqn{\smash{\mathop{\vphantom{\Bigl(}}\limits_{\,\sim}
\atop \ }}{\cal N}}

\newcommand{\SA}{{\cal A}}

\newcommand{\SSA}{{\bf A}}

\newcommand{\nd}{{\cal N}}

\newcommand{\pr}[1]{\pi^{(#1)}}

\def\npm{n^\pm}

\newcommand{\gb}{{\rm g}}

\newcommand{\gx}{{\rm g}}
\newcommand{\gl}{g}
\newcommand{\Db}{{\bf D}}

\def\xb{\mathfrak{a}}
\def\eb{\mathbf{e}}

\def\Int{\CI}
\def\inter{i}

\def\sgn{\mathrm{sgn}}
\def\lp{\ell_\text{Pl}}
\def\Tr{\text{Tr}}
\def\Pexp{\overrightarrow{\exp}}
\def\Hk{\mathcal{H}_\text{kin}}

\def\Hp{\mathcal{H}_\text{phys}}

\newcommand{\f}{\frac}

\newcommand{\SU}{\text{SU}}
\newcommand{\Uq}{\text{U}_q}
\newcommand{\ISU}{\text{ISU}}
\newcommand{\SO}{\text{SO}}
\newcommand{\SL}{\text{SL}}
\newcommand{\Sp}{\text{Spin}}
\newcommand{\su}{\mathfrak{su}}
\newcommand{\isu}{\mathfrak{isu}}
\newcommand{\so}{\mathfrak{so}}
\def\Pexp{\overrightarrow{\exp}}

\def\pD#1{\CD #1}
\def\lmul{\lambda}
\newcommand{\ClG}[6]{{{C\lefteqn{\vphantom{\overline{A}_-}}^{\smash{#1}}_{\smash{#4}}}^{\smash{#2}}_{\smash{#5}}}^{{#3}}_{{#6}}}

\begin{document}

\allowdisplaybreaks

\renewcommand{\thefootnote}{$\star$}

\renewcommand{\PaperNumber}{055}

\FirstPageHeading

\ShortArticleName{Spin Foams and Canonical Quantization}

\ArticleName{Spin Foams and Canonical Quantization\footnote{This
paper is a contribution to the Special Issue ``Loop Quantum Gravity and Cosmology''. The full collection is available at \href{http://www.emis.de/journals/SIGMA/LQGC.html}{http://www.emis.de/journals/SIGMA/LQGC.html}}}

\AuthorNameForHeading{S.~Alexandrov, M.~Geiller and K.~Noui}

\Author{Sergei ALEXANDROV~${}^{\dag^1\dag^2}$, Marc GEILLER~${}^{\dag^3}$, Karim NOUI~${}^{\dag^4\dag^3}$}

\Address{${}^{\dag^1}$~Universit\'e Montpellier 2, Laboratoire Charles Coulomb UMR 5221,\\
\hphantom{${}^{\dag^1}$}~F-34095, Montpellier, France}
\EmailDD{\href{mailto:salexand@univ-montp2.fr}{salexand@univ-montp2.fr}}

\Address{${}^{\dag^2}$~CNRS, Laboratoire Charles Coulomb UMR 5221, F-34095, Montpellier, France}

\Address{${}^{\dag^3}$~Laboratoire APC, Universit\'e Paris Diderot Paris 7, 75013 Paris, France}
\EmailDD{\href{mailto:mgeiller@apc.univ-paris7.fr}{mgeiller@apc.univ-paris7.fr}}

\Address{${}^{\dag^4}$~LMPT, Universit\'e Fran\c cois Rabelais, Parc de Grandmont, 37200 Tours, France}
\EmailDD{\href{mailto:karim.noui@lmpt.univ-tours.fr}{karim.noui@lmpt.univ-tours.fr}}

\ArticleDates{Received January 30, 2012, in f\/inal form August 12, 2012; Published online August 19, 2012}

\Abstract{This review is devoted to the analysis of the mutual consistency of the spin foam and canonical loop quantizations
in three and four spacetime dimensions.
In the three-dimensional context, where the two approaches are in good agreement,
we show how the canonical quantization \`a la Witten of Riemannian gravity with a positive cosmological constant
is related to the Turaev--Viro spin foam model, and how the Ponzano--Regge amplitudes are related
to the physical scalar product of  Riemannian loop quantum gravity without cosmological constant.
In the four-dimensional case, we recall a Lorentz-covariant formulation of loop quantum gravity
using projected spin networks, compare it with the new spin foam models, and identify interesting relations
and their pitfalls. Finally, we discuss the properties which a spin foam model is expected
to possess in order to be consistent with the canonical quantization,
and suggest a new model illustrating these results.}

\Keywords{spin foam models; loop quantum gravity; canonical quantization}

\Classification{83C45; 83C05; 83C27}

\vspace{-2mm}

\tableofcontents

\renewcommand{\thefootnote}{\arabic{footnote}}
\setcounter{footnote}{0}

\section{Introduction}

Canonical loop quantization and spin foam models are two complementary approaches to quantum gravity
\cite{Alexandrov:2010un,Ashtekar:2004eh,Perez:2004hj, Rovelli:2004tv,Thiemann:2001yy}.
The former is designed to describe a {\it quantum space} in the frozen time formalism, via
a (kinematical) Hilbert space on which certain constraints encoding the dynamics of the theory are imposed.
In contrast, the latter is supposed to be a version of the discretized path integral for gravity, or a sum over histories
where every history represents a~{\it quantum spacetime}.
Each of the approaches has its own advantages and shortcomings.
For example, the canonical approach is well grounded and relies on a solid mathematical construction,
but at the same time in this framework all the dynamical questions look extremely complicated.
On the other hand, spin foam models bring powerful geometric methods and suggest a simple
way to recover the dynamics of the theory, but their derivations are much less robust than
the corresponding canonical constructions.

In this situation, it is of primary importance to consider spin foam and canonical quantizations in parallel,
in order to be able to learn some lessons in one approach and use them in the second one.
If such an exchange of ideas and results is successful, it would allow to overcome the problems arising in each particular
quantization scheme and, if we are lucky, to f\/ind a~model for quantum gravity
resulting from both approaches. This implies that the results derived from a~spin foam model, which is claimed to be
a possible candidate for quantum gravity, must be in agreement with the canonical quantization.
If the agreement is found, this can be considered as a serious argument in favor of the model.
If not, then either the spin foam model or the canonical quantization (if not both) should be reconsidered.

Thus, the study of mutual consistency of the loop and spin foam quantizations
is crucial for getting new insights and eventually a better understanding of the existing constructions.
This issue will be the main subject of this review, as our goal is to present
the current understanding of the relations between the results of these two approaches.
In particular, we would like to show an amazing interplay between the canonical and covariant
quantizations of three-dimensional gravity, possibly coupled to point-like particles,
and to analyze the current situation in the four-dimensional case.

Let us recall that a close relation between loops and spin foams
follows already from the very basic features of these two approaches. Both the kinematical states in loop quantum gravity (LQG)
and the boundary states between which one evaluates transition amplitudes in spin foam models
are represented by spin networks, i.e.\ by graphs colored with group theoretic data such as representations
and intertwiners. Moreover, in their seminal work \cite{Reisenberger:1996pu}, Reisenberger and Rovelli
provided a precise link between the dif\/ferent physical quantities associated to these spin network states.
Namely, they showed that an expansion in powers of a Hamiltonian operator
of the physical scalar product between two kinematical states in LQG
generates a sum over spin foams with weights given by matrix elements of this Hamiltonian.

In principle, this result suggests a direct way to construct a spin foam model fully consistent with LQG.
In three dimensions, although the spin foam models have been derived originally in a dif\/ferent way,
their link with the canonical approach is established using the same set of ideas.
In fact, there are two rather dif\/ferent frameworks in which this can be done.
The f\/irst one relies on the work of Witten \cite{Witten:1988hf}, who proposed the f\/irst complete and consistent
quantization of three-dimensional Riemannian gravity with a positive cosmological constant.
In this quantization, the path integral is represented by the scalar product between physical states.
On the other hand, Turaev and Viro \cite{Turaev:1992hq}
proposed a state sum representation of the same path integral,
which was the f\/irst mathematically well-def\/ined spin foam model.
The second framework is provided by Riemannian gravity with vanishing cosmological constant.
The corresponding spin foam quantization was found by Ponzano and Regge \cite{Ponzano:1968dq},
and then in \cite{Noui:2004iy} it was shown that the Ponzano--Regge amplitudes are reproduced
by evaluating the physical scalar product between spin network kinematical states of the loop quantization.

Thus, in the case of three-dimensional gravity (at least for Riemannian signature)
there are already solid results conf\/irming the agreement between the spin foam and canonical quantizations.
In contrast, the agreement in the four-dimensional case is less obvious.
The dif\/f\/iculties associated with the Hamiltonian operator and the evaluation of its matrix elements
have led people to reject the strategy which consists in f\/inding the spin foam representation from the canonical side.
Instead, the spin foam quantization has developed its own,
completely independent methods. As a result, although at the qualitative level the models derived in this way
produced a picture similar to LQG, quantitatively for a long time there was a striking disagreement.
Since spin foam models represent a completely covariant quantization of gravity in the f\/irst order formulation,
their boundary states are labeled by the group theoretic data of the unbroken gauge group in the tangent space,
which is either the Lorentz group or~$\Sp(4)$ depending on the signature. On the other hand, LQG is based on
the $\SU(2)$ group obtained by a~partial gauge-f\/ixing at the classical level.
A longstanding problem was to realize how a similar reduction to the compact subgroup
can take place in spin foam models.

The progress came from two sides. On one hand, it was understood how to represent
LQG in a Lorentz-covariant form~\cite{Alexandrov:2002br}.
In particular, this brought into play the so-called projected spin networks,
which constitute a certain generalization of the usual spin networks.
As it turned out, they are equally relevant for the spin foam approach
since the boundary states of all existing spin foam models of four-dimensional gravity
can be expressed as their linear combinations. On the other hand, some new spin foam models
were introduced \cite{Engle:2007wy, Freidel:2007py}, with the hope that they could
potentially overcome various problems encountered with the previous ones.
Furthermore, it was argued that the new models implement the reduction
to the $\SU(2)$ gauge group mentioned above,
which is a necessary step if one requires agreement with the canonical quantization.
These f\/indings suggest that we are now in a much better position than just a few years ago,
and that the issue of the mutual consistency might be soon (if not already) resolved.

In this review, we reconsider this issue in detail and
provide an analysis of the relationship between the spin foam and canonical quantizations
in three and four spacetime dimensions.
We start by presenting the general ideas of the spin foam approach in Section~\ref{sec_SFgen}.
In particular, we recall its relation to the physical scalar product of the canonical loop quantization
established in \cite{Reisenberger:1996pu} and explain the main strategy used to derive spin foam models
in four dimensions.

In Section~\ref{sec_3d} we concentrate on Riemannian three-dimensional gravity.
After brief\/ly reviewing the classical aspects of pure gravity and gravity coupled to point particles,
we f\/irst recall the quantization \`a la Witten \cite{Witten:1988hf},
which relies on the Chern--Simons formulation of Riemannian gravity with a positive cosmological constant.
In this context, we show how the physical Hilbert space can be constructed using the so-called
combinatorial quantization and how the physical scalar product
is related in a very precise way to the Turaev--Viro spin foam model.
Then we turn to Riemannian gravity without cosmological constant.
After reviewing its loop quantization and the Ponzano--Regge spin foam model,
we show, following \cite{Noui:2004iy}, how the spin foam representation of the path integral can be obtained
from the kinematical Hilbert space of the canonical quantization
by projecting states onto the physical Hilbert space.
This provides an example of the concrete realization of the ideas of \cite{Reisenberger:1996pu}.
We also present some attempts to generalizing this result to the case of a positive
cosmological constant.

\looseness=1
In Section~\ref{sec_4d} we study the four-dimensional case.
First, we review the basic notions of LQG and its Lorentz-covariant reformulation.
For this purpose, we introduce the projected spin networks, which turn out to play
also an important role in spin foam models. Second, we recall the canonical structure
of the Plebanski formulation of general relativity which is at the heart of the spin foam approach
in four dimensions.
After this, we def\/ine the most relevant spin foam models studied in the literature
(BC~\cite{Barrett:1999qw, Barrett:1997gw}, EPRL~\cite{Engle:2007wy} and FK~\cite{Freidel:2007py})
and present their main features. The EPRL model is then compared in detail to LQG.
This comparison suggests some exciting relations, but at the same time a thorough analysis opens various problems
challenging some of the conclusions which have been taken for granted.
In particular, it reveals a tension between the derivation of the new models and the canonical quantization.
Taking seriously the lessons obtained in the canonical theory, we forget about the existing models and discuss
restrictions imposed on the spin foam quantization in four dimensions by
the requirement of consistency with the canonical approach.
In particular, we address the issues of the imposition of constraints,
the path integral measure and the form of the vertex amplitude.

Then, in Section~\ref{sec-new}, we describe a new model proposed recently in \cite{geiller-noui}.
Although this model is three-dimensional, it is designed
to test the implementation of the simplicity constraints appearing in the four-dimensional case.
Its analysis shows that implementing the constraints following the method of either the BC or the EPRL model,
one does not arrive to the expected vertex amplitude. Instead, the correct result can be recovered
by incorporating the ideas presented at the end of Section~\ref{sec_4d} and modifying
the standard prescription for evaluating the vertex amplitudes from the simplex boundary states.
In the construction of~\cite{geiller-noui} however, the weak imposition of the simplicity constraints is an essential ingredient
which enables to recover the expected amplitude.
The f\/inal section contains our conclusions.

Some comments about our notations.
The signature of spacetime will be encoded in a sign function which is $\sigma=1$ in the Riemannian
case and $\sigma=-1$ in the Lorentzian case.
The corresponding isometry algebras of the f\/lat Euclidean or Minkowskian space, $\so(4)$ and $\so(3,1)$,
both admit $\su(2)$ as a subalgebra. Notations are such that
$\mu,\nu,\dots$ refer to spacetime indices, $a,b,\dots$ to spatial
indices, $I,J,\dots$ to $\so(4)$ or $\so(3,1)$ indices, $i,j,\dots$ to
$\su(2)$ indices, and $(\cdot \,\cdot)$ (or $[\cdot \,\cdot]$) to the (anti-)symmetrization taken with weight 1/2.
$\Tr$ and $\tr$ will denote traces over $\so(4)$ or $\so(3,1)$ and $\su(2)$ indices, respectively.
The star $\star$ is the Hodge operator acting on bivectors
(or equivalently in the adjoint representation of the isometry algebra)
as $(\star B)^{IJ}=\hf\,{\eps^{IJ}}_{KL} B^{KL}$.
In the spin foam context we will use $\sigma$, $\tau$, $t$ to label four-simplices, tetrahedra and triangles
of a~triangulation~$\Delta$; $v$, $e$, $f$ for vertices, edges and faces of the dual triangulation~$\Delta^*$;
and~$n$,~$\ell$ for nodes and links of the boundary graphs $\Gamma$.
For the quantities which have a~meaning for both, say~$\Delta^*$ and $\Gamma$, the indices may be used interchangeably.
For example, the holonomy associated to a face~$f$ bounding the link~$\ell$, can be denoted either as~$g_f$ or~$g_\ell$.
Note also that the intertwiners are considered in this review as generic tensors in the tensor product of representations,
{\it not necessarily invariant}. Instead, invariant tensors are usually specif\/ied explicitly
as {\it invariant intertwiners}.
Finally, we will assume that the $d$-dimensional spacetime
manifold $\CM$ is topologically $\Sigma\times\mathbb{R}$,
where $\Sigma$ is a $(d-1)$-dimensional manifold without boundaries.

\section{Spin foams -- general concepts}
\label{sec_SFgen}

In this introductory section, we expose the main concepts behind the spin foam approach to quantum gravity.
We f\/irst give the def\/inition of a spin foam and a spin foam model, then explain the qualitative
relations between the spin foam and canonical approaches, and f\/inally present the general strategy
used to derive spin foam models in three and four spacetime dimensions.

\subsection{Structure of spin foam models}
\label{subsec_concepts}

A spin foam $\mathcal{F}=(\Delta^*,\l_f,\CI_e)$ is def\/ined as a two-dimensional
cellular complex\footnote{A two-dimensional cellular complex is a combinatorial
structure consisting of faces $f$ meeting at edges $e$, in turn
meeting at vertices~$v$.} $\Delta^*$ colored with group representations $\l_f$
on its faces and intertwiners $\CI_e$ on its edges. One can think of this coloring as some
kinematical information inherited from the boundary states, which are one-dimensional
cellular complexes (graphs) obtained by intersecting the spin foam with codimension-one surfaces.
These graphs are colored with representations $\l_\ell$ on their links and with intertwiners
$\CI_n$ on their nodes. If we think of them as encoding quantum spatial geometry,
the analogy allows us to see a spin foam as a representation of quantum spacetime.
In this way, the sum over dif\/ferent spin foams can be viewed as
a discrete path integral over possible geometries interpolating
between given boundary information.
Note that in this picture the dynamics is encoded at the vertices of the spin foam,
which therefore carry no labels.
\begin{figure}[t]
\centering
\includegraphics[scale=0.44]{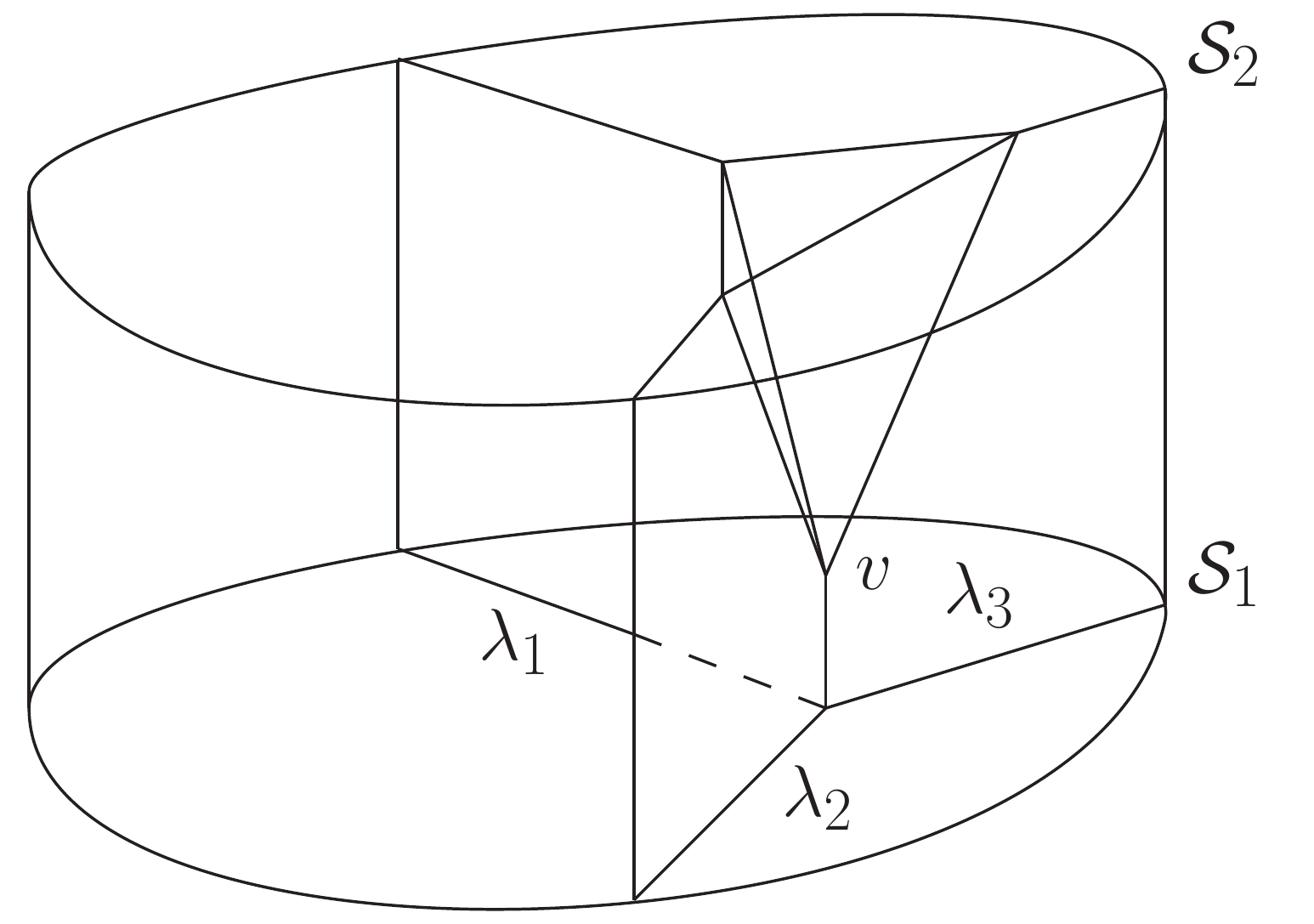}
\caption{A spin foam def\/ined on a two-complex interpolating between two boundary
states~$\CS_1$ and~$\CS_2$. A section def\/ines an intermediate state in the middle of the evolution.}
\label{spin foam}
\end{figure}

A spin foam model is def\/ined by an assignment of complex amplitudes $A_f(\l_f)$,
$A_e(\l_{f\supset e},\CI_e)$ and $A_v(\l_{f\supset v},\CI_{e\supset v})$
to the faces, edges and vertices of $\mathcal{F}$ respectively.
These amplitudes are local in the sense that $A_f$ depends only on
the representation coloring the face $f$, $A_e$ on the
representations coloring the faces $f\supset e$ meeting at $e$ and the
intertwiner coloring $e$, and $A_v$ on the representations and
intertwiners carried by the faces $f\supset v$ and edges $e\supset v$ meeting at the vertex
$v$. Given two boundary states $\CS_1$ and
$\CS_2$, and a two-complex $\Delta^*$ interpolating between their graphs, one can def\/ine the transition amplitude
\begin{gather}
\label{SF amplitude}
W(\Delta^*)=\langle\CS_1|\CS_2\rangle_{\Delta^*}\equiv
\int\de\mu_{\{\l_f\}}\int\de\mu_{\{\CI_e\}}
\prod_fA_f(\l_f)\prod_eA_e(\l_{f},\CI_e)\prod_vA_v(\l_{f},\CI_{e}).
\end{gather}
For the sake of generality, we have written this amplitude with arbitrary integration
measures, $\de\mu_{\{\l_f\}}$ and $\de\mu_{\{\CI_e\}}$, depending on the group-theoretic
data coloring the two-complex. If the spin foam model is built with a compact Lie group,
these continuous integrals become discrete sums over all possible colorings
with representations and intertwiners of all the internal faces and edges of the two-complex.

Two comments are in order. First of all, being similar to quantum amplitudes associated
with Feynman diagrams in quantum f\/ield theory, the spin foam amplitude \eqref{SF amplitude}
may diverge. Given this analogy, it is natural to ask the following question: if the amplitude does diverge,
can we understand the nature of this divergence and f\/ind a procedure to regularize it?
The answer will turn out to crucially depend on the dimension of spacetime
and on the particular model under consideration.

Second, the amplitude (\ref{SF amplitude}) represents a contribution of only one particular two-complex.
Recalling again the analogy with Feynman diagrams, it is natural to expect
that the full transition amplitude should be given by a sum over
cellular complexes interpolating between the boundary graphs.
The same expectation is suggested by the relation to canonical quantization presented in the next subsection.
However, in the important case of topological theories, it turns out that the contribution of one spin foam is suf\/f\/icient.
In particular, we are going to see that this is the case for general relativity in
three dimensions, where the amplitude (\ref{SF amplitude}) becomes independent
of the choice of cellular decomposition. This was in fact the original motivation
for introducing spin foam models from the mathematical point of view, since
it enables one to compute topological invariants of three-manifolds.

Now, in order to understand the geometric meaning of a spin foam,
let us consider a simplicial decomposition $\Delta$ of the four-dimensional spacetime manifold $\CM$.
Every such triangulation appears to be dual to a cellular two-complex
$\Delta^*$ whose vertices are all f\/ive-valent. Indeed, the dual two-complex can be constructed in the following way.
To each four-simplex of $\Delta$, we assign a point in its interior, which
corresponds to a vertex $v$ of $\Delta^*$. Then, to each of the f\/ive
tetrahedra corresponding to the boundary of this four-simplex, we
assign an edge $e\in\Delta^*$ which is connected to the vertex $v$.
Finally, to each of the four triangles corresponding to the boundary of a
tetrahedron, we assign a face $f$ in $\Delta^*$ whose boundary is
the edge $e$. Spin foams with vertices of valence higher than f\/ive (or those whose boundary graphs
have nodes of valence higher than four) can be obtained from discretizations which are more general
than the simplicial ones considered above. The spin foam represented in Fig. \ref{spin foam}
is def\/ined on a simplicial triangulation of a three-dimensional manifold.

Given this duality between spin foams and triangulations,
one can view a spin foam model as a proposal for a
regularized version of the path integral for gravity. Namely, given a manifold~$\CM$
with boundaries $\Sigma_1$ and $\Sigma_2$, and a dif\/feomorphism class of metrics $q_1$
on $\Sigma_1$ and $q_2$ on $\Sigma_2$, the spin foam model is supposed to give a meaning to
the transition amplitude
\begin{gather*}
\langle(\Sigma_1,q_1)|(\Sigma_2,q_2)\rangle_\text{phys}
=\int_{g|_\Sigma=q}\pD{g}\,\exp(\I S),
\end{gather*}
where the left-hand side denotes the physical scalar product in the yet to be def\/ined
physical Hilbert space of the theory. The integration is performed
over the dif\/feomorphism class of metrics~$g$ on~$\CM$
which agree with the metrics~$q_1$ and~$q_2$ on the boundaries~$\Sigma_1$ and~$\Sigma_2$, respectively. The idea is therefore to
give a meaning to the physical inner product between boundary states of quantum geometry.

\subsection{Relation to canonical quantization}

We saw that the boundary states of spin foam models are represented by graphs
colored with representations and intertwiners of a certain group. Such objects are known
as spin networks and generically appear as the so-called kinematical states
in the canonical loop quantization of f\/irst order gravity\footnote{Here we discuss only the general structure.
The precise relationship between groups, representations and intertwiners coloring the graphs in the canonical
and spin foam approaches will be discussed in the following sections.}.
These states solve the constraints generating gauge transformations in the tangent space
and spatial dif\/feomorphisms. However, it remains to solve the so-called Hamiltonian
constraint which implements the dynamics of the theory. This can be done by mapping the kinematical states into
the kernel of the quantum Hamiltonian operator~$\hat{H}$.
As was f\/irst realized in~\cite{Reisenberger:1996pu}, this directly leads, at least formally, to the spin foam picture.
We recall here the main steps of this construction.

The idea of Reisenberger and Rovelli is to construct the projection
operator using the path integral representation of the $\delta$-distribution,
\begin{gather}
\label{projection}
\CP_H=\prod_{x\in\Sigma}\delta\big(\hat{H}(x)\big)
=\int\pD{N}\,\exp\big(\I\hat{H}(N)\big),
\end{gather}
where
\begin{gather*}
\hat{H}(N)=\int_\Sigma \de^3x\,N(x)\hat{H}(x).
\end{gather*}
If we forget for the moment about the mathematical obstructions to
def\/ining such an operator, we can think of this formula as a group
averaging map which enables to def\/ine the physical inner product as
\begin{gather*}
\langle\CS_1|\CS_2\rangle_\text{phys}
\equiv\langle\CS_1|\CP_H\CS_2\rangle_\text{kin},
\end{gather*}
where $\CS_1$ and $\CS_2$ are two kinematical states. Using an expansion of the exponential in
(\ref{projection}), this inner product becomes
\begin{gather}
\label{series}
\langle\CS_1|\CP_H\CS_2\rangle_\text{kin}
=\int \pD{N}\,\langl\CS_1\left|\sum_{n=0}^\infty\f{\I^n}{n!}
\big(\hat{H}(N)\big)^n\right|\CS_2\rangl_\text{kin},
\end{gather}
where the right-hand side is def\/ined as
$n$ actions of the Hamiltonian constraint on the state
$\CS_2$. A~key property of the Hamiltonian constraint is
that it acts only at the nodes of a graph, creating new
links and nodes, and thereby changing the geometry encoded in the state.
Therefore, we can rewrite~(\ref{series}) in the following form
\begin{gather*}
\langle\CS_1|\CP_H\CS_2\rangle_\text{kin}
=\sum_{n=0}^{\infty}\f{\I^n}{n!}
\sum_{\mathcal{F}_n\,:\,\CS_1\rightarrow\CS_2}\prod_vA_v(\l_{f},\CI_{e}),
\end{gather*}
where the (divergent) integral over the lapse function was formally included in the def\/inition
of the amplitude~$A_v$. In this formula, the second sum runs over the
spin foams~$\mathcal{F}_n$ with $n$ vertices interpolating between
the two boundary states. The contribution of each~$\mathcal{F}_n$ corresponds to~$n$ actions of the Hamiltonian
constraint on these states, each creating a~vertex~$v$ and producing a~factor given by~$A_v$.

Given this construction, it is clear that the kinematical structure
of canonical quantum gravity, together with a suitably
regularized Hamiltonian constraint, should lead to a certain spin
foam model. However, the dif\/f\/iculties in dealing with the quantum
Hamiltonian constraint (see \cite{Bonzom:2011jv} for a recent overview) have forced people to address the problem
from the other side. Namely, one usually starts by constructing a
spin foam model along the lines explained in the following
subsection, and then tries to compare it with the discrete
Hamiltonian dynamics built from the canonical theory. Thus, on a general level,
the canonical and covariant approaches should be considered as two completely dif\/ferent paths.
On the other hand, if at some point the results obtained in the spin foam context and in the canonical quantization
do converge in the same direction, this might be considered as a non-trivial consistency check of
the two approaches.

\subsection{Strategy}
\label{subsubsec_strategy}

Spin foam models for quantum gravity were originally introduced in
three spacetime dimensions. In this case, pure gravity is a
topological f\/ield theory, i.e.\ it has no propagating local degrees
of freedom (no gravitons). It is one of the reasons why it is possible to discretize
the theory in a~way which does not depend on
the choice of discretization. Because of this crucial property, the
amplitudes (\ref{SF amplitude}) do not depend on $\Delta^*$, and the
sum over dif\/ferent two-complexes (for a f\/ixed topology) becomes irrelevant\footnote{However,
the question of summing over topologies is very interesting but still
unsolved. The group f\/ield theory formulation of spin foam models of\/fers a nice
framework to study this question~\cite{Freidel:2005qe,Oriti:2009wn}.}.
The f\/irst two models for quantum
Riemannian general relativity which were built along these lines are
the Ponzano--Regge model \cite{Ponzano:1968dq} (which has a vanishing cosmological
constant) and the Turaev--Viro model~\cite{Turaev:1992hq} (which has a positive
cosmological constant). We will review this construction in the next
section.

In four spacetime dimensions, the situation is drastically
dif\/ferent, since the theory is not topological anymore. However, it
is possible to exploit the fact that, in any dimension, gravity can
be written as a constrained topological f\/ield theory. The
topological theory itself, known as BF~theory and introduced for the
f\/irst time in \cite{Horowitz:1989ng}, is built using an
$\so(4)$- or $\so(3,1)$-valued $(d-2)$-form $B$ and
the curvature two-form $F$ of a connection $\omega$. Its action is
given by
\begin{gather}
\label{BF action}
S_\text{BF}[\omega,B]=\int_\CM\Tr(B\wedge F).
\end{gather}
The equations of motion, $DB=0$ (where $D$ is the covariant dif\/ferential associated to $\omega$)
and $F=0$, imply that all the solutions are locally pure gauge.
General relativity (in dimensions $d>3$) is obtained from BF theory
by requiring that the $B$ f\/ield is not arbitrary, but rather constructed from {\it vierbein} one-forms $e^I$.
In four dimensions this implies that it is constrained to
satisfy the relation
\begin{gather*}
B^{IJ}=\star\(e^I\wedge e^J\),
\end{gather*}
in which case (\ref{BF action}) becomes the usual
Hilbert--Palatini action. To achieve this, one can modify the BF
action to obtain the Plebanski action~\cite{Capovilla:1991qb,DePietri:1998mb, Plebanski:1977zz}
\begin{gather*}
S_\text{Pl}[\omega,B,\Phi]
=S_\text{BF}[\omega,B]+\int_\CM\Tr\big(\Phi\cdot B\wedge B\big).
\end{gather*}
The f\/ield $\Phi$ has 20
independent components generating constraints which ensure that the
$B$ f\/ield comes from the wedge product of two tetrad f\/ields. These
are the so-called simplicity constraints, reducing the topological
f\/ield theory to general relativity (plus additional sectors that we will discuss later on).

Most of the spin foam models for four-dimensional quantum gravity
rely on the Plebanski formulation and use the same strategy, but translated to the quantum level.
Namely, instead of quantizing the phase space of the full Plebanski action, obtained
by imposing the simplicity constraints at the classical level,
one f\/irst discretizes the classical theory on a cellular complex,
quantizes its topological part given by the discrete BF action,
and f\/inally imposes the simplicity constraints at the quantum level.
The last step is implemented via certain conditions on the group-theoretic data associated to the spin foam
quantization of BF theory, and is supposed to convert the trivial dynamics of a topological spin foam model
into the dynamics of quantum general relativity.

\section{Three-dimensional quantum gravity}
\label{sec_3d}

Three-dimensional classical general relativity
admits two  equivalent f\/irst order formulations. In the BF formulation,
the dynamical variables are a triad~$e$ and an $\hmat$-valued spin connection~$\omega$,
where the Lie algebra $\hmat$ of the gauge group is $\su(2)$ in Riemannian signature
and $\su(1,1)$ in the Lorentzian case.
In the Chern--Simons formulation, the dynamical variable is a $\gmat$-valued connection~$A$, where $\gmat$ is a Lie algebra which is totally and uniquely determined by
the signature of the spacetime and the sign of the cosmological
constant (which will be denoted by $\Lambda$ in what follows)
according to the following table:
\begin{center}
\begin{tabular}{c|c|c|c}
 & $\Lambda<0$ &  $\Lambda=0$ &  $\Lambda>0$
\\[2pt]
\hline
\rule{0pt}{15pt}
$\sigma=1$ & $\so(3,1)$ & $\isu(2)$ & $\so(4)$
\\[3pt]
\hline
\rule{0pt}{15pt}
$\sigma=-1$ & $\so(2,2)$ & $\isu(1,1)$ & $\so(3,1)$
\end{tabular}\label{page-tablegroups}
\end{center}

In spite of the apparent dif\/ferences, the BF and Chern--Simons formulations
are equivalent at the classical level.
More precisely, the Chern--Simons connection $A$ appears as a linear
combination of the triad one-form $e$ and the spin connection $\omega$.
Furthermore, the gauge algebra of the Chern--Simons theory is nothing but the symmetry
algebra of the associated BF formulation. This comes from the fact that a three-dimensional BF theory,
associated to a gauge algebra $\hmat$, admits a larger symmetry algebra whose nature depends
on the sign of the cosmological constant according to the classif\/ication that we have given above.
In all cases, $\gmat$ contains $\hmat$ as a~subalgebra and the remaining symmetries describe translations
in the coset space~$G/H$, where~$G$ and~$H$ are universal coverings of the Lie groups
associated to $\gmat$ and $\hmat$, respectively.

\looseness=-1
As we have already argued, the BF and Chern--Simons formulations posses the same space of classical solutions.
However, in spite of this
equivalence at the classical level, the two theories naturally lead
to dif\/ferent quantization schemes for three-dimensional gravity.
If it is clear that both formulations
can be quantized canonically and covariantly, it turns out that the covariant
quantization is much simpler with the BF formulation, whereas Chern--Simons theory
of\/fers a very powerful framework for the canonical
quantization (regardless of the sign of the cosmological constant).
To make this point more concrete, one can develop
an analogy with the quantization of the harmonic oscillator in quantum
mechanics. Indeed, the harmonic oscillator admits
two well-known f\/irst order formulations. In the f\/irst one, the dynamical variables
are the position $\mathtt{q}$ and the momentum $\mathtt{p}$, whereas
in the second one the dynamical variables are the chiral variables
$a$ and $a^\dagger$, which are in turn complex
linear combinations of $\mathtt{q}$ and $\mathtt{p}$. The position and momentum variables
are clearly more suited for a covariant quantization,
whereas it is much simpler and much more powerful
to use the chiral ones for the canonical quantization. Thus, $e$~and~$\omega$
can be seen as being equivalent to $\mathtt{q}$ and $\mathtt{p}$,
whereas the Chern--Simons connection can be interpreted
as being analogous to $a$. As a consequence,
establishing a direct link between the covariant and canonical
quantizations is {\it a priori} a non-trivial task. Nevertheless,
this has been achieved for the Chern--Simons formulation of
three-dimensional gravity, and led to amazing and deep relationships
between topological invariants, knot invariants, conformal
f\/ield theories and quantum gravity. In particular, these discoveries have opened a new
way towards the understanding of knot polynomials.
On the other hand, the link between the covariant and canonical quantizations of
BF theory leads to a clear understanding of the relation between LQG and
spin foam models in three dimensions (at least in the case of a vanishing cosmological constant).

\looseness=-1
We are going to present these aspects in much more details in this section.
The f\/irst subsection is devoted to the classical theory.
We will present the two f\/irst order formulations mentioned above,
perform their canonical analysis and discuss the inclusion
of point particles. In the second subsection, we will focus on
the canonical quantization of Chern--Simons theory using the so-called
combinatorial quantization scheme. This method is very powerful
and allows to understand clearly how quantum groups appear in
three-dimensional quantum gravity.
Then, we will review the covariant quantization \`a la Witten and explain how
it provides a bridge between topological and knot invariants on one side,
and Riemannian quantum gravity with a positive
cosmological constant on the other side. The third subsection is devoted
to the quantization of BF theory. We will mainly concentrate on the case
where the gauge group is $G=\SU(2)$ and where there is no cosmological constant.
In this framework, we will present the canonical loop quantization
and derive the spin foam representation as a realization of the Reisenberger--Rovelli path integral.
This will enable us to construct the physical scalar product
and to show the relationship between the canonical quantum theory
and the Ponzano--Regge spin foam model. We will present some observations
about the theory in the presence of a positive cosmological constant.

\subsection{Classical theory}
\label{subsec_class3d}

In the standard second order formulation, gravity is a theory of the metric
$g_{\mu\nu}$ of the spacetime manifold $\cal M$ and
its dynamics is governed by the Einstein--Hilbert action
\begin{gather}\label{EH 3D}
S_\text{EH}[g]=\frac{1}{8\pi G_\text{N}}\int_{\cal M}\de^3x\,\sqrt{|g|} (R-2\Lambda),
\end{gather}
where $\de^3x\,\sqrt{|g|}$ is the volume form on $\cal M$, $\Lambda$ is the cosmological constant,
$R$ is the scalar curvature, and $G_\text{N}$ is the three-dimensional Newton constant.
As in four dimensions, the resulting Euler--Lagrange equations
of motion are the well-known Einstein equations
\begin{gather}\label{E equations}
R_{\mu\nu}-\frac{1}{2} g_{\mu\nu}R=-\Lambda g_{\mu\nu}.
\end{gather}
However, the solutions are very dif\/ferent from the solutions in the four-dimensional case. Indeed, in three dimensions,
the components of the Riemann tensor are linear combinations of the components of the Ricci tensor, and therefore
the Weyl tensor vanishes identically. But the Weyl tensor is responsible for the presence of local degrees of freedom,
and in particular gravitational waves in four dimensions. As a consequence, there are no gravitational waves
in three dimensions, there are no local degrees of freedom, all solutions are locally of constant curvature, and the only non-trivial
degrees of freedom are topological. In other words, the spacetime manifold has to be topologically non-trivial for the theory to admit
non-trivial and physically interesting solutions.

The absence of local degrees of freedom gives a priori
the feeling that three-dimensional gravity is too trivial
to be interesting in itself, and cannot be a good toy model
to test ideas and principles of classical and quantum gravity in
four dimensions. But this viewpoint is too naive, f\/irstly because
the quantization of three-dimensional gravity is far from being trivial,
and in fact turns out to be mathematically extremely rich,
and also because it has proven to be a very insightful
toy model to understand important aspects of quantum gravity in four dimensions~\cite{MR1637718}.
For instance, this is the context in which spin foam models have originally been
introduced.

\looseness=-1
In this subsection we perform
the classical analysis of the two f\/irst order formulations of three-dimensional
gravity mentioned previously, which we take as a starting point to explain the main quantization schemes.
In a f\/irst part, we will present the BF formulation, and in a second part the Chern--Simons formulation.
Then, we will study the inclusion of point-like particles,
which are crucial ingredients to understand the link between
quantum gravity and knot invariants like the Jones polynomial~\cite{MR830613}.
In the last part, we will perform the canonical analysis and construct
the classical physical phase space in both formulations, with and without particles.

\subsubsection{The BF formulation}

In the BF formulation of (Riemannian) three-dimensional gravity, the
basic f\/ield variables are a triad $e$ and a spin connection $\omega$.
Both f\/ields can be viewed as $\su(2)$-valued one-forms in the sense that
$e=e^i_\mu\tau_idx^\mu$ and $\omega=\omega^i_\mu\tau_idx^\mu$, where
$\{\tau_i\}$ forms a basis of $\su(2)$ def\/ined by the commutation relations $[\tau_i,\tau_j]=\eps_{ij}{}^k\tau_k$,
with $\eps_{ijk}$ the totally antisymmetric tensor such that $\eps_{123}=+1$.
Indices are raised and lowered with the f\/lat metric $\delta_{ij}$. In the spinorial (two-dimensional) representation,
the $\su(2)$ generators are obtained from the Pauli matrices $\{\sigma_i\}$ via $\tau_i=-\I\sigma_i/2$, and in the
rest of this work we will often identify the abstract generators $\tau_i$ with their two-dimensional representation.

In the BF formulation, the triad variables are related to the spacetime metric by the relation
$g_{\mu\nu}=e_\mu^i \delta_{ij} e_\nu^j$, and the spin connection is taken to be an independent f\/ield.
When expressed in terms of these variables, the Einstein--Hilbert action reduces to the topological BF action
\begin{gather}
\label{3d action}
S_{\rm 3d}[e,\omega]=\frac{1}{4\pi G_\text{N}}\int_\CM\left(\tr(e\wedge F[\omega])
- \frac{\Lambda}{6}\tr(e\wedge e \wedge e)\right),
\end{gather}
where $\tr$ denotes the trace in the spinorial representation of $\su(2)$
and $F=D\omega=\de\omega+\omega\wedge\omega$ is the curvature two-form corresponding to the connection $\omega$.
Varying  this action with respect to the triad and the spin connection leads to the following two equations:
\begin{subequations}\label{flatcond}
\begin{gather}
 F_{\mu \nu}[\omega]\equiv \partial_\mu\omega_\nu - \partial_\nu \omega_\mu +[\omega_\mu, \omega_\nu] =
-\Lambda [e_\mu,e_\nu],\\
 T_{\mu \nu}[e,\omega] \equiv \partial_\mu e_\nu - \partial_\nu e_\mu  + 2[\omega_\mu, e_\nu] = 0.
\end{gather}
\end{subequations}
The second equation is the torsion-free condition and is equivalent to the requirement that the triad
$e$ be covariantly constant with respect to $\omega$. When the triad (viewed as a square
three-dimensional matrix) is invertible, the torsion-free equation
implies that~$\omega$ is the unique connection compatible
with the triad which can be expressed explicitly in terms of~$e$.
Substituting the expression~$\omega(e)$ into the BF action~(\ref{3d action}), one ends up with the initial
second order Einstein--Hilbert action~(\ref{EH 3D}). This establishes the
equivalence between the BF formulation and the second order formulation of three-dimensional gravity,
when the triad is assumed to be invertible.

Note that some of the classical equations \eqref{flatcond} do not involve time derivatives
and therefore should be considered as constraints on the dynamical variables.
As we are going to see in the Hamiltonian framework, these constraints are
f\/irst class and generate the gauge symmetries of the theory.
Understanding these symmetries is essential in order to describe properly the physical solutions.
However, we postpone the canonical analysis of BF theory after the introduction and
the study of the Chern--Simons formulation of gravity for the sake of clarity and simplicity.

\subsubsection{The Chern--Simons formulation}

To go from the BF formulation to the Chern--Simons formulation of three-dimensional
gravity, one combines the triad and the spin connection into a single gauge f\/ield: the Chern--Simons connection~$A$.
To see how this works explicitly, it is necessary to introduce a family of Lie algebras
$\gmat_\Lambda$ labeled by the cosmological constant~$\Lambda$.
$\gmat_\Lambda$ is an extension of the algebra $\hmat=\su(2)$, in the sense that it is generated by
the three generators of $\su(2)$, denoted $\{\tau_i\}$, supplemented by three extra generators
$\{P_i\}$. They satisfy the following commutation relations:
\begin{gather}\label{poincbrack}
[\tau_i,\tau_j]=\eps_{ij}{}^k\tau_k,\qquad
[P_i,P_j]=-\sgn(\Lambda)\eps_{ij}{}^k \tau_k,\qquad
[\tau_i,P_j]=\eps_{ij}{}^k P_k,
\end{gather}
where $\sgn(x)$ is the sign function def\/ined such that $\sgn(0)=0$. These relations def\/ine three
dif\/ferent algebras characterized by the sign of the cosmological constant and given
in the f\/irst line of the table on p.~\pageref{page-tablegroups}.
These algebras admit a Killing form def\/ined by the pairing
\begin{gather}\label{sblf}
\langle\cdot\,,\cdot\rangle  : \  \gmat_\Lambda\times \gmat_\Lambda \longrightarrow \Rmat,
\qquad
\langle P_i,\tau_j\rangle=\delta_{ij},
\end{gather}
all the other pairings being trivial\footnote{Notice that we use the notation $\Tr$ and $\tr$ exclusively for the
bilinear pairing of $\so(4)$ or $\su(2)$ elements respectively.}.

\looseness=-1
These algebras have an immediate geometric
interpretation as the isometric algebras of the three-dimensional spaces of constant scalar curvature.
More precisely, $\isu(2)$ is the isometry algebra of the three-dimensional Euclidean space $\mathbb E^3$
(of vanishing scalar curvature), $\so(4)$ is the isometry algebra
of the three-dimensional sphere $\mathbb S^3$ (of positive scalar curvature),
and $\so(3,1)$ is the isometry algebra of the three-dimensional
hyperbolic space $\mathbb H^3$ (of negative scalar curvature).
Locally, all the solutions to the Einstein equations~(\ref{E equations})
look like one of these three spaces, depending on the
sign of the cosmological constant.
It is therefore not a surprise to see that the corresponding  isometry algebras will play
an important role in three-dimensional gravity.
This role is particularly important and explicit in the Chern--Simons
formulation. From now on, in order to simplify the notations, we will omit the index~$\Lambda$
in the Lie algebras~$\gmat_\Lambda$ when there is no ambiguity.

The Chern--Simons connection is def\/ined by
\begin{gather*}
A=  \omega^i\tau_i+\f{1}{\ell_c}e^iP_i,
\end{gather*}
where the cosmological length $\ell_c$ is a function of the cosmological constant $\Lambda$ given by
$\ell_c(0)=1$ and $\ell_c(\Lambda)=|\Lambda|^{-1/2}$. The gauge f\/ield $A$ is a one-form with values in the Lie algebra
$\gmat$. It is shown in \cite{MR974271} that the f\/irst order action for three-dimensional gravity
can then be rewritten as a Chern--Simons  action
\begin{gather}\label{CS action}
S_\text{CS}[A(e,\omega)]=\frac{k}{4\pi}\int_{\cal M}\de^3x \, \eps^{\mu \nu \rho}
\left(\langle A_\mu, \partial_\nu A_\rho \rangle + \frac{1}{3} \langle A_\mu, [A_\nu, A_\rho] \rangle\right),
\end{gather}
where $\langle\cdot\,,\cdot\rangle$ is the  bilinear form
(\ref{sblf}) and $k=\ell_cG_\text{N}^{-1}$ is called the level of the Chern--Simons
action. Using (\ref{poincbrack}),
it is easy to check that this action is indeed equivalent
to~(\ref{3d action})  up to a~boundary term for $\partial\cal M\neq \emptyset$,
which does not af\/fect the equations of motion.

Before going any further, let us discuss in more detail the change
of variables to go from the BF variables $e$ and $\omega$ to the
Chern--Simons gauge f\/ield $A$. As we explained in the introduction
to this section, this is not a simple change of variables, but rather
a kind of ``complexif\/ication'' of the initial f\/ields.
Indeed, everything works as if we had extended
the f\/ield (in the mathematical sense) of rotations to
a ``complex'' f\/ield which involves the rotations but also the translations.
In other words, we have doubled the number of dimensions of the initial f\/ield.
Then, we have combined these two types of variables, $e$~and~$\omega$, into a single ``complexif\/ied'' connection~$A$
whose rotational part is the spin connection, and whose translational part
(in the complex direction) is the triad.
In this sense, $A$ is analogous to the chiral variable $a$
for the harmonic oscillator, this later being related to the phase space variables $\mathtt{q}$ and $\mathtt{p}$ by the
complexif\/ication $a=\mathtt{q}+\I\mathtt{p}$.

Varying  the Chern--Simons action (\ref{CS action}) with respect to the gauge f\/ield
results in a f\/latness condition on the $\gmat$-valued Chern--Simons connection $A$.
As expected, this f\/latness condition on~$A$ combines the torsion-free and constant curvature
conditions for the spin connection $\omega$:
\begin{gather*}
F_{\mu\nu}[A] =0\quad\Longleftrightarrow\quad
\begin{cases}
  F_{\mu \nu}[\omega] =-\Lambda[e_\mu,e_\nu],
 \\
  T_{\mu \nu}[e,\omega]= 0.
\end{cases}
\end{gather*}
The consequences of this unif\/ication of the variables $e$ and $\omega$ into a single gauge f\/ield $A$
are manifold. First, the Euler--Lagrange equations are encoded into a single, simple and beautiful equation, which
is the requirement of f\/latness for the connection $A$. Second, the gauge symmetry group is now appearing explicitly,
whereas in the BF formulation a part of these symmetries was somehow hidden.
Third, it makes obvious that the theory admits no local degrees of freedom since
it is always possible to f\/ind a gauge transformation that sends locally every f\/lat connection to
the trivial connection.

As the Chern--Simons formulation of three-dimensional gravity is a gauge theory
with local symmetry group~$G$, its action admits an inf\/inite-dimensional symmetry group given by
${\cal G}=C^\infty({\cal M},G)$. This is the group of $G$-valued smooth functions on $\cal M$, and its action
on the connection reads as
\begin{gather}\label{gauge symmetry}
\forall \, g \in {\cal G}  \qquad A\longmapsto A^g  = g^{-1}Ag + g^{-1}\de g.
\end{gather}
The invariance of the Chern--Simons action (\ref{CS action}) with respect to the inf\/initesimal version of
the transformations (\ref{gauge symmetry}) is an immediate consequence of the  invariance of the bilinear
form $\langle\cdot\,,\cdot\rangle$ under the adjoint action of $G$.
To see how these transformations encode dif\/feomorphisms, let us consider an inf\/initesimal version
of the transformation \eqref{gauge symmetry} generated by $\phi\in\gmat$.
If we decompose the Lie algebra element $\phi$ into its translational and rotational parts according to
$\phi=u^iP_i+v^i\tau_i$, we can use the commutation relations~(\ref{poincbrack}) to show that~$e$ and~$\omega$
transform according to the following rules:
\begin{gather*}
\delta e=  \ell_c (\de u  + [\omega,u]) + [e,v],\qquad
\delta \omega =  \de v + [\omega,v]+ \frac{1}{\ell_c}[e,u].
\end{gather*}
Setting  $u=\xi^\mu e_\mu$ and $v=\xi^\mu\omega_\mu$, we can then express these
transformations in terms of the Lie derivatives ${\cal L}_\xi$
along the vector f\/ield $\xi=\xi^\mu \partial_\mu$. This leads to
\begin{gather}
\delta e_\mu =  {\cal L}_\xi e_\mu+ \xi^\nu T_{\mu \nu}[e,\omega],\qquad
\delta \omega_\mu  =  {\cal L}_\xi \omega_\mu  + \xi^\nu (F_{\mu \nu}[\omega] +\Lambda [e_\mu,e_\nu]).\label{infinitesimal symmetries}
\end{gather}
As a result, when the gauge f\/ields satisfy
the equations of motion, the inf\/initesimal gauge transformations~(\ref{infinitesimal symmetries}) are in fact given by dif\/feomorphisms,
which establishes the on-shell equivalence between inf\/initesimal
dif\/feomorphisms and inf\/initesimal Chern--Simons gauge transformations.
However, this equivalence applies only to
gauge transformations and dif\/feomorphisms that are connected to the identity,
whereas the status of large (i.e.\ not inf\/initesimally generated) gauge transformations
is more subtle~\cite{MR1335350,Meusburger:2003wk}.

In fact, the Chern--Simons action is not invariant under the large gauge transformations~(\ref{gauge symmetry}).
Instead, one has the relation
\begin{gather}
S_\text{CS}[A^g]-S_\text{CS}[A]\equiv w[{\cal M},g]=
\frac{k}{6\pi} \int_{\cal M} \langle g^{-1}\de g \wedge g^{-1}\de g  \wedge  g^{-1}\de g \rangle.
\label{anomaly}
\end{gather}
The quantity on the right hand side, which is known as the Wess--Zumino--Witten term, is a~priori not null.
It is a topological invariant of the principal $G$-bundle
over $\cal M$, and as such it is a~constant which does not modify the classical Euler--Lagrange equations.
In this sense, it is fair to claim that the Chern--Simons action is invariant under gauge transformations.
Note that when the gauge group is compact, it is possible to show that
the topological invariant is $w[{\cal M},g]=2\pi N k$,
where $N$ is an integer and $k$ is the level of the Chern--Simons action.

\subsubsection{Inclusion of particles and Wilson loops}

\looseness=-1
In three-dimensional spacetime, there is no gravitational
interaction between massive particles. Indeed, from a classical
(non-relativistic) point of view, it is possible to show that massive
particles do not create any gravitational f\/ield in three dimensions.
From the  point of view of general relativity, particles do not
deform the spacetime manifold. Locally, in the vicinity of the
particle, spacetime remains f\/lat (if $\Lambda=0$) or of constant
positive (if $\Lambda>0$) or negative curvature (if $\Lambda<0$).
This property contrasts drastically with what happens in four dimensions,
where the matter f\/ields have in general a strong ef\/fect on the geometry of spacetime.
Therefore, at f\/irst sight, it might seem useless to study particles in three
dimensions since they have no ef\/fect.

However, it turns out that a particle in three dimensions has a non-local ef\/fect on
spacetime~\cite{MR748720,MR734213}. It does not modify the geometry but the topology of
spacetime. More precisely, at any time, a massive (non-spinning) particle creates
a conical singularity at the place where it is located, so that
space looks like a cone around the particle. The angle $\alpha$
of the cone is proportional to the mass $m$ of the particle in Planck units.
In the presence of several
particles, space looks like a multi-cone and its dynamics
is far from being trivial~\cite{Matschull:2001ec}. For instance, there is a close and deep
relationship between the properties of the dif\/fusion of particles
and the so-called exotic statistics~\cite{Bais:2002ye}. Furthermore, when the particles
possess a non-trivial spin, the topology becomes even more
complicated. In particular, time-helical structures appear
and lead to the presence of closed timelike curves~\cite{Gott:1990zr}.

These examples show that in the end the dynamics of point particles
in three dimensions is quite complicated, even if there are
just modif\/ications of spacetime topology, and not geometry.
Unfortunately, we cannot review all these features in the present
work. Instead, we are going to focus only on the aspects which are directly
related to our subject. The reason why we are interested in studying
particles in three-dimensional quantum gravity is that they are
intimately related to the Jones polynomials. To see this, we f\/irst need to understand
how to describe a~point particle coupled to gravity at the classical level.
This can be done using the so-called algebraic action.
Then the coupling to the gravitational f\/ield reduces
to gauging the global symmetries of this action with the Chern--Simons gauge f\/ield.

The algebraic formalism is adapted to describe
a free point particle evolving in any homogeneous three-dimensional spacetime.
In the Riemannian three-dimensional case, the homogeneous manifolds are
the Euclidean space $\mathbb E^3$, the sphere $\mathbb S^3$, and the hyperbolic space~$\mathbb H^3$.
They correspond to the maximally symmetric solutions to Einstein equations
with vanishing, positive or negative cosmological constant, respectively.
Each of these spaces can be obtained as the coset~$G/H$, where $H=\SU(2)$
and $G$ is the associated isometry group.
In particular, $G=\ISU(2)$ for~$\mathbb E^3$, $G=\Sp(4)$ for~$\mathbb S^3$ and $G=\SL(2,\mathbb{C})$ for
$\mathbb H^3$. As the action of $H$ is transitive, any element $\mathtt{g} \in G$ can be written
as the product\footnote{Note that this factorization is not necessarily unique.}
$\mathtt{g} =\mathtt{q}\mathtt{p}$ of an element $\mathtt{q}\in G/H$ with an element
$\mathtt{p}\in H$. This is summarized by the following map:
\begin{gather*}
G \ \longrightarrow \ G/H \times  H,
\qquad
\mathtt{g}=\mathtt{q}\mathtt{p} \ \longmapsto\ (\mathtt{q},\mathtt{p}).
\end{gather*}
In the algebraic formulation, we consider $\mathtt{g}\in G$ as a variable describing
the dynamics of the point particle evolving in the corresponding homogeneous space.
We identify $\mathtt{q}\in G/H$ with the position of the particle in spacetime, whereas the
element $\mathtt{p}\in H$ is viewed as its momentum.

In this context, relativistic particles at rest are classif\/ied
by the unitary irreducible representations of the isometry group $G$.
For any value of $\Lambda$, these representations are labeled
by a~couple $(m,s)$ of real parameters f\/ixing the values
of the two Casimir operators of (the enveloping algebra of)
$\gmat$ in the given representation:
\begin{gather*}
C_G^{(1)} =P^iP_i  -  \sgn(\Lambda) \tau^i\tau_i  = m^2-\sgn(\Lambda) s^2,
\qquad
C_G^{(2)} = P^i \tau_i = ms.
\end{gather*}
Even if the spacetime is taken to be Riemannian here,
we will interpret $m$ as the mass and $s$ as the spin of the particle.
When $G$ is compact, which corresponds to $G=\Sp(4)$
in the present case, the two parameters are
half-integers. If $G$ is the non-compact group $\ISU(2)$
or $\SL(2,\mathbb{C})$, the mass $m$ is any real number and the spin $s$ is a half-integer.

The algebraic action for a relativistic particle of mass $m$ and spin $s$ evolving in the homogeneous space
$G/H$ is given by
\begin{gather*}
S[\mathtt{g} ]=\int_\gamma\langle\chi(m,s),\mathtt{g}^{-1}\de\mathtt{g}\rangle,
\end{gather*}
where $\chi(m,s)=m\tau_1+sP_1\in\gmat$ characterizes the particle at rest and $\gamma$
denotes its world line. Such an action has been introduced in~\cite{Balachandran:1983pc}
and studied in the context of Chern--Simons theory in~\cite{MR1080700,MR1025431, Witten:1988hf}.
We refer to these articles for more details.
Here we will just recall the main results.
\begin{itemize}\itemsep=0pt
\item
The algebraic action is a f\/irst order action for the particle since
it involves both position~$\mathtt{q}$ and momentum $\mathtt{p}$ through the variable $\mathtt{g}$.

\item
The equations of motion are easy to obtain and are simply given by
\begin{gather}\label{equation free particle}
\left[\chi(m,s),\mathtt{g}^{-1}\frac{\de\mathtt{g}}{\de t}\right]= 0.
\end{gather}
The general solution is $\mathtt{g}(t)=\mathtt{g}_0\mu(t)$, where $\mathtt{g}_0$ is a constant
and $\mu(t)$ is any function in the maximal two-dimensional
Cartan torus of $G$ generated by $\exp(\tau_1)$ and $\exp(P_1)$.
These equations imply that $\mathtt{q}$  satisf\/ies the geodesic equations on $G/H$.

\item
The action is clearly invariant under the global left
action of the isometry group on itself, $\mathtt{g}\longmapsto{}^g\mathtt{g} =g\mathtt{g}$ for any
$g \in G$. Physically, this symmetry represents the invariance
of the theory under a change of inertial frame. It implies the
existence of Noether charges which are nothing but the momentum
$\text{p}_i$ of the particle and its total angular momentum~$\text{j}_i$.
These conserved quantities are shown to be Dirac observables.
They satisfy the Poisson algebra
\begin{gather*}
\{\text{j}_i,\text{j}_j\}=\eps_{ij}{}^k\text{j}_k,\qquad
\{\text{p}_i,\text{p}_j\}=-\sgn(\Lambda)\eps_{ij}{}^k \text{j}_k,\qquad
\{\text{j}_i,\text{p}_j\}=\eps_{ij}{}^k\text{p}_k,
\end{gather*}
as well as the quadratic Casimir relations
\begin{gather}\label{casimir particle}
\text{p}^2-\sgn(\Lambda)\text{j}^2 = m^2 - \sgn(\Lambda)s^2,\qquad
\text{p}\cdot \text{j} = ms.
\end{gather}
This is the reason why the algebraic action correctly captures the dynamics
of the free particle of mass $m$ and spin $s$ evolving in
the homogeneous space $G/H$.

\item
Besides, the action is invariant under the local two-dimensional (inf\/initesimal) gauge symmetry
$\mathtt{g}\longmapsto\mathtt{g}(\alpha\tau_1+\beta P_1)$.
This symmetry is closely related to the invariance of the action
under time reparametrizations. As a consequence, the particle is  determined at rest by its mass and
its spin, and dynamically by the group element $\mathtt{g}$ up to this local symmetry.
In other words, the conf\/iguration spaces for the particles in three dimensions are
in one to one correspondence with the coadjoint orbits
\begin{gather}\label{coadjoint}
{\cal C}_{m,s}\equiv\big\{\mathtt{g}=\gx\chi(m,s)\gx^{-1} \; | \; \gx \in G \big\} \subset \gmat
\end{gather}
of the Lie algebra $\gmat$. In this formulation, $\gx$
is the group element which sends the dynamical particle
characterized by $\mathtt{g}$ to its rest frame.
Furthermore, the coadjoint orbits of $\gmat$ are canonically endowed with a
symplectic structure given by the Kirillov--Poisson bracket which coincides with
the symplectic structure following from the Hamiltonian analysis of the algebraic action.
\end{itemize}

Coupling the relativistic particle to the gravitational f\/ield simply amounts to gauging the global
symmetry (i.e.\ the left action of the group $G$) of the free particle action by the Chern--Simons gauge f\/ield~$A$.
This is possible because the connection $A$ is valued in the Lie algebra $\gmat$ of the same group $G$.
As a consequence, the particle is minimally coupled to gravity and the action of the coupled system
reads
\begin{gather*}
S[\mathtt{g},A]=S_\text{CS}[A]
+\frac{k}{4\pi}\int_\gamma \langle\chi(m,s),\mathtt{g}^{-1}\de\mathtt{g}+\mathtt{g}^{-1}A\mathtt{g}\rangle.
\end{gather*}
Note that one can always add or absorb the factor $k/4\pi$
in front of the action by a simple redef\/inition of the
mass~$m$ and the spin~$s$.
It is also easy to rewrite this action in the BF formulation.
To this end, one replaces~$A$ by its expression in terms of
$e$ and $\omega$ and, up to a~boundary term, the action becomes
\begin{gather*}
S[\mathtt{g},e,\omega]= S_\text{BF}[e,\omega] +\f{k}{4\pi}\int_\gamma (e^i\text{j}_i + \omega^i\text{p}_i),
\end{gather*}
where $\mathtt{g}$ appears only in the def\/inition of the momentum $\text{p}^i$ and the total angular momentum $\text{j}^i$,
which are the Noether charges of the free particle satisfying the quadratic
Casimir relations~(\ref{casimir particle}) and def\/ining its mass $m$ and spin $s$.

The equations of motion for the degrees of freedom $\mathtt{g}$
of the particle are easily obtained by replacing the derivative $\de\mathtt{g}$ in~(\ref{equation free particle}) by the covariant derivative
$D\mathtt{g}=\de\mathtt{g}+A_t\mathtt{g}$, where $A_t$ is the component of the connection in the direction of~$\gamma$.
$A_t$ is not a dynamical variable of the theory in the sense that the coupled
action does not involve time derivatives of $A_t$. It is a Lagrange multiplier which can be set to zero with a gauge choice.
As a result, the particle coupled to gravity evolves as a free particle, which is consistent with the fact that
particles are not sensitive to the gravitational f\/ield in three-dimensional spacetimes.

\subsubsection{The classical phase space}

On manifolds of topology  ${\cal M}=\Sigma \times \Rmat$,
one can give a Hamiltonian formulation of the theory.
For simplicity, we f\/irst focus on the case where $\Sigma$
is an oriented two-surface of arbitrary genus. The case of a surface
with punctures representing massive spinning particles
is a straightforward generalization which will be discussed in a second time.
A detailed discussion can be found in~\cite{MR1637718} and in references therein.

\subsubsection*{Canonical analysis of pure gravity}

Decomposing the gauge f\/ield
$A=A_0\de t+A_a\de x^a$
into a time component $A_0$ and a gauge f\/ield $A_\Sigma=A_a\de x^a$ on the spatial surface,
we can rewrite
the Chern--Simons action (\ref{CS action})  as
\begin{gather*}
S_\text{CS}[A]= \f{k}{4\pi}\int_{\Rmat} \de t \int_\Sigma \de^2x \, \eps^{ab}
\big( - \langle A_a, \partial_0 A_b \rangle + \langle A_0, F_{ab}[A]\rangle \big),
\end{gather*}
where $\eps^{ab}\equiv\eps^{0ab}$.
This implies that the phase space coincides with the space of $G$-connections
on the surface $\Sigma$, which we denote by ${\cal A}(G,\Sigma)$.
The canonical Poisson bracket on this phase space is given by
\begin{gather}
\label{Poisson bracket}
\{A^a_{I}(x) , A^b_{J}(y)\} =\f{2\pi}{k}\eps^{ab} \,\langle \xi_I ,\xi_J \rangle\, \delta^2(x-y),
\end{gather}
where $\xi_I \in \{\tau_i,P_j\}_{i,j=0,1,2}$ are the generators of the Lie algebra $\gmat$ and
$\delta^2(x-y)$ is the  delta distribution on $\Sigma$.  The time components
$A_0$ of the gauge f\/ield act as Lagrange multipliers
which impose the six primary constraints
\begin{gather*}
 F^I \equiv \eps^{ab}F_{ab}^I[A]\approx 0.
\end{gather*}
It is easy to check that these constraints are f\/irst class and that the system
admits no more constraints. In terms of the smeared curvature $F(\alpha)=\int_\Sigma \de^2x \, \alpha_I F^I$ def\/ined
for any $\alpha=\alpha^I\xi_I \in C^\infty(\Sigma,\gmat)$, their algebra is given by
\begin{gather*}
\{F(\alpha_1)  , F(\alpha_2)\}= \frac{2\pi}{k} F([\alpha_1,\alpha_2]),
\end{gather*}
and the Hamiltonian action of these constraints on the Chern--Simons connection,
\begin{gather*}
\{F(\alpha) ,  A\} = D\alpha = \de\alpha  + [A,\alpha],
\end{gather*}
reproduces as expected the inf\/initesimal gauge transformations.

When expressed in terms of the BF variables $e$ and $\omega$, the only non-trivial
Poisson brackets in \eqref{Poisson bracket} are the ones which pair the components of the triad and spin connection,
\begin{gather}\label{BF Poisson bracket}
\{e_a^i(x), \omega_b^j(y)\}=\f{2\pi}{k}\delta^{ij} \eps_{ab}\delta^2(x-y),
\end{gather}
so that the triad $e$ and the connection $\omega$ are canonically conjugated variables.
Moreover, the f\/irst class constraints can be grouped into the two sets
\begin{gather}\label{continuous constraints}
\tilde{F} \equiv \eps^{ab}F_{ab}[\omega] + \Lambda \eps^{ab}[e_a,e_b] \approx 0,
\qquad
T  \equiv  \eps^{ab} T_{ab}[e,\omega]\approx 0.
\end{gather}

\subsubsection*{The physical phase space}

To give a simple description of the physical phase space,
it is advantageous to work with the Chern--Simons formulation of the theory.
Let us recall that solutions to the constraints span the inf\/inite-dimensional af\/f\/ine
space of f\/lat $G$-connections on $\Sigma$, which we denote by
\begin{gather*}
{\cal F}(G,\Sigma)  \equiv  \{A \in {\cal A}(G,\Sigma) \; \vert \; F[A]=0\}.
\end{gather*}
This space inherits the Poisson bracket (\ref{Poisson bracket})
from the Chern--Simons action and the action of the gauge symmetry
(\ref{gauge symmetry}).
The physical phase space, denoted ${\cal P}(G,\Sigma)$,
is the moduli space of f\/lat $G$-connections modulo gauge transformations on the spatial surface $\Sigma$:
\begin{gather*}
{\cal P}(G,\Sigma) \equiv  {\cal F}(G,\Sigma)/{\cal G}_\Sigma,
\qquad
{\cal G}_\Sigma=C^\infty(\Sigma,G).
\end{gather*}
It inherits a symplectic structure from
the Poisson bracket on ${\cal F}(G,\Sigma)$ and, remarkably, is  of f\/inite dimension.
More specif\/ically, the physical phase space ${\cal P}(G,\Sigma)$
can be parametrized by the holonomies along curves
on $\Sigma$ and is isomorphic to the space
$\text{Hom}(\pi_1(\Sigma),G)/ G$, where the quotient is taken
with respect to the action of $G$ by simultaneous conjugation.
The physical observables are, by def\/inition, functions on ${\cal P}(G,\Sigma)$.
A basis can be constructed using the
notion of spin networks on $\Sigma$. Alternatively, one can work with conjugation-invariant
functions of the holonomies along a set of curves on $\Sigma$ representing
the elements of its fundamental group $\pi_1(\Sigma)$. The simplest elements of this type are the Wilson loops
\begin{gather*}
W_\ell^{(j)}(A) = \text{tr}\big[ \pi^{(j)}\big(g_\ell(A)\big)\big],
\end{gather*}
which are given by the trace of the holonomy $g_\ell(A)$ of the connection along the closed loop $\ell$
evaluated in a f\/inite-dimensional representation $\pi^{(j)}:G\longrightarrow\text{End}\(\CH^{(j)}\)$ of the group $G$.
The fact that the representation is f\/inite-dimensional ensures that
the Wilson loop is well-def\/ined. The Poisson bracket between two
such observables was f\/irst described by Goldman \cite{MR762512}.

It is easy to see that the dimension of ${\cal P}(G,\Sigma)$
is given by $(2\text{g}-2)\dim G$, where $\text{g}$ is the genus of the surface.
It is in general very dif\/f\/icult to exhibit an explicit basis of the physical phase space. We know how to do it for the torus
and the genus-two surface, but simple parametrizations for higher genus surfaces have not yet been found.
As a consequence, a concrete description of the physical phase space is not yet available
and this is an obstacle for its direct quantization, i.e.\ after the implementation of all constraints.
Due to this, the only possible strategy will consist in quantizing before implementing the constraints.
But before focusing on the quantum aspects, let us brief\/ly describe how one can include point particles
in this scenario.

\subsubsection*{The physical phase space with point particles}

To account for the presence of point particles, let us consider a Riemann surface $\Sigma$ punctured with $p$
particles at the points $\{z_1,\dots,z_p\}$. The particle at the point $z_i$ has a mass $m_i$ and a~spin~$s_i$
in the sense described in the previous subsection.
We know that the connection $A$ is f\/lat everywhere, except at the locations $z_i$
of the particles where it is singular. The singularity is completely encoded in the conjugacy class of the
holonomy $g_{\ell_i}(A)$ of the connection along a small closed loop $\ell_i$ around the particle (small means
that the loop surrounds only the particle $z_i$). We have shown that each particle $z_i$ comes with a phase space
${\cal C}_{m_i,s_i}$ \eqref{coadjoint}, which is a coadjoint orbit of $\gmat$ endowed with the Kirillov--Poisson bracket.
Furthermore, the action of the gauge group ${\cal G}_\Sigma=C^\infty(\Sigma,G)$
on the space ${\cal A}(\Sigma,G)$ of regular connections
on the surface $\Sigma$ extends on the coadjoint orbits (by the gauge principle) and acts by the adjoint action
on each ${\cal C}_{m_i,s_i}$. This action is Poissonian, which means that it is compatible with the
Kirillov--Poisson bracket. For the sake of simplicity, we keep the notation ${\cal G}_\Sigma$ for the gauge group on the
punctured surface~$\Sigma$. As a consequence, the generalization of the phase space
to the case of punctured surfaces is now immediate and is given by
\begin{gather}\label{phase space 2 3D}
{\cal P}(G,\Sigma) \equiv \left(\{A \in {\cal A}(G,\Sigma) \; \vert \; F[A]=0 \} \times {\cal C}_{m_1,s_1}
\times \cdots \times {\cal C}_{m_p,s_p}\right)/{\cal G}_\Sigma.
\end{gather}
This phase space is the cartesian product between the space of
f\/lat connections and the $p$ coadjoint orbits, up to the simultaneous
action of the gauge group ${\cal G}_\Sigma$.
It is the action of ${\cal G}_\Sigma$ which ``couples'' the coadjoint orbits with
the space of f\/lat connections. As in the case without particles, the phase space is related
to the space $\text{Hom}(\pi_1(\Sigma),G)$ where $\Sigma$ is now the punctured surface.
If one denotes by $\ell_i$, $i=1,\dots,p$, the $p$ generators of $\text{Hom}(\pi_1(\Sigma),G)$
corresponding to the loops around the punctures, then ${\cal P}(G,\Sigma)$ is isomorphic to
\begin{gather}\label{phase space 3D}
\{\phi \in \text{Hom}(\pi_1(\Sigma),G) \; \vert \; \phi(\ell_i) \in {\cal O}_{m_i,s_i} \}/G,
\end{gather}
where ${\cal O}_{m,s} \subset G$ are now conjugacy classes\footnote{Notice
that the elements of the conjugacy classes ${\cal O}_{m_i,s_i}$ are
def\/ined as the ``exponentiated'' elements of the coadjoint orbits
${\cal C}_{m_i,s_i}$ def\/ined in (\ref{coadjoint}).}
of~$G$ and
the quotient is taken again with respect to the action of $G$ by simultaneous conjugation.

This completes the description of the classical theory.
However, we emphasize that we did not propose any explicit parametrization of
the physical phase space (or equivalently the space of classical solutions). Such a parametrization is very dif\/f\/icult to obtain
in general (for any punctured Riemann surface) and is not very convenient for the quantization. For these reasons,
we ended the Hamiltonian analysis with the abstract def\/inition~(\ref{phase space 3D}), which is the starting point of the
combinatorial quantization. Instead, in loop quantum gravity one starts with the def\/inition~(\ref{phase space 2 3D}), written
in terms of the BF variables.  Of course, as we have already emphasized, the two def\/initions are totally
equivalent, the latter being the inf\/inite-dimensional version of the former coset.

\subsection{Quantum gravity in the Chern--Simons formulation}
\label{subsec-3dtop}

To build the quantum theory, we are going to perform the quantization before implementing the gauge symmetries or,
in other words, before imposing the f\/irst class constraints generating these symmetries.
Thus, the main dif\/f\/iculty in the quantization process is to understand the symmetries at the quantum level.
Exactly the same issue arises in four-dimensional loop quantum gravity,
where the implementation of the quantum scalar constraint
is certainly the most important unresolved issue. In three dimensions,
we do know how this can be done at least in some cases.

\looseness=-1
The f\/irst complete quantization of three-dimensional gravity was done by Witten \cite{Witten:1988hf}, who
found amazing and deep relationships between quantum gravity and the theory of knots and  manifold invariants.
This discovery has opened a new way towards the understanding of topological invariants
and of\/fered in particular a simple and beautiful framework to unify
the dif\/ferent knot invariants like the Jones \cite{MR830613}
and the HOMFLY (Hoste--Ocneanu--Millett--Freyd--Lickorish--Yetter) \cite{MR776477} polynomials. Even the more
``mysterious'' Vassiliev invariants~\cite{MR1089670} are incorporated into this approach
(see~\cite{Labastida:2000hw} for a general review of these aspects).
Unfortunately, we will not consider these aspects here in detail. It is nonetheless important to emphasize that
any other approach to quantum gravity in three dimensions must reproduce at least the results obtained by Witten.
In this sense, we have a good test for any candidate of quantum gravity in three dimensions.

For the present discussion, the most interesting point in Witten's work is the clear relationship between
the canonical and the covariant quantizations of three-dimensional gravity.
It is indeed possible to perform both quantizations
in the case of Riemannian gravity with a positive cosmological constant.
The reason for this is that it corresponds to the only case in which
the gauge group $G$ of the Chern--Simons theory
is compact\footnote{The other Riemannian cases (vanishing and negative cosmological constant)
are much more complicated. It  remains possible
to perform the canonical quantization and to construct
the physical Hilbert space~\cite{Buffenoir:2002tx,Meusburger:2010bc},
but the covariant quantization is still poorly understood.}.
The construction of a bridge between the covariant and canonical
quantizations was crucial for Witten to establish the link between knot invariants and quantum gravity.
After him, Turaev and Viro~\cite{Turaev:1992hq},
and Reshetikhin and Turaev~\cite{Reshetikhin:1991tc} made this link even more precise,
because they settled the basis for the construction of the physical Hilbert space, on one hand,
and proposed a state sum formulation of the path integral, which is known as the Turaev--Viro model,
on the other hand. In their construction,
the link between the canonical and the covariant quantizations is very clear.
This can therefore help us to clarify the relationship
between the two quantization schemes in other approaches like loop quantum gravity
and spin foams in three and four dimensions. For all these reasons,
we think that it is really useful to recall the basis and the main results
of the quantization initiated by Witten and established more rigorously by Reshetikhin and Turaev.

\looseness=-1
This section is structured as follows. First, we review the main ideas behind Witten's construction.
We will see how particles are naturally associated to knots in
spacetime, and how the computation of the path integral of Riemannian gravity with a positive cosmological
constant coupled to particles leads to knot polynomials. In fact, the techniques involved
mix both the covariant and canonical quantizations, and require a precise knowledge of the physical Hilbert space.
The second part will be devoted to the construction of the physical Hilbert space using the combinatorial
quantization, which is a Hamiltonian quantization of Chern--Simons theory where
quantum groups play a central role. This method is powerful because it is very general. In particular, it
works for all punctured Riemann surfaces and with all
the gauge groups~$G$ of three-dimensional gravity,
the only exception being $\SO(2,2)$, which corresponds to Lorentzian gravity
with a negative cosmological constant.
The third part will deal with the covariant quantization.
We will show that it is possible to properly formulate
the path integral of Riemannian gravity with a positive
cosmological constant in the presence of point particles as a state sum model.
This state sum, which generalizes the Turaev--Viro model, does not suf\/fer
from any divergences and is therefore mathematically well-def\/ined. Furthermore,
it is shown to give (or to be given by) the scalar product between physical states
of quantum gravity. This property is known as the equivalence between Witten--Reshetikhin--Turaev
and Turaev--Viro invariants.
We will conclude with a summary and a discussion.

\subsubsection{Quantization \`a la Witten and topological invariants}
\label{subsubsec-Witten}

The goal of this quantization is to give a meaning and to compute
explicitly the path integral for three-dimensional Riemannian gravity
with a positive cosmological constant, coupled to point particles $M_1,\dots,M_p$.
Each particle $M_i$ is characterized by its mass $m_i$, its spin $s_i$
and its worldline $\gamma_i$ corresponding to a closed loop in a Riemannian spacetime.
Using the Chern--Simons formulation, the path integral can be formally written as
\begin{gather}\label{CS partition}
\CZ({\cal M};M_1,\dots,M_p) = \int \pD{A}\left( \prod_{i=1}^p\pD{\mathtt{g}_i}\, \exp\left( \I S[A,\mathtt{g}_i] \right) \right),
\end{gather}
where $S[A,\mathtt{g}_i]$ is the $\Sp(4)$ Chern--Simons action coupled to the group elements $\mathtt{g}_i$
which are the f\/irst order dynamical variables of the particles. Due to fact that $\Sp(4)\simeq\SU(2) \times\SU(2)$,
the partition function can be written as the product of
an $\SU(2)$ Chern--Simons partition function with its complex conjugate\footnote{The fact that we have the complex conjugate
is due to the expression of the non-degenerate bilinear symmetric form which def\/ines the $\Sp(4)$ Chern--Simons action. Indeed,
when decomposing the Chern--Simons action into self-dual and anti self-dual parts, the associated levels come with an opposite sign.}.
Therefore, it is suf\/f\/icient to consider the partition function~(\ref{CS partition})
with the gauge group $\SU(2)$. In this case, ``half-particles''
at rest are now characterized by only one parameter,
which corresponds to a choice of a unitary irreducible representation
of $\SU(2)$. We will denote it by $m_i \in \mathbb N/2$ and abusively refer to it as the mass of the
``half-particle''. The ``half-particle'' will  also be abusively referred to as particle.

\subsubsection*{Pure gravity and quantization of the level}

Let us start with pure gravity. While studying the symmetries of the Chern--Simons action,
we have mentioned the presence of an anomaly (\ref{anomaly}) in the large gauge transformations.
This anomaly has no ef\/fect in the classical theory, but its consequences at the quantum level are important.
Indeed, if one requires that the formal measure in the path integral (\ref{CS partition})
be gauge invariant under the action of the gauge group,
then the following identity holds for any $g\in C^\infty({\cal M},\SU(2))$:
\begin{gather*}
\CZ({\cal M}) = \int\pD{A}\,e^{\I S_\text{CS}[A]}
=\int\pD{A^g}\,e^{\I S_\text{CS}[A^g]}
= e^{\I w[{\cal M},g]}\int\pD{A}\,e^{\I S_\text{CS}[A]},
\end{gather*}
where the winding number $w[{\cal M},g]$ introduced in (\ref{anomaly}) is non-trivial when we the gauge transformations
are not connected to the identity. For this relation to make sense, $w[{\cal M},g]$
must be an integer modulo $2\pi$. As the group $\SU(2)$ is compact, for any $g$ and every closed manifold $\cal M$,
$w[{\cal M},g]=2\pi N k$ where $N$ is an integer. Since $N$ can take in particular the value $1$,
$k$ must be an integer as well.

Let us emphasize that the path integral quantization is the most suitable framework to see that the level~$k$ is discrete.
We are not aware of any argument for the discreetness of~$k$ coming from the canonical quantization. This is
an important input from the covariant quantization and, as we will see in the remaining of this work,
it is a good illustration of the complementarity between the covariant and canonical approaches.

\subsubsection*{Integrating out the particle degrees of freedom}

Now we turn to gravity coupled to point particles.
In order to compute the partition function~(\ref{CS partition}), it is
easier to f\/irst integrate out the degrees of freedom of the particles.
The integrals over the variables $\mathtt{g}_i$ decouple.
They are in fact all identical and given by
\begin{gather}\label{particle measure}
W_\gamma^{(m)}[A] \equiv \int \pD{\mathtt{g}}\,
\exp\( \I \int_\gamma \langle \chi(m),\mathtt{g}^{-1}\de\mathtt{g}+
\mathtt{g}^{-1}A\mathtt{g}\rangle \),
\end{gather}
where $m \in \mathbb N/2$ is the representation of $\SU(2)$
corresponding to the mass of the particle, $\gamma$ is a closed loop in the manifold $\cal M$,
and $\chi(m)\in\su(2)$ is a given element in a Cartan subalgebra $\mathfrak{u}(1) \subset \su(2)$.
As the full theory (gravity coupled to particles)
is invariant under the action of the gauge group,
it is natural to require that the measure $\int\pD{\mathtt{g} }$ in (\ref{particle measure}) be also
invariant under the gauge transformations. They reduce in the present case
to the right action of an element $g \in C^{\infty}(\gamma,\SU(2))$
on the group variable $\mathtt{g} $. An immediate consequence of the requirement of invariance
is that
\begin{gather} \label{invariance path particle}
W_\gamma^{(m)}[A^g]= W_\gamma^{(m)}[A] .
\end{gather}
Therefore,\looseness=-1 as it was suggested by the notation, $W_\gamma^{(m)}[A]$
is in fact the Wilson loop of the connection $A$ along
the closed loop $\gamma$ evaluated in the representation of spin $m$.
To be more precise, equation (\ref{invariance path particle})
implies that $W_\gamma^{(m)}[A]$ is proportional to the Wilson loop~\cite{MR1025431, Witten:1988hf}, but
one can always choose a normalization of the measure
$\int \pD{\mathtt{g} }$ in such a way that the proportionality factor is trivial.
As a consequence, the partition function of gravity coupled
to point particles can be expressed as the expectation value
of a product of Wilson loop observables in Chern--Simons theory:
\begin{gather*}
\CZ({\cal M};M_1,\dots,M_p) =
\int \pD{A}\left( \prod_{i=1}^p W_{\gamma_i}^{(m_i)}[A] \right) \exp\left( \I S_\text{CS}[A]\right) .
\end{gather*}

\subsubsection*{Complementarity of the covariant and canonical quantizations}

Now we have all the ingredients to see how the canonical and covariant
quantizations are related to one another. In particular, we will show that a
complete description of the physical Hilbert space inherited from
the canonical quantization of the theory provides a machinery
to compute explicitly the path integral.

Following \cite{Witten:1988hf}, let us decompose the spacetime manifold $\cal M$
into two pieces ${\cal M}_1$ and ${\cal M}_2$ sharing the same boundary
$\Sigma=\partial {\cal M}_1=\partial {\cal M}_2$. Furthermore, let us assume that the common boundary
has the topology of a two-sphere $\mathbb{S}^2$.
When particles are present in the theory, their worldlines cross this boundary,
which can then be interpreted as a punctured two-sphere.

In the previous subsections, we have seen that any punctured surface
is endowed with a~certain symplectic structure.
Therefore, the canonical quantization of the theory on these punctured surfaces
associates a Hilbert space ${\cal H}$ to $\mathbb{S}^2$.
In the next subsection we provide a precise description of this Hilbert space.
For the moment, we are only going to recall some of its properties:
\begin{itemize}\itemsep=0pt
\item
When there are no punctures on $\mathbb{S}^2$, $\cal H$ is one-dimensional.
\item
When there is one puncture on $\mathbb{S}^2$, $\cal H$ is zero-dimensional.
This follows from the fact that on a closed manifold without boundaries
the wordlines of particles are closed loops and the number of punctures of $\mathbb{S}^2$ is necessarily even.
It is therefore only with an even number of punctures that the dimension of the Hilbert space is non-zero.
\item
When there are two punctures on $\mathbb{S}^2$, $\cal H$ is one-dimensional.
\item
When there are more than two punctures, the dimension of the Hilbert space will depend non-trivially
on the representations carried by the punctures, but will always remain f\/inite-dimensional\footnote{There is a nice
way to understand physically the f\/initeness of the Hilbert space~\cite{Witten:1988hf}. It is easy to show that the classical
phase space has (generically) a f\/inite volume computed with the symplectic volume form when the gauge group is compact. As the
Hilbert space is a discretization of the phase space into quantum cells of microvolume~$\hbar$,
it is natural to expect that it has a f\/inite dimension.}.
For instance, when there are four punctures colored with the spin $m=1/2$, $\cal H$ is two-dimensional.
\end{itemize}

The path integral is directly related to the Hilbert space $\cal H$ in the sense that it can be written as a scalar product
\begin{gather}
\CZ({\cal M};M_1,\dots,M_p) = \langle v_1 , v_2 \rangle
\label{pathont-scpr}
\end{gather}
between two vectors $v_1$ and $v_2$ in $\cal H$. More precisely, $v_1$ is def\/ined as an element of the dual
space~${\cal H}^*$, which we have implicitly identif\/ied with~$\cal H$.
Formally, $v_1$ and $v_2$ correspond to the vectors given by the path integral~(\ref{CS partition}) restricted to the manifolds~$\mathcal{M}_1$  and~$\mathcal{M}_2$.
This is a~standard property of any topological quantum f\/ield theory.

\begin{figure}[t]
\centering
\includegraphics[scale=0.45]{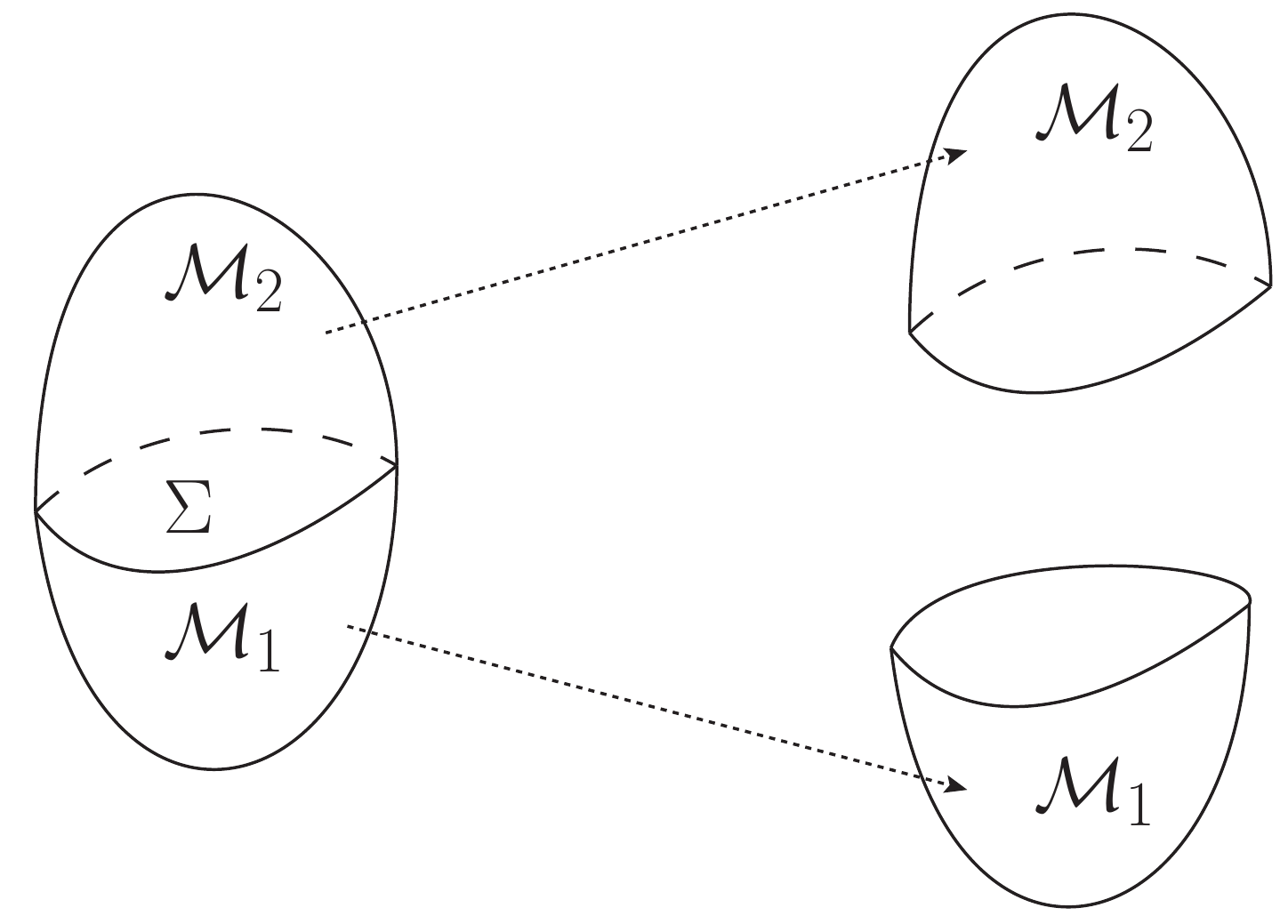}
\caption{Decomposition of the manifold $\cal M$ into two pieces ${\cal M}_1$ and ${\cal M}_2$ on the left.
On the right,  $\overline{{\cal M}_i}$ is obtained as the connected sum of ${\cal M}_i$ with a three-ball.}
\label{factorization}
\end{figure}

Furthermore, let us assume that there are no punctures. Then, $\cal H$ is one-dimensional and the path integral
$\CZ({\cal M})$ is still given by the scalar product as in~\eqref{pathont-scpr}.
Let $w_1$, $w_2$ be the corresponding vectors for $\mathcal M=\mathbb{S}^3$, which are {\it a priori}
dif\/ferent from $v_1$ and $v_2$.
Then one can use the ``marvelous'' property \cite{Witten:1988hf}
\begin{gather*}
\langle v_1,v_2\rangle  \langle w_1,w_2 \rangle = \langle v_1,w_2\rangle \langle w_1,v_2 \rangle,
\end{gather*}
which follows trivially from $\dim \mathcal{H}=1$. If one translates this identity
in the language of partition functions, one immediately obtains the factorization property
\begin{gather*}
\CZ({\cal M}) \CZ(\mathbb{S}^3) =
\CZ(\overline{{\cal M}_1}) \CZ(\overline{{\cal M}_2}) \quad
\Longleftrightarrow \quad \frac{\CZ({\cal M})}{\CZ(\mathbb{S}^3) }=
\frac{\CZ(\overline{{\cal M}_1})}{\CZ(\mathbb{S}^3) } \frac{\CZ(\overline{{\cal M}_2})}{\CZ(\mathbb{S}^3)},
\end{gather*}
where $\overline{{\cal M}_i}$ is the connected sum of ${\cal M}_i$
with a three-ball, which is illustrated in Fig.~\ref{factorization}.

It is remarkable how simple arguments coming from both the canonical and the covariant quantizations allow
to derive such a non-trivial property. With the same construction, we can extend this property to the case where particles
$M_1,\dots,M_p$ are present. If one assumes that the loops~$\gamma_i$ associated to the particles~$M_i$
are not knotted one to the other, then one has
\begin{gather*}
\frac{\CZ({\cal M};M_1,\dots,M_p)}{\CZ(\mathbb{S}^3)} = \prod_{i=1}^p \frac{\CZ({\cal M};M_i)}{\CZ(\mathbb{S}^3)}.
\end{gather*}

\looseness=-1
Proving these factorization properties is certainly enough to illustrate the complementarity between the
covariant and the canonical quantization schemes. The path integral gives the geo\-metric viewpoint, whereas the
physical Hilbert space brings the algebraic tools. However, we want to go further and show
how to establish the relationship with the Jones polyno\-mials.

\begin{figure}[t]
\centering
\includegraphics[scale=0.45]{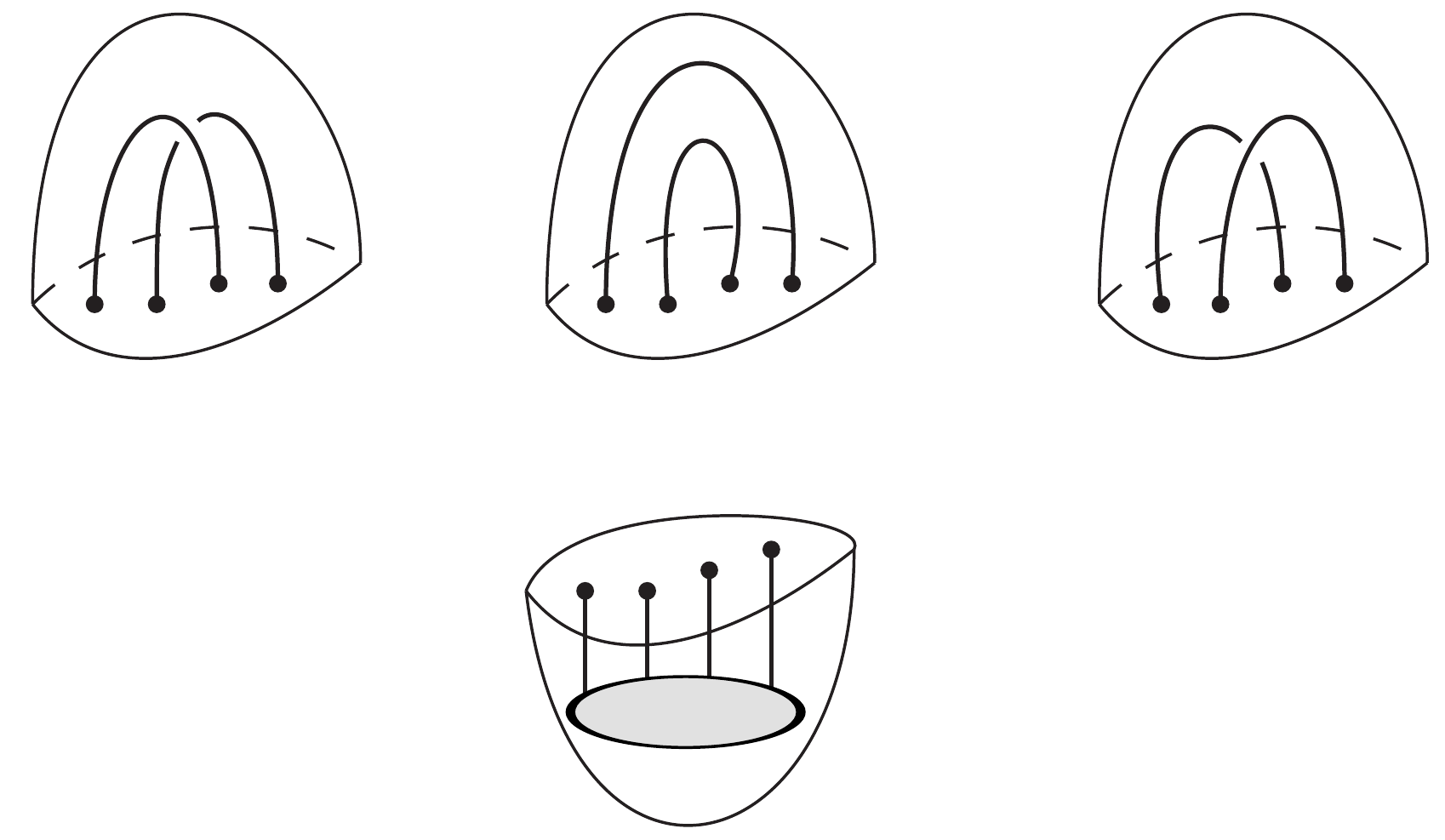}
\caption{Examples of knots in $\mathbb{S}^3$. They
coincide in the ``southern hemisphere'' of $\mathbb{S}^3$ but dif\/fer in the ``northern hemisphere''.}
\label{knots}
\end{figure}

\begin{figure}[t]
\centering
\includegraphics[scale=0.45]{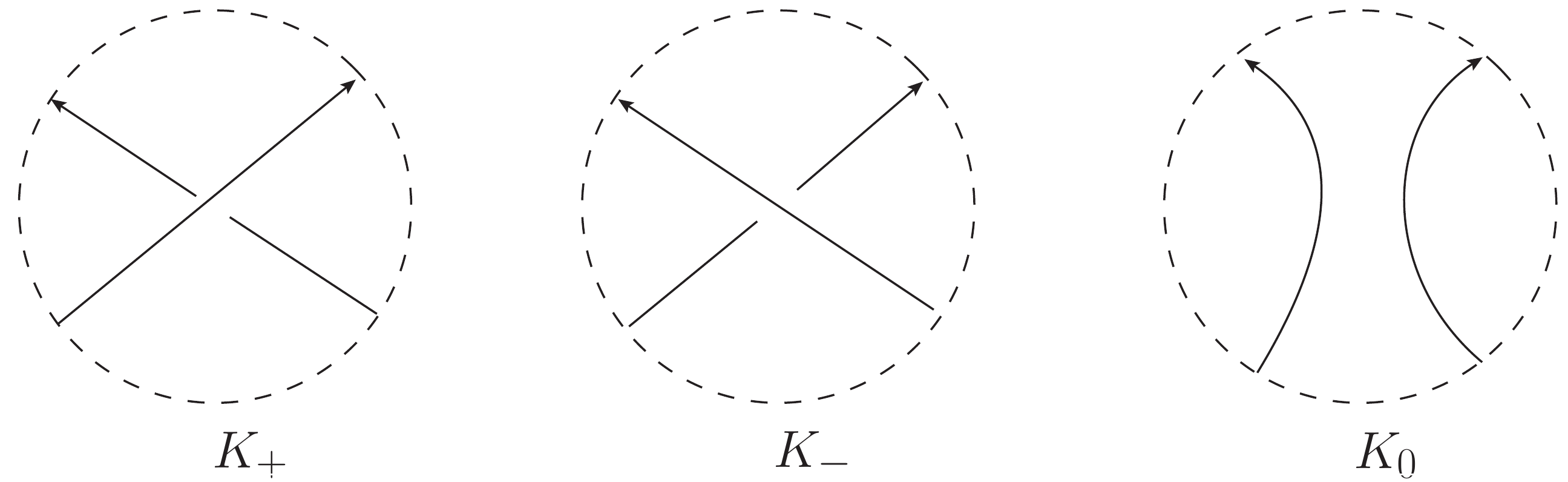}
\caption{The three ``crossings'' appearing in the def\/inition of the skein relation.}
\label{knots2}
\end{figure}

In order to do so, let us consider the situation depicted in Fig.~\ref{knots}.
It shows three knots in~$\mathbb{S}^3$ which coincide in the ``southern hemisphere'' but dif\/fer in the ``northern hemisphere''.
The shape of the knots in the southern hemisphere is not important for the moment. According to the previous discussion, the
knots in the northern hemisphere of $\mathbb{S}^3$ def\/ine three vectors $v_1$, $v_2$ and $v_3$ in the Hilbert space $\cal H$ associated to
the sphere with four punctures. If we assume for simplicity that the knots are all colored with the $1/2$ representation,
then $\dim \mathcal{H}=2$. As a consequence, $v_1$, $v_2$ and $v_3$ are not independent, and there are three real
numbers~$\alpha^i$, $i=1,2,3$, such that $\alpha^iv_i=0$. The portion of the knots lying in the southern hemisphere
def\/ines a vector $w$ in the (dual of the) Hilbert space $\cal H$ such that
\begin{gather*}
\langle v_i ,  w \rangle = \CZ\big(\mathbb{S}^3,M_i\big),
\end{gather*}
where $M_i$ denotes the particle whose worldline is the knot obtained
by gluing the northern part of $\mathbb{S}^3$ associated to $v_i$
with the southern part associated to~$w$. As a consequence, we obtain the following relation:
\begin{gather*}
\alpha^1 \CZ\big(\mathbb{S}^3,M_1\big)+ \alpha^2 \CZ\big(\mathbb{S}^3,M_2\big)+ \alpha^3 \CZ\big(\mathbb{S}^3,M_3\big)= 0.
\end{gather*}
This relation is important because it is closely related to the skein relation def\/ining the Jones polynomial.
To see that this is indeed the case, we need to compute the explicit values of the coef\/f\/icients $\alpha^i$, and in order to do so
the concrete description of the physical Hilbert space~$\cal H$ which will be developed in the next subsection is required.
Using this description, we f\/ind that
\begin{gather}\label{skein1}
q \CZ\big(\mathbb{S}^3,M_1\big)+  \big(q^{1/2} - q^{-1/2}\big)\CZ\big(\mathbb{S}^3,M_2\big)  -  q^{-1}\CZ\big(\mathbb{S}^3,M_3\big)  = 0,
\end{gather}
where $q=\exp(2\I \pi/(k+2))$ is a root of unity because the level $k$ is an integer.
Before going further, let us point out that the values of $\alpha^i$ depend on the choice of a framing
corresponding to a thickening of the knot into a ribbon.
There is no canonical choice of framing, but two dif\/ferent choices
lead to scalar products which dif\/fer only by a phase. In our example,
a canonical framing exists (it is the trivial one), and with this choice~(\ref{skein1}) is exactly the skein relation
\begin{gather*}
q J(q,K_+) +  \big(q^{1/2} - q^{-1/2}\big)J(q,K_0)-  q^{-1} J(q,K_-)= 0,
\end{gather*}
which def\/ines the Jones polynomials $J(q,K)$ for any knot $K$ embedded in $\mathbb{S}^3$,
up to a global normalization \cite{MR830613}. In this relation, $J(q,K)$ is a Laurent polynomial
in the variable $q$, and $K_+$, $K_-$ and $K_0$ are three dif\/ferent knots
which coincide with a knot $K$ everywhere but in a~small region around a crossing, where they dif\/fer
according to the representation of Fig.~\ref{knots2}.
In order to evaluate the Jones polynomial of a knot, one needs to give it an orientation, but in the end
the evaluation does not depend on this choice. As a consequence, the path integral
of three-dimensional Riemannian gravity with a positive cosmological constant coupled to a point particle
is closely related to the Jones polynomial of the knot associated to the worldline of the particle.
One can make this statement more precise, but this would require an extended discussion about the role of the framing,
and we want to avoid this for the sake of simplicity.

This construction can be generalized to compute the path integral of the Chern--Simons theory with compact gauge group
in any closed three-dimensional spacetime using the surgery representation of three-manifolds~\cite{Witten:1988hf}.
When the gauge group is~$\SU(2)$ and the knot is colored with a spin-$m$ representation, the path integral gives
the generalized Jones polynomial. When the gauge group is~$\SU(N)$, it can be related to the so-called
HOMFLY polynomial, which is a~two-parameter generalization of the Jones polynomial.

\subsubsection{The combinatorial quantization -- emergence of quantum groups}
\label{subsubsec-combin}

This part is devoted to the construction of the Hilbert space ${\cal H}$ associated to a punctured Riemann surface
$\Sigma$.
To our knowledge, the most powerful technique to perform the canonical quantization of Chern--Simons theory
is the combinatorial quantization. It is based on the seminal work of Fock and Rosly~\cite{Fock:1998nu}, and has been used in
\cite{Alekseev:1994pa,Alekseev:1994au,Alekseev:1995rn} to quantize Chern--Simons theory with the gauge group $G=\SU(2)$,
in~\cite{Buffenoir:2002tx} when $G$ is the Lorentz group
and in \cite{Meusburger:2005in,Meusburger:2003wk,Meusburger:2003ta,Meusburger:2003hc}
when~$G$ is the Euclidean  group.
These three cases correspond respectively to Riemannian
gravity with a positive, negative, or vanishing cosmological constant. The Lorentzian case with $\Lambda=0$
has been studied as well, but we restrict ourselves only to the Riemannian case.

The combinatorial quantization relies on the fact that the quantization turns classical symmetry groups into quantum groups.
In this approach, quantum groups are not introduced artif\/icially by hand, but appear naturally
in a very beautiful way during the quantization process.
Let us start by presenting the quantum groups which are at the heart of this combinatorial quantization.
The classical (universal coverings of the) symmetry groups of Riemannian gravity are the Euclidean group $\ISU(2)$,
the Lorentz group $\SL(2,\mathbb C)$, or the rotational group $\Sp(4)$ depending on the value of the scalar curvature.
These groups are turned into quantum groups that can all be described as Drinfeld doubles \cite{CP}:
$\text{D}\SU(2)$, $\text{D}\Uq(\su(2))$ for $q \in \mathbb R$,
and $\text{D}\Uq(\su(2))$ with~$q$ a~root of unity, respectively,
where $\Uq(\gmat)$ denotes a deformation of the associated enveloping algebra $\text{U}(\gmat)$
of a Lie algebra $\gmat$. In fact, $\text{D}\Uq(\su(2))$ for~$q$ real or root of unity can be shown
to be a~deformation of either SL(2,$\mathbb C$) or $\Sp(4)$, and therefore these groups are also denoted by
$\Uq(\mathfrak{sl}(2,\mathbb C))$ and $\Uq(\so(4))$.
The deformation parameter $q$ is the same as the one which appears
in the skein relation~(\ref{skein1}) and is related to the level $k$ of the Chern--Simons theory.
When~$q$ is a~root of unity, one has $q=\exp(2\I\pi/(k+2))$, whereas for $q$ real the relation
is known only at f\/irst order in $1/k$ and reads $q=1+2\pi/k +\circ(1/k)$.
Thus, the quantization leads to two important physical ef\/fects: the level $k$ becomes an integer and
classical groups are turned into quantum groups.
As a result, the mass and the spin of the particles~$M_i$ are, after the quantization process,
given by unitary irreducible representations of the quantum groups instead of classical groups. In this sense,
the mass and the spin become quantized.

In order to make all the previous statements more precise, let us assume for simplicity
that~$\Sigma$ is a sphere with~$p$ punctures colored with $\SU(2)$ representations
$m_i$ for $i=1,\dots,p$.
The starting point of the combinatorial quantization is
the def\/inition (\ref{phase space 3D}) of the physical phase space, which reduces to
\begin{gather}\label{phase space 3 3D}
\bigg\{a\in {\cal O}_{m_1} \otimes \dots \otimes {\cal O}_{m_p} \; \vert \; \prod_{i=1}^p a_i =1\bigg\}/G,
\end{gather}
where ${\cal O}_m \subset G$ is a conjugacy class in $\SU(2)$ characterized by the spin $m$.
Usually, conjugacy classes are labeled by an angle
$\theta \in [0,\pi]$, but in the quantum theory
the angle is discrete and related to $m$ by
\begin{gather}\label{cuttoff}
\theta_m=2\pi m/k,
\end{gather}
where $k$ is the level. This relation suggests that $m \leq k/2$,
which has an immediate interpretation in the quantum theory as we will see very soon.
If we denote by $\tau_1$ one generator of $\su(2)$, then a~representative element of ${\cal O}_m$
is given by $h(\theta_m)=\exp(\theta_m \tau_1)$ and the conjugacy class is explicitly def\/ined by
\begin{gather}\label{orbits htheta}
{\cal O}_m = \big\{h \in\SU(2) \; \vert \; h=gh(\theta_m) g^{-1},\ \forall \,g \in\SU(2)\big\}  \subset \SU(2).
\end{gather}
It is clear from this def\/inition that ${\cal O}_m \simeq \mathbb{S}^2$ and that any of its elements $h$ satisfy
$\chi_j(h)=\chi_j(h(\theta_m))$,
where $\chi_j(h)$ is the character of $h$ evaluated in the representation~$j$.
Thus, the space~(\ref{phase space 3 3D}) is a f\/inite-dimensional symplectic manifold whose Poisson bracket
is inherited from the canonical analysis of Chern--Simons theory.
The equation
\begin{gather}\label{flatness on the sphere}
F=\prod_{i=1}^p a_i =1
\end{gather}
expresses the f\/latness of the connection, and $G$
acts by simultaneous conjugation, i.e.\ $a_i \longmapsto g^{-1}a_i g$ for any $i \in [1,p]$
and any $g \in \SU(2)$. Furthermore, $F$ is the Hamiltonian generator of this symmetry.

The combinatorial quantization proceeds by quantizing the space of connections of the punctured surface $\Sigma$,
\begin{gather*}
{\cal A}(\Sigma)  = {\cal O}_{m_1} \times\dots\times {\cal O}_{m_p}
\end{gather*}
and then by imposing the f\/latness condition (\ref{flatness on the sphere}) at the quantum level.
Any element $a_i$ in the space of connections can be viewed as the holonomy of the Chern--Simons connection starting at a
given base point and turning around the puncture~$i$. The problem is that such a space does not possess a regular Poisson
structure, and the one inherited from the canonical analysis is ill-def\/ined due to the following reason:
Although one knows how to compute the Poisson bracket between two holonomies associated
to paths which cross each other, the Poisson bracket between holonomies
associated to paths which end at the same point but do not cross is not def\/ined.
This is precisely our case since the paths associated to the elements~$a_i$ of ${\cal A}(\Sigma)$
do not cross, they just end or start at the same base point.
As a consequence, ${\cal A}(\Sigma)$ is not naturally endowed with a~Poisson bracket
and one needs to introduce a certain regularization.
This is exactly what Fock and Rosly did in a very clever way~\cite{Fock:1998nu}.
They have shown that if one endows the space~${\cal A}(\Sigma)$ with a linear ordering (which amounts to saying
that $a_1$ comes f\/irst, $a_2$
second, and so on~\dots), then it is possible to endow it with a quadratic Poisson--Lie structure schematically
given by a~formula of the type
\begin{gather}\label{FR Poisson}
\{a_1 \otimes 1,1 \otimes a_2\}_\text{FR} = \frac{2\pi}{k}(r\, a_1 \otimes a_2 + a_1 \otimes a_2\, {}^t r),
\end{gather}
where $r$ is called a classical $r$-matrix (${}^tr$ is its transpose)
and satisf\/ies the so-called classical Yang--Baxter equation as a consequence
of the Jacobi identity. In~(\ref{FR Poisson}), $a_1$ and $a_2$ are viewed as
group elements\footnote{To make the Poisson--Lie structure more concrete, it is useful
to evaluate the holonomies $a_i$ in f\/inite-dimensional representations of $\SU(2)$ so that the variables of the phase
space become the matrix elements of~$\pi^{(j)}(a_i)$.}
(in fact, they are holonomies), and $r$ is an element of~$\mathfrak{g} \otimes \mathfrak{g}$.
Similar Poisson brackets hold for any couples $(a_i,a_j)$ of
${\cal A}(\Sigma)$.
This prescription possesses the following important features:
\begin{itemize}\itemsep=0pt
\item
First of all, it reduces to the standard Poisson bracket on the physical phase space (\ref{phase space 3 3D}) after imposing
the f\/latness condition, or equivalently the gauge invariance.
\item
Second, its quantization is well-known and leads to the appearance of quantum groups.
In particular, the $r$-matrix becomes the classical limit
of a quantum $R$-matrix which satisf\/ies in turn the quantum Yang--Baxter equation. This
is a clear sign of the quantum group structure behind this construction.
\end{itemize}

Indeed, when one proceeds to the quantization of the (non-physical) phase space $\mathcal{A}(\Sigma)$
according to the standard rules of quantum mechanics, the resulting quantum algebra, called the graph algebra
${\cal L}(\Sigma)$, is generated by the quantization $\hat{a}_i$ of the classical variables $a_i$
and is def\/ined by commutation relations of the form\footnote{More precisely, the graph algebra is generated by the quantization of
the matrix elements $\pi^{(j)}(a_i)$ which can be denoted  $\pi^{(j)}(\hat{a}_i)$, but with $j$ labelling a representation of the
quantum group instead of the classical group.}
\begin{gather*}
R(\hat{a}_1 \otimes 1) R^{-1} (1 \otimes \hat{a}_2)  = (\hat{a}_2 \otimes 1) R (1 \otimes \hat{a}_1) R^{-1},
\end{gather*}
where $R$ is the quantum $R$-matrix. As we already said, the presence of $R$
is a strong indication of the presence of quantum groups.
This is in fact what Alekseev has shown in a beautiful theorem~\cite{MR1290818} stating that
\begin{gather*}
{\cal L}(\Sigma)\simeq \Uq(\su(2))^{\otimes p},
\end{gather*}
i.e.\ that ${\cal L}(\Sigma)$ is isomorphic to $p$ copies of the quantum group $\Uq(\su(2))$.
This theorem (which can be generalized to any Riemann surface) allows to construct the
kinematical states of the theory. They are given by
unitary irreducible representations of ${\cal L}(\Sigma)$ which are obtained from the tensor product of~$p$ unitary
irreducible representations $j_i$ of $\Uq(\su(2))$,
\begin{gather}\label{kinematical Com}
{\cal H}_0(j_1,\dots,j_p) \equiv \bigotimes_{i=1}^p\CH^{(j_i)},
\end{gather}
where $\CH^{(j)}$ is the module of the representation labeled by $j \in [0,k/2]$.
The cut-of\/f on the number of representations is a result of the quantization and is totally consistent with
the classical relation (\ref{cuttoff}) that we have obtained above. As in the classical case, the dimension
of the representation of the quantum group is $d_j=2j+1$.

Note that ${\cal H}_0$ is not the Hilbert space of physical states. Rather, it is analogous
to the kinematical Hilbert space in loop quantum gravity. For this reason, we have called its elements
kinematical states.
To extract the physical Hilbert space out of ${\cal H}_0$, one still has to impose two types of constraints.
The f\/irst one is the quantum analogue of $a_i \in {\cal O}_{m_i}$ and
f\/ixes the representations\footnote{It is possible to
show that for any representation $j$, the elements
\begin{gather*}
W^{(j)}_i = \text{Tr}_q^{(j)}(\hat{a}_i) \equiv \text{Tr}^{(j)}(\mu \hat{a}_i),
\end{gather*}
where the quantum trace $\text{Tr}_q$ is def\/ined from the ribbon element $\mu \in \Uq(\su(2))$,
are in the center of the graph algebra ${\cal L}(\Sigma)$. Therefore, these elements form
an Abelian subalgebra and they are scalar operators in ${\cal H}_0$.
The quantum version of the constraint
$a_i \in {\cal O}_{m_i}$ f\/ixes the value of each element $W^{(j)}_i$ to
\begin{gather*}
W^{(j)}_i = \f{[d_j m_i]_q }{[m_i]_q}=\f{\sin\(\f{2\pi}{k+2}d_jm_i\)}{\sin\(\f{2\pi}{k+2}m_i\)},
\end{gather*}
where the quantum number $[x]_q$ is def\/ined in (\ref{q number}).
In fact, $W^{(j)}_i$ is the quantum analogue of the charac\-ter~$\chi_{j}(m_i)$, and it is easy to show that
the quantum constraint f\/ixes $j_i=m_i$.}
in (\ref{kinematical Com}) to $j_i=m_i$.
Therefore, among all the Hilbert spaces ${\cal H}_0$, only one of them is admissible.
The second constraint is the quantum f\/latness condition $\hat F=0$.
To impose this constraint, one has to understand how the action of the symmetry group $G$
translates at the quantum level. As we know, the classical action of $G$
is turned into the action of $\Uq(\su(2))$ on ${\cal H}_0$
viewed as a tensor product of $\Uq(\su(2))$ modules. As a consequence, the physical Hilbert space
is simply given by
\begin{gather*}
{\cal H} = \text{Inv}\big(\CH^{(m_1)}\otimes \dots \otimes\CH^{(m_p)}\big),
\end{gather*}
where the notation $\text{Inv}$ denotes the subspace invariant under the action of $\Uq(\su(2))$.

\begin{figure}[t!]
\centering
\includegraphics[scale=0.45]{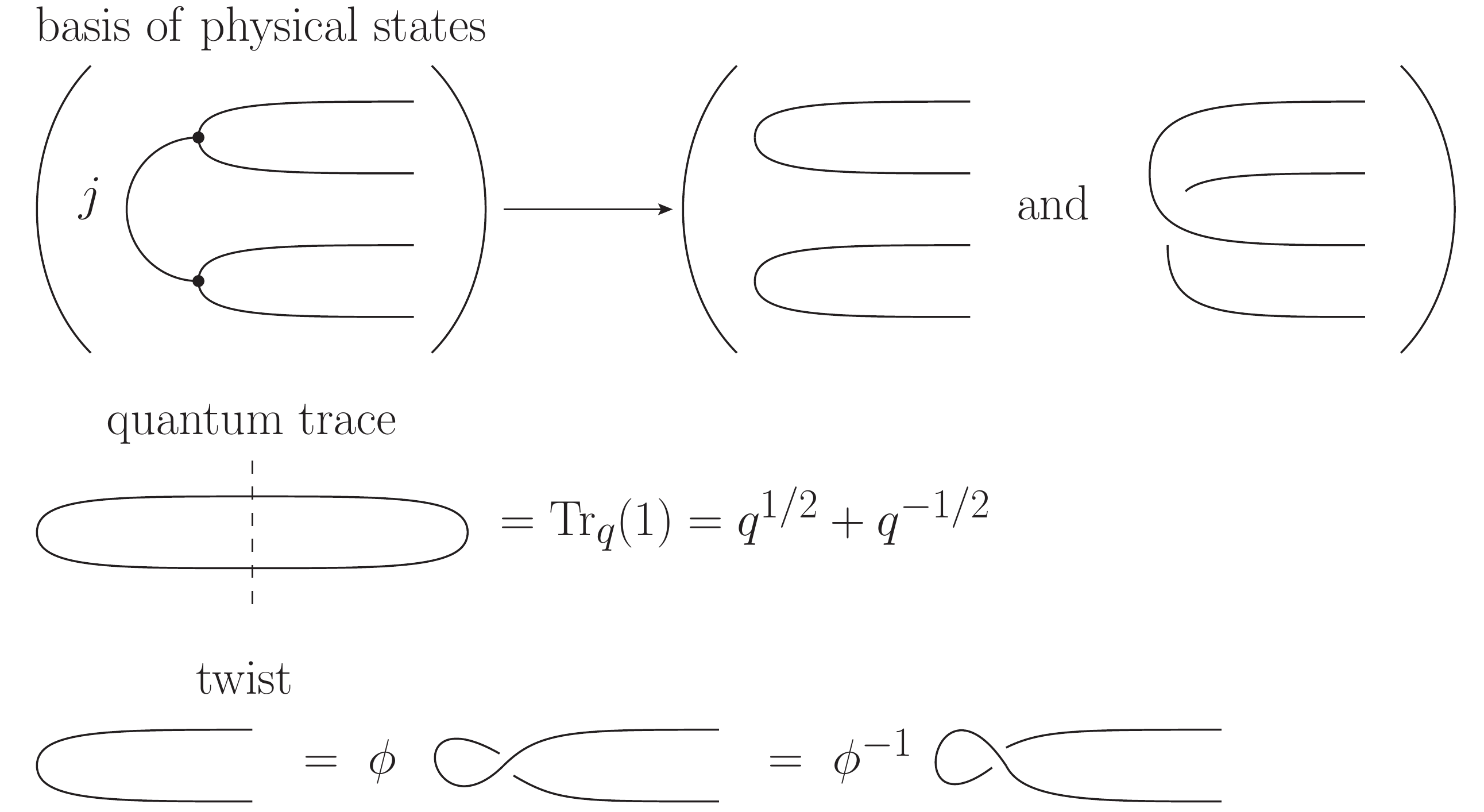}
\caption{Graphical representation of the physical states on the sphere with four punctures labeled with $j=1/2$.
On the top, we have given two dif\/ferent basis. The one on the left is described in terms
of intertwiners ($j\in \{0,1\}$), while the one on the right has two dif\/ferent vectors (one of them being def\/ined
using the quantum $R$-matrix). In the middle, we have represented the quantum trace. At the bottom, we have given
some properties related to the framing. The links have to be understood as ribbons, and therefore the twist has a non-trivial ef\/fect
on the evaluation of the quantum scalar product. The phase is given by $\phi=-q^{-3/2}$.}
\label{qspinnets}
\end{figure}

\begin{figure}[t!]
\centering
\includegraphics[scale=0.45]{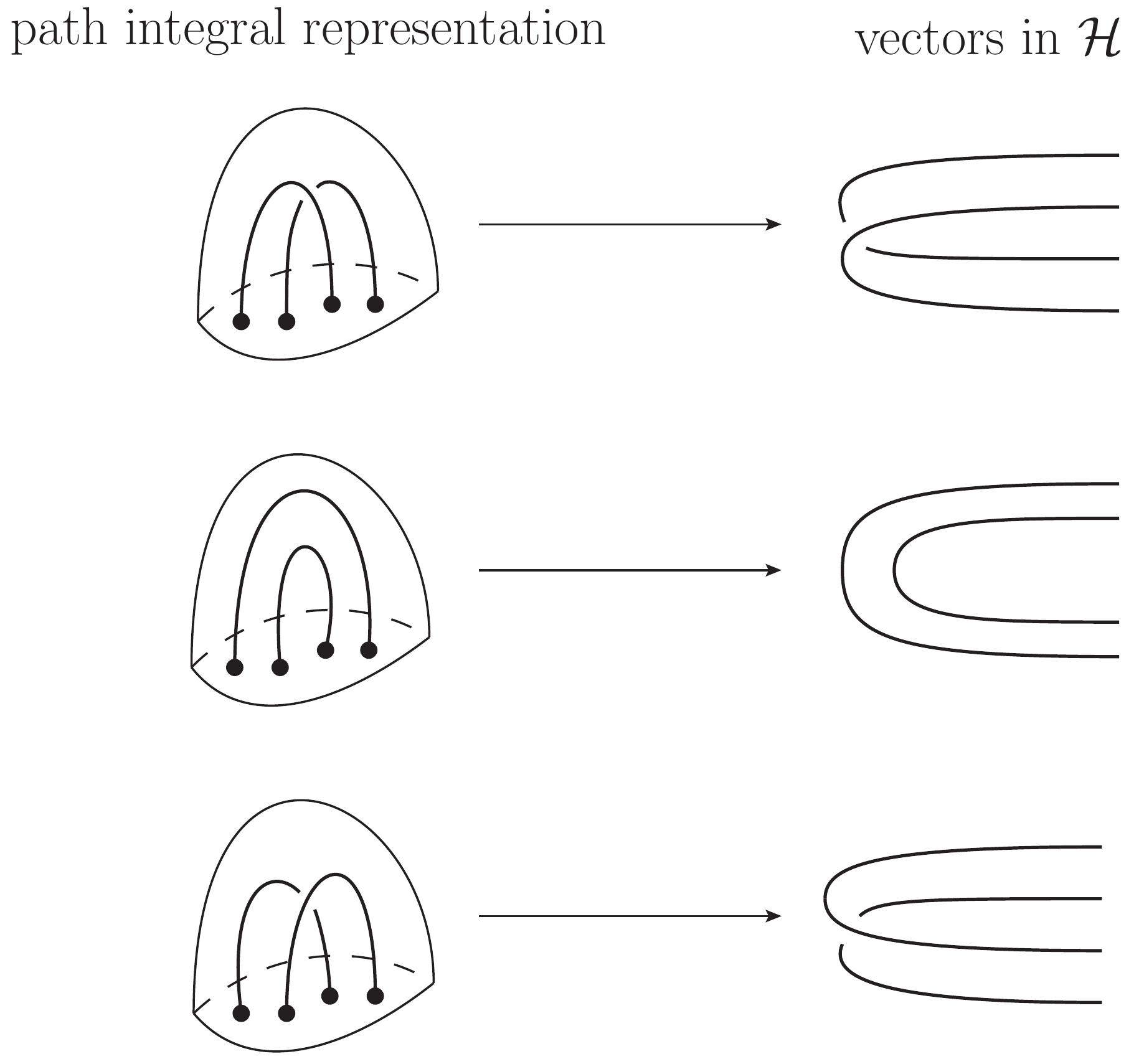}
\caption{A map from the path integral representation of the physical states to the elements of the physical
Hilbert space $\cal H$.}
\label{path H}
\end{figure}

Due to the $\Uq(\su(2))$ recoupling theory, it is clear that $\dim \mathcal {H}$ is equal to one
when there are no punctures, and to one or zero when there are three punctures at most. When there are more than four
punctures, the dimension becomes a non-trivial function of the spins $m_j$. In the case in which there are four punctures
colored with the spin $1/2$ representation, one has
\begin{gather*}
\big(\CH^{(1/2)}\big)^{\otimes 4} = 2\CH^{(0)} \oplus 3\CH^{(1)} \oplus \CH^{(2)},
\end{gather*}
and $\dim \mathcal{H}=2$ as was claimed previously. In order to justify the results that we have used
to compute the skein relation (\ref{skein1}), we now give a more precise description of $\cal H$
and show how the vectors in $\cal H$ are related to the path integral on the sphere with boundaries.
These vectors and some of their properties can be given a nice pictorial representation as in Fig.~\ref{qspinnets}.
The interpretation of this graphical representation is the same as in the classical case, the only important dif\/ference being
that now there is a non-trivial braiding.
Using this representation, it is immediate to relate the three vectors $v_i$ def\/ined by the path integral
in Fig.~\ref{knots} to the elements of $\cal H$. This relation is given in Fig.~\ref{path H}.
The vectors $v_i$ are viewed as elements of $\cal H$ in an obvious way.
Indeed, one identif\/ies the projection of their graph onto a two-dimensional
surface with a quantum spin network. This projection is not canonical and depends on the choice of the framing,
but the ``physical results'' are independent of these choices.
Once this identif\/ication is made, the calculation of the coef\/f\/icients $\alpha^i$ (up to a global factor) is straightforward
and one ends up with the skein relation.

This example is a good illustration of the interplay between the covariant and canonical quantizations of Chern--Simons
theory with a compact gauge group. This construction can be generalized to any arbitrary manifold $\cal M$
using its surgery representation. On the other hand, the generalization to non-compact gauge groups is still an open
problem~\cite{Witten:2010cx}.

\subsubsection{The state sum quantization -- spin foams}
\label{subsubsec-TVmodel}

The previous quantization, \`a la Witten--Reshetikhin--Turaev, relies technically on the canonical quantization.
Even if the use of the path integral was necessary to establish the relation with knot invariants,
physical amplitudes were explicitly computed using the scalar product on the physical Hilbert space.
Therefore, this quantization appears to be more canonical than path integral-like.

After the seminal paper of Witten, the path integral quantization of Chern--Simons theory
or three-dimensional gravity has been extensively studied.
The perturbative quantization of Chern--Simons theory was initiated soon after
the work of Witten and led to a better understanding of the Vassiliev invariants. The non-perturbative quantization
was also studied and led to the f\/irst mathematically well-def\/ined spin foam model, the so-called Turaev--Viro
model, which we will now brief\/ly review.

The original idea was to construct some manifold and knot invariants in three-dimensional
spacetime in a totally covariant way, by taking into account an important indication coming from the canonical quantization,
namely the fact that quantum groups play a fundamental role. The construction is as follows.
Let $\cal M$ be a closed three-dimensional spacetime, $\Delta$ a triangulation of $\cal M$, and $\Delta^*$
its dual triangulation. Elements of $\Delta^*$ consists of a set of
vertices $v$ (dual to tetrahedra $\tau$ in $\Delta$), edges $e$
(dual to triangles $t$ in $\Delta$) and faces $f$ (dual to links $\ell$ in $\Delta$).
To each face $f\in\Delta^*$ (or equivalently to each link $\ell$ of the triangulation),
we assign a unitary irreducible representation of $\Uq(\su(2))$ labeled by a spin $j_f$.
Then one can def\/ine the following quantity
\begin{gather}\label{turaev viro}
\mathcal{Z}_\text{TV}(\Delta^*)
=\left( -\frac{q-q^{-1}}{2(k+2)}\right)^{n_0} \sum_{j\rightarrow f}\prod_{f\in\Delta^*}[2j_f+1]_q\prod_{v\in\Delta^*}
\left\{\begin{array}{ccc}
         j_1 & j_2 & j_3\\
         j_4 & j_5 & j_6\end{array}\right\}_q,
\end{gather}
where $n_0$ denotes the number of vertices of $\Delta$,
the sum is taken over all possible assignments of irreducible representations of $\Uq(\su(2))$ to the faces,
and the two products are taken over all the faces and vertices of the triangulation, respectively.
The labels $j_1,\dots,j_6$ are associated to the six links $\ell$ bounding
a tetrahedron $\t\in\Delta$. Finally the weights associated to the faces are the quantum dimensions
\begin{gather}\label{q number}
[x]_q=\f{q^x - q^{-x}}{q-q^{-1}},
\end{gather}
with $x=2j_f+1$, and the weights associated to the vertices
are the totally symmetrized quantum~$\{6j\}$ symbols.
The quantity def\/ined by~\eqref{turaev viro} is f\/inite and was shown to be independent on the choice of the triangulation.
Therefore, it gives rise to a well-def\/ined topological invariant of the manifold $\cal M$ and we can write that
\begin{gather*}
\mathcal{Z}_\text{TV}(\Delta^*) = \mathcal{Z}_\text{TV}(\mathcal{M}) .
\end{gather*}
It was realized later on, after the introduction of the Ponzano--Regge model and the study of its
asymptotics, that the Turaev--Viro model
is in fact a quantization of Riemannian three-dimensional gravity with a positive cosmological
constant. As the action for gravity is the product of two uncoupled Chern--Simons actions with opposite levels $k$ and
$-k$, one can formally write that
\begin{gather*}
\mathcal{Z}_\text{TV}(\mathcal M) = \vert \mathcal{Z}_\text{WRT}(\mathcal M)\vert^2,
\end{gather*}
where $\mathcal{Z}_\text{WRT}(\mathcal M)$ is the canonical evaluation
\`a la Witten--Reshetikhin--Turaev of the path integral,
as described in the previous subsection.
This relation has been rigorously proved very recently~\cite{virelizier},
and it establishes a precise link between the Turaev--Viro model and the physical
scalar product of the canonical quantization. Even if this relation seems natural
from the physical point of view, the rigorous mathematical proof is much
more involved.

The construction of the covariant path integral can be generalized to accommodate for the
presence of particles~\cite{Barrett:2008wh, Freidel:2004vi,Freidel:2004nb}.
Let us recall that particles in Riemannian gravity with a positive cosmological constant
are classif\/ied by the unitary irreducible representations of~$\so(4)$ and are therefore characterized by two half-integers.
When the particles are assumed to be without spin, they are classif\/ied by their masses only (which label
the so-called simple representations,
see Section~\ref{subsubsec-BC}) and are characterized by one
half-integer denoted $m$. After coupling to gravity, $m$ becomes a label for the (simple) representation of
the quantum group $\Uq(\so(4))$ and as such it is bounded by~$k/2$.

In order to include massive particles in the framework developed by Turaev and Viro, we consider once again
a triangulation $\Delta$ of $\cal M$ in which particles live on the links (one-cells) $\ell\in\Delta$ and have trajectories
describing closed loops $\gamma_i$ in $\Delta$. The generalized Turaev--Viro amplitude,
which now depends on the data $M_i$ (the mass $m_i$ of the $i^\text{th}$ particle and the closed loop $\gamma_i$), is given by
\begin{gather*}
 \mathcal{Z}_\text{TV}(\Delta^*;M_1,\dots,M_p)
 =  \left( -\frac{q-q^{-1}}{2(k+2)}\right)^{n_0}
\sum_{j\rightarrow f}\prod_{f\in\Delta_{(g)}^*}[2j_f+1]_q\nonumber\\
\hphantom{\mathcal{Z}_\text{TV}(\Delta^*;M_1,\dots,M_p) =}{}
\times
\prod_{f\in\Delta_{(p)}^*}\frac{[(2j_f+1)(2m_f+1)]_q}{[2m_f+1]_q}
\prod_{v\in\Delta^*}
\left\{\begin{array}{ccc}
         j_1 & j_2 & j_3\\
         j_4 & j_5 & j_6\end{array}\right\}_q,
\end{gather*}
where $\Delta_{(p)}^*$ denotes the set of faces dual to the links $\ell$ where the particles live, and $\Delta_{(g)}^*$
denotes the set of faces dual to the links $\ell$ where there are no particles.
It has been shown that $\mathcal{Z}_\text{TV}(\Delta;M_1,\dots,M_p)$
is a knot invariant which does not depend on the choice of the triangula\-tion~$\Delta$ for~$\cal M$, and also that
\begin{gather*}
\mathcal{Z}_\text{TV}(\Delta^*;M_1,\dots,M_p) = \vert \mathcal{Z}_\text{WRT}(\mathcal{M};M_1,\dots,M_p)\vert^2.
\end{gather*}
This provides a clear relation between the covariant and canonical quantizations and
demonstrates their mutual consistency in the present case.

\subsubsection{Discussion}

Before going to the loop and spin foam quantizations of three-dimensional gravity, we would like to end the review of the
quantization of Chern--Simons theory with a short discussion. First, let us emphasize once again the important role played by
quantum groups in the canonical and covariant quantizations. They appear in a very natural way in the combinatorial
quantization and were used to def\/ine the Turaev--Viro state sum. The question whether they are fundamental symmetries
of quantum gravity or simply mathematical tools to construct the physical Hilbert space is still open. Indeed,
quantum groups can be viewed as an intermediate step towards the construction of physical states, which are at the end of the day
invariant under the action of quantum groups. In this sense, we do not see their ef\/fect on the
physical states.

\looseness=-1
Here we have considered exclusively Riemannian gravity with a positive cosmological constant. The reason is that it is the only case
relevant for gravity which involves a compact group. All the other cases (Riemannian or Lorentzian)
are formulated using non-compact gauge groups and they are much more dif\/f\/icult to deal with. It is nonetheless possible to perform
the canonical quantizations for Riemannian gravity with a vanishing or negative cosmological constant. However, the
spin foam quantizations of these cases lead to state sums that need to be regularized. Usually, the
regularization breaks the topological invariance and this makes these models less interesting.

\looseness=-1
Thus, we conclude by emphasizing that Riemannian three-dimensional gravity with a positive cosmological constant
admits a complete quantization, both from the canonical and the covariant point of view.
It is therefore a good ground to test the ideas and the techniques of loop quantum gravity.
In particular, LQG should reproduce the results obtained by Witten to claim that it is a~serious candidate for
quantizing gravity. We will see in the next section if this is indeed the case.

\subsection{Quantum gravity in the BF formulation}
\label{subsec_LQGcov}

Loop quantum gravity has been introduced as an attempt to quantize
gravity in four dimensions. Soon after the foundations of LQG were established, it was applied to
the quantization of three-dimensional gravity in order to test its ideas and techniques
\cite{Ashtekar:1989qd,Ashtekar:1994uy,Freidel:2002hx,Marolf:1993au,Noui:2004iy,Thiemann:1997ru}.
Although many aspects of three-dimensional quantum gravity can be recovered from the LQG point of view,
the results presented in the previous subsection are still out of reach.
For instance, in the Riemannian case without cosmological constant
it is possible to perform the construction of the physical Hilbert space \cite{Noui:2004iy} and
to see that the Drinfeld double of $\SU(2)$ (the relevant quantum group in the combinatorial quantization
scheme) plays a central role, even if it is somehow hidden \cite{Noui:2006ku}.
On the other hand, the construction of the physical Hilbert space in the presence of a cosmological constant
is still an open problem in LQG.
As a consequence, in the context of LQG, one does not see the emergence of quantum groups as
deformations of classical gauge symmetry algebras, and the link with knot invariants and the Jones polynomial
remains a mystery. Note however that the preliminary results of \cite{Noui:2011im}
strongly suggest that the Jones polynomials could appear in LQG,
but many things remain to be done in order to make this statement more precise.

The story with spin foam models is a bit dif\/ferent. In contrast to LQG, they were f\/irst introduced in three dimensions,
and then generalized to four dimensions.
The f\/irst spin foam model was proposed by Ponzano and Regge \cite{Ponzano:1968dq}
as an attempt to compute non-perturbatively the path integral of Riemannian gravity with vanishing cosmological
constant. However, the original model suf\/fers from divergences and requires a suitable regularization.
Although it is possible to do this without modifying the original theory,
a simple and nice regularization is provided by the inclusion of the cosmological constant
and leads to the Turaev--Viro state sum model \cite{Turaev:1992hq}.
The cosmological constant compactif\/ies the symmetry gauge group already in the classical theory,
which has the ef\/fect of curing the divergences\footnote{Note that this procedure
has been used as an attempt to regularize four-dimensional spin foam models as well~\cite{Fairbairn:2010cp,Han:2010pz, Noui:2002ag}.}.
Thus, in three dimensions we have at least two well-def\/ined spin foam models.

\looseness=-1
This section is devoted to reviewing some aspects of loop quantum gravity and spin foam models in three dimensions.
We would like to focus on the main results and the main open questions arising in this context.
Both quantizations are based on the BF formulation of three-dimensional gravity instead of the Chern--Simons
one. We start with a brief overview of the Ponzano--Regge model. In the second part,
we review the loop quantization approach. Then, we make a contact between the canonical and
covariant quantizations. Finally, we explain the preliminary results obtained in the LQG framework
for the case of a positive cosmological constant.

\subsubsection{Ponzano--Regge model}
\label{sec_3dSF}

The idea leading to the spin foam quantization of gravity is to use a discretization of
the spacetime manifold in order to def\/ine a regularization of the formal partition function
\begin{gather}
\mathcal{Z}_{\rm 3d}=\int\pD{e}\pD{\omega}\,\exp\left(\I\int_\CM\tr(e\wedge F[\omega])\right).
\label{Z3d}
\end{gather}
To this end, we choose an arbitrary
cellular decomposition $\Delta$ of the three-manifold~$\CM$,
together with its dual two-complex~$\Delta^*$.
Since the triad $e^i_a$ is a one-form, it is natural to integrate
it along the one-cells $\ell$ of $\Delta$ (dual to $f\in\Delta^*$)
to form the $\su(2)$-valued elements $X_f$.
The connection of the gauge group $G=\SU(2)$ is discretized by computing its holonomy
$g_e$ along the edges $e$ of $\Delta^*$, so that the discrete curvature is associated to the faces
and is given by the product of holonomies along the edges lying on the boundary of~$f$:
\begin{gather*}
g_f=\prod_{e\subset f}g_e.
\end{gather*}

With these variables, the discretization of the action appearing in \eqref{Z3d}
can be written as
\begin{gather*}
S[X_f,g_f]=\sum_{f\in\Delta^*}\tr(X_fg_f),
\end{gather*}
and the partition function becomes
\begin{gather}
\mathcal{Z}_{\rm PR}(\Delta^*)=\left(\prod_{f\in\Delta^*}
\int_{\su(2)}\de X_f\right)\left(\prod_{\vphantom{f}e\in\Delta^*}
\int_{\SU(2)}\de g_e\right)\exp\left(\I\sum_{f\in\Delta^*}\tr(X_fg_f)\right),
\label{initPRpf}
\end{gather}
where $\de X_f$ is the Lebesgue measure on $\su(2)\simeq\mathbb{R}^3$
and $\de g_e$ the Haar measure on $\SU(2)$.
The integral over $X_f$ can be trivially performed to f\/ind
\begin{gather}
\mathcal{Z}_{\rm PR}(\Delta^*)=\left(\prod_{\vphantom{f}e\in\Delta^*}\int_{\SU(2)}\de g_e\right)\prod_{f\in\Delta^*}\delta(g_f),
\label{ZBF3delta}
\end{gather}
where the delta distribution over $\SU(2)$ imposes the f\/latness of the connection\footnote{Technically,
the integration over the variables~$X_f$ leads to a delta function
over $\SO(3)$ and not $\SU(2)$, but we will ignore this minor complication in the following.
In fact, the model can be slightly modif\/ied in order to immediately yield the $\SU(2)$ delta function
\cite{Freidel:2004vi,MR2476216}.}.
Using the Peter--Weyl decomposition, one can write
\begin{gather}
\label{pf}
\mathcal{Z}_{\rm PR}(\Delta^*)
=\sum_{j\rightarrow f}\prod_{f\in\Delta^*}(2j_f+1)
\left(\prod_{\vphantom{f}e\in\Delta^*}\int_{\SU(2)}\de g_e\right)
\prod_{f\in\Delta^*}\chi_{j_f}\left(\prod_{e\subset f}g_e\right),
\end{gather}
where the sum is taken over all possible $\SU(2)$ representations $j$
labeling the set of faces $f\in\Delta^*$. For arbitrary cellular decompositions $\Delta$,
let us call $\text{F}$ the number of faces meeting at every edge $e\in\Delta^*$.
In~(\ref{pf}) we will have an integral over $g_e$ of~$\text{F}$ products
of its representation matrices.
Such an integral can be evaluated as follows:
\begin{gather*}
\int\de g_e\,\Db^{(j_1)}(g_e)\cdots \Db^{(j_{\text{F}})}(g_e)=\sum_{i_e}i_e^{(\vec\jmath)}i_e^{(\vec\jmath)*},
\end{gather*}
where $\vec{\jmath}=(j_1,\dots,j_{\text{F}})$ denotes the collection of spins coloring the faces meeting at the edge~$e$,
$\Db^{(j)}(g_{\ell})$ is the spin-$j$ representation matrix of the group element $g_\ell$,
and the sum runs over an orthonormal basis in the space $\text{Inv}^{(\vec\jmath)}_{\SU(2)}\equiv
\text{Inv}\big( \mathcal{H}_{\SU(2)}^{(j_1)}\otimes\dots\otimes\mathcal{H}_{\SU(2)}^{(j_{\text{F}})}\big) $
of invariant intertwiners between the representations $\vec \jmath$.
Then, all the intertwiners meeting at a vertex $v$ can be contracted
to def\/ine a vertex amplitude $A_v(j_{f\supset v},i_{e\supset v})$.
Finally, the partition function can be written as a~sum over colorings decorating the dual two-complex
of the amplitudes constructed from~$A_v$:
\begin{gather}
\mathcal{Z}_{\rm PR}(\Delta^*)=\sum_{j\rightarrow f}\sum_{i\rightarrow e}
\prod_{f\in\Delta^*}(-1)^{2j_f}(2j_f+1)\prod_{v\in\Delta^*}A_v(j_{f},i_{e}).
\label{ZBF3final}
\end{gather}

To clarify the meaning of this formula, let us assume that the cellular decomposition
$\Delta$ is simplicial (i.e.\ three-dimensional). In this case, vertices in $\Delta^*$
are four-valent, while edges are three-valent. The vertex amplitude
is therefore given by a contraction of four (normalized) three-valent intertwiners,
which is the so-called totally symmetric $\{6j\}$ symbol, and~\eqref{ZBF3final} takes the form~\cite{Ponzano:1968dq}
\begin{gather}\label{3d partition function}
\mathcal{Z}_{\rm PR}(\Delta^*)
=\sum_{j\rightarrow f}\prod_{f\in\Delta^*}(-1)^{2j_f}(2j_f+1)\prod_{v\in\Delta^*}
\left\{\begin{array}{ccc}
         j_1 & j_2 & j_3\\
         j_4 & j_5 & j_6\end{array}\right\},
\end{gather}
where the labels $j_1,\dots,j_6$ are associated to the six links bounding
the tetrahedron dual to the vertex $v\in\Delta^*$, and we absorbed all possible sign factors associated
to the edges into the normali\-za\-tion of the
$\{6j\}$ symbol (see~\cite{Barrett:2008wh}).
Notice that the sum over intertwiners
has now disappeared since there is a unique (up to normalization)
three-valent intertwiner, i.e.\
$\dim \big(\text{Inv}^{(\vec\jmath)}_{\SU(2)}\big)=1$.

Some important comments concerning the result \eqref{3d partition function} are in order:
\begin{itemize}\itemsep=0pt
\item
An extremely important observation is that the partition function~(\ref{3d partition function})
does not depend on the choice of discretization $\Delta$, which is a manifestation
of the topological nature of the quantum theory. More precisely, given two dif\/ferent cellular decompositions
$\Delta$ and $\tilde{\Delta}$ and using the Biedenharn--Elliot relation, one can prove the following relation:
\begin{gather}
\text{s}^{-n_0}\CZ_{\rm PR}(\Delta^*)=\text{s}^{-\tilde{n}_0}\CZ_{\rm PR}(\tilde{\Delta}^*),
\label{rel3dspinfoam}
\end{gather}
where $\text{s}=\sum_j(2j+1)$ is a divergent factor and $n_0$ is the number of zero-cells in $\Delta$,
i.e.\ the number of bubbles of $\Delta^*$. Thus, \eqref{3d partition function} def\/ines a formal (due to divergences)
three-manifold invariant and, as a result, one can formally write that
\begin{gather*}
\mathcal{Z}_{\rm PR}(\Delta^*) = \mathcal{Z}_{\rm PR}({\cal M}).
\end{gather*}
In particular, this implies that
an appropriate choice of a single two-complex is suf\/f\/icient to provide the full transition amplitude
and the sum over triangulations should be omitted in its def\/inition.

\item
Although the sum over dif\/ferent two-complexes in the Ponzano--Regge model is redundant,
it is meaningful to ask the question of the sum over topologies.
This question has been partially studied in~\cite{Freidel:2002tg} and~\cite{Magnen:2009at}
using the group f\/ield theory formalism for spin foam models. It was shown that the sum over triangulations
in three-dimensional quantum gravity can be tamed rigorously and paves the way for the renormalization program in
group f\/ield theory~\cite{Freidel:2009hd}.

\item
The divergence of the sum over spins in $\mathcal{Z}_{\rm PR}(\Delta^*)$ and in the def\/inition of the factor $\text{s}$
makes the relation \eqref{rel3dspinfoam} rather formal.
In fact, this divergence has a physical origin and arises due to a remnant of
the gauge symmetry at the discrete level~\cite{Freidel:2004vi}. Indeed, the spin foam state sum is
a version of the path integral where no gauge-f\/ixing was done. On the other hand, the classical gauge group
in the present case is the non-compact (Euclidean) Lie group $\ISU(2)$ and the divergences appear
as a manifestation of its inf\/inite volume. They can be removed by partially f\/ixing the gauge freedom
remaining after discretization, which amounts to eliminating the bubbles from~$\Delta^*$.
Another way to understand these divergences is to see that
they result from the ill-def\/ined product of distributions.
This product can be regularized by replacing in~(\ref{ZBF3delta}) the $\delta$-distributions over the group $\SU(2)$
by the heat kernel. This allows to explicitly identify the divergent part of the amplitudes and to relate
it to the bubbles in~$\Delta^*$ \cite{Bonzom:2010ar,Bonzom:2010zh,Bonzom:2011br}.
Then the physical convergent part turns out to produce the well-known Reidemeister torsion.
In fact, this was already expected from the work of Witten \cite{Witten:2010cx},
and was proved in a less general case in~\cite{Barrett:2006ru}.
As we are going to see in what follows, the canonical quantization enables one to
avoid the infrared divergencies as well.
\end{itemize}

Finally, let us brief\/ly discuss the inclusion of particles into the model.
In the language of the quantization \`a la Witten,
the particles correspond to Wilson loop observables in the Chern--Simons theory with the non-compact gauge group
$\ISU(2)$. They are supposed to have closed trajectories describing knots $\gamma_i$ in spacetime.
For the particles with masses $m_i$ and with no spin, the partition function
of the coupled system is formally given by \cite{Freidel:2004vi,Freidel:2004nb}
\begin{gather}
\mathcal{Z}_{\rm PR}(\Delta^*;M_1,\dots,M_p)
=\sum_{j\rightarrow f}\prod_{f\in\Delta_{(g)}^*}(-1)^{2j_f}(2j_f+1)\nonumber\\
\hphantom{\mathcal{Z}_{\rm PR}(\Delta^*;M_1,\dots,M_p)=}{}\times
\prod_{f\in\Delta^*_{(p)}} (-1)^{2j_f} \chi_{j_f}(m_f)\prod_{v\in\Delta^*}
\left\{\begin{array}{ccc}
         j_1 & j_2 & j_3\\
         j_4 & j_5 & j_6\end{array}\right\},
\label{PRstatesum-particle}
\end{gather}
where $M_i=(m_i,\gamma_i)$ is the data of each particle,
$\Delta^*_{(p)}$ are the faces in the cellular decomposition
that are crossed by the trajectories $\gamma_i$,
and $\Delta^*_{(g)}$ the faces that are not crossed.
A face $f\in\Delta^*$, initially colored by a representation~$j_f$ and
crossed by a particle of mass $m$, inherits a new color~$m_f$
which is precisely the mass of the particle $m_f=m$. As a result,
its weight $W_f$ in the state sum is no longer $(2j_f+1)$ but
\begin{gather}\label{particle weight}
W_f = \chi_{j_f}(m_f) = \frac{\sin((2j_f+1)m_f)}{\sin m_f}.
\end{gather}
This factor has a clear group-theoretic interpretation in terms of characters.
Although the state sum~\eqref{PRstatesum-particle} is still formal, in some cases it can be regularized.

Note that the inclusion of particles makes it clear that Riemannian quantum
gravity without cosmological constant is related to the quantum double $\text{D}\SU(2)$,
since the particles are shown to be classif\/ied by unitary irreducible representations
of $\text{D}\SU(2)$ and not by those of $\ISU(2)$. This fact can be illustrated in two ways.
The f\/irst one is that the masses of the particles are bounded by $2\pi$,
as can be seen from the expression (\ref{particle weight}), where the mass is def\/ined mod~$2\pi$ only.
The physical origin of this bound can be traced back to the conical singularity created by the particle
after its coupling to gravity. The def\/icit angle of the cone coincides (in some units) with the mass
of the particle, and therefore the mass cannot exceed the value~$2\pi$. And we know that the unitary
irreducible representations of $\text{D}\SU(2)$ are labeled by a couple $(m,s)$ with $m\le 2\pi$, whereas
this is not the case for representations of $\ISU(2)$.
The second argument is based on the way in which interactions between particles are described, since it
can be shown to follow the rules of $\text{D}\SU(2)$ recoupling theory~\cite{Koornwinder:1998xg}.
This latter is drastically dif\/ferent from the recoupling theory of~$\ISU(2)$.

The coupling to particles of\/fers the possibility to build a quantum f\/ield theory
over a quantum background, which is often referred to as a third quantization.
Let us explain how this comes about.
It is not possible to fully quantize a matter (or gauge)
f\/ield coupled to gravity using the standard techniques.
Indeed, the standard approach consists in f\/irst quantizing
the f\/ield on a f\/lat background, and then quantizing perturbatively
the gravitational f\/ield around the f\/lat metric.
We know that this does not work because of the non-renormalizability of perturbative gravity.
Another idea would be to quantize the gravitational
f\/ield f\/irst, while keeping the matter (or gauge) degrees of freedom classical.
In four-dimensional gravity, such an idea seems to be unrealistic,
because there is no regime in which spacetime
is quantum-mechanical while matter remains classical.
In three dimensions however, this idea does make sense because
matter does not really interact with gravity.
This was realized explicitly and f\/irst done in the context of
spin foam models~\cite{Freidel:2005me,Freidel:2005bb}, for matter f\/ields that are massive point particles.
The result is very interesting. Integrating out the gravitational
degrees of freedom while keeping the particles classical turns
the classical f\/lat geometry into a non-commutative geometry.
In other words, quantizing particles coupled to gravity, is equivalent
to quantizing particles without gravity but leaving in a non-commutative spacetime.
Afterwards, it was realized \cite{Joung:2008mr, Noui:2006ku,Noui:2006kv} that the emergence of a non-commutative space
has a very simple physical interpretation. Indeed, functions on this non-commutative
space form a non-commutative algebra which is covariant under
the action of the quantum double $\text{D}\SU(2)$. More precisely,
the classical isometry algebra $\ISU(2)$ of the f\/lat Euclidean space
$\mathbb E^3$ is deformed at the Planck scale and becomes
the Drinfeld double $\text{D}\SU(2)$. This result is very exciting because
it is an explicit realization of the old idea that quantum
gravity turns classical geometry into quantum geometry.
It can be generalized to the case
of Riemannian gravity with a non-vanishing cosmological constant.
When $\Lambda >0$, the classical isometry algebra
$\so(4)$ is quantized into $\text{D}\Uq(\su(2))$ with $q$ a root of unity,
and when $\Lambda<0$, the classical isometry algebra
$\so(3,1)$ is deformed into $\text{D}\Uq(\su(2))$ with $q$ real.
As expected, these quantum groups are the ones appearing
in the context of the combinatorial quantization.
The associated non-commutative spaces are the quantum analogues of the
three-sphere $\mathbb S^3$ (for positive $\Lambda$) and
of the three-hyperbolic space $\mathbb H^3$ (for negative $\Lambda$).

\subsubsection{Canonical loop quantization with $\Lambda=0$}
\label{subsec-canq3d}

Now let us look at the canonical loop quantization of three-dimensional Riemannian gravity.
The starting point is the classical phase space def\/ined by the Poisson brackets (\ref{BF Poisson bracket})
together with the constraints (\ref{continuous constraints}). We do not consider particles for the moment
and we assume that there is no cosmological constant, i.e.~$\Lambda=0$.
In contrast to the combinatorial quantization, we start with an inf\/inite-dimensional phase space
and take its quotient with an inf\/inite-dimensional gauge group generated by the inf\/inite set
of constraints~(\ref{continuous constraints}). The result is a f\/inite-dimensional phase space,
as follows from the classical analysis.
This strategy is chosen in order to mimic the situation in four dimensions.
In this way, one can use three-dimensional gravity as a toy model to test the ideas of the loop quantization.

As in the case of the combinatorial quantization, the canonical quantization proceeds in two steps.
The f\/irst one is to promote the classical variables to an operator algebra and to determine its
unitary irreducible representations, which def\/ines the space of quantum states.  In loop quantum gravity,
the state space is obtained by completion of the so-called space of cylindrical functions,
which itself is obtained by taking the projective limit of the spaces $\text{Cyl}(\Gamma)=L^2\(G^{\text{L}_\Gamma}\)$
associated with particular graphs $\Gamma$.  $\text{Cyl}(\Gamma)$ is the space of functions
of the $G$-valued holonomies assigned to the links $\ell$ of the graph,
with the scalar product def\/ined by
\begin{gather}
\label{physscalprod}
\langle \Psi_{\Gamma,f_1}, \Psi_{\Gamma,f_2}\rangle = \int_{G^{\text{L}_\Gamma}} \prod_{\ell=1}^{\text{L}_\Gamma} \de g_\ell\,
\overline{f_1(g_1,\ldots, g_{\text{L}_\Gamma})} f_2(g_1,\ldots,g_{\text{L}_\Gamma}),
\end{gather}
where $f_1$ and $f_2$ represent the functions on the group and
$\text{L}_\Gamma$ denotes the cardinal of the set of oriented links of $\Gamma$.
Note that the topological nature of the theory in three dimensions allows one
to work on a single graph, as long as it is suf\/f\/iciently ref\/ined to resolve the topology of $\Sigma$.
The resulting quantum theory will be independent of the choice of the graph. Therefore, in
three dimensions, the projective limit is redundant.
The second step is to construct the kinematical and physical Hilbert spaces.
This can be done by promoting the constraints (\ref{continuous constraints})
to operators acting on the space of cylindrical functions $L^2\(G^{\text{L}_\Gamma}\)$.
Schematically,  kinematical states span the kernel of the quantum operator
associated to the discretized version of the torsion $T(x)$,
and physical states are kinematical states that are in the kernel
of the operator corresponding to the curvature $F(x)$.

\subsubsection*{The kinematical Hilbert space}

The basic discrete variables of loop quantum gravity  are the holonomies $g_\ell$ of the connection
$\omega$ along links $\ell$ of $\Gamma$
and the ``f\/luxes'' $E_\ell$ of the triad f\/ield along the links $\ell^*$ dual to the links $\ell$.
They are def\/ined by
\begin{gather*}
g_\ell =\Pexp\int_\ell \omega,\qquad
E^i_\ell = \int_{\ell^*} e^i,
\end{gather*}
where $\Pexp$ denotes the path ordered exponential\footnote{Notice that the f\/luxes def\/ined in this way are not gauge-covariant
under the action of gauge transformations of the group~$G$. This can be f\/ixed by def\/ining their smearing with
an additional holonomy dependance~\cite{freidelgz11, thiemannQSDVII}, but this is however not important for the general discussion since
we will only be interested in the action of the f\/lux operators on functionals of the holonomies.}.
On the space~$\text{Cyl}(\Gamma)$ the holonomy acts by multiplication and the f\/lux by derivation.
These actions are completely characterized by the relations
\begin{gather}\label{LQG representation}
{g_{\ell'}} \rhd g_{\ell}=g_{\ell'}g_\ell,
\qquad
{E}^i_{\ell'} \rhd g_\ell
=-\I \delta_{\ell,\ell'} \, \lp \,  g_{\ell_1} \tau^i g_{\ell_2},
\end{gather}
where $\lp=\hbar G$ is the three-dimensional Planck length, $\ell_1$ is the part of the oriented link $\ell$ before
the intersection $\ell \cap (\ell')^*$, and $\ell_2$ is the remaining part.
This formula is a direct quantization of the Poisson bracket~(\ref{BF Poisson bracket}). It follows that the operators
$E_\ell$ act as vector f\/ields on the space of cylindrical functions and the action (\ref{LQG representation}) provides
a unitary representation of the algebra of quantum operators on $\text{Cyl}(\Gamma)$. Note that
the f\/lux operator is not gauge covariant
as it is written, however its action on a holonomy is gauge covariant and therefore it is well-def\/ined.

The kinematical Hilbert space $\Hk(\Gamma)$ is obtained as the set of solutions to
the quantum Gauss constraint. They appear as elements of $L^2\(G^{\text{L}_\Gamma}\)$
that are invariant under the action of the gauge group $G$ at the nodes $n$ of the graph:
\begin{gather*}
\Hk(\Gamma) = L^2\big(G^{\text{L}_\Gamma}\big)/G^{\text{N}_\Gamma} .
\end{gather*}
Due to the left and right invariance of the Haar measure on $G$, the norm
(\ref{physscalprod}) is compatible with the action of the gauge group and therefore
induces a norm on $\Hk(\Gamma)$, whereas
the representation  def\/ined in \eqref{LQG representation} provides a unitary
representation of kinematical operators on this space.

In our case, $G=\SU(2)$, and a dense basis of $\Hk(\Gamma)$ is provided by the spin network functions, obtained by
assigning a representation of $G$ to each link $\ell$ and an intertwiner to each node
$n$. More precisely,
harmonic analysis over the group $\SU(2)$ enables one to expand any cylindrical
function in terms of unitary irreducible representations as
\begin{gather*}
\Psi_{\Gamma,f}[\omega]=\sum_{\vec{\jmath}}f^{(\vec{\jmath})}\Db^{(j_1)}(g_1)
\cdots \Db^{(j_L)}(g_{\text{L}_\Gamma}),
\qquad
f^{(\vec{\jmath})}\in\left(\bigotimes_{\ell=1}^{\text{L}_\Gamma}\mathcal{H}_{\SU(2)}^{(j_\ell)}\right)
\otimes\left(\bigotimes_{\ell=1}^{\text{L}_\Gamma}\mathcal{H}_{\SU(2)}^{(j_\ell)}\right)^*.
\end{gather*}
With this decomposition of the states,
it is easy to see that imposing the Gauss constraint amounts to requiring that the intertwiners,
attached to the nodes of the graph and used to contract the indices of the representation matrices,
be $\SU(2)$-invariant:
\begin{gather*}
i_n^{(\vec\jmath)}\in\text{Inv}\left(\bigotimes_{\ell\supset n}\mathcal{H}_{\SU(2)}^{(j_\ell)}\right).
\end{gather*}
An orthonormal basis of the gauge-invariant
Hilbert space $\Hk(\Gamma)$ is then given by the so-called spin network states
\begin{gather}
\CS_{(\Gamma,\vec{\jmath},\vec{\imath})}[g_\ell]
\equiv\bigotimes_\ell \Db^{(j_\ell)}(g_\ell)\cdot\bigotimes_n i_n^{(\vec\jmath)},
\label{defSpinNet}
\end{gather}
where the dot denotes the contraction between the indices of the intertwiners and the representation matrices.

The construction of the kinematical Hilbert space can be easily generalized to
account for the presence of particles~\cite{Noui:2004iz}.
Let us recall that particles in Euclidean space are classif\/ied by unitary irreducible representations of the isometry group
$\ISU(2)$. Thus, the physical Hilbert space for a massive particle of mass $m$ and spin $s$ is nothing but the module
$\CH^{(m,s)}$ of the associated representation of $\ISU(2)$, which is def\/ined by~\cite{Koornwinder:1996uq}
\begin{gather*}
\CH^{(m,s)}=\big\{ \phi \in L^2(\SU(2)) \; \vert \; \phi(xh(\theta)) = e^{\I\theta s} \phi(x),\; \forall \, x \in \SU(2) \big\},
\end{gather*}
where $h(\theta) \in\SU(2)$ is an element of a given maximal Cartan torus of $\SU(2)$.
This latter is isomorphic to $\text{U}(1)$ and can be chosen such that $h(\theta)= e^{\theta\tau_1}$,
as in~(\ref{orbits htheta}). As $\CH^{(m,s)}$ is
a subvector space of $L^2(\SU(2))$, it is naturally endowed with the $L^2$ measure on $\SU(2)$ and therefore inherits
a Hilbert space structure. Let us remark that $\CH^{(m,0)}\simeq L^2(\SU(2)/\text{U}(1))$, which makes the study
of particles without spin a bit simpler.
Applying the Peter--Weyl
decomposition theorem, it is straightforward to show that $\phi \in\CH^{(m,s)}$ can be decomposed
in terms of  $\SU(2)$ representations as follows:
\begin{gather*}
\phi(x)= \sum_j\sum_{n=-j}^j \phi^{j}_n\Db^{(j)}_{ns}(x) .
\end{gather*}
Therefore, the set $\{\Db^{(j)}_{ns}, \; j \in \mathbb N/2, \; n \in [-j,j]\}$ gives an orthogonal basis
of $\CH^{(m,s)}$. Elements of this set are the point particle analogues of spin networks.

Now we have all the ingredients to construct the kinematical Hilbert space for gravity coupled to point particles.
In order to do so, we choose a graph $\Gamma$, suf\/f\/iciently ref\/ined to resolve the topology of $\Sigma$,
with particles sitting at its nodes. We distinguish the nodes $\text{N}_\Gamma^{(p)}$
where particles are present and its complement $\text{N}_\Gamma^{(g)}$.
Then the space of cylindrical functions for the coupled
system is def\/ined by the tensor product
\begin{gather*}
\text{Cyl}(\Gamma;M_1,\dots,M_p) = L^2\big(G^{\text{L}_\Gamma}\big) \otimes \(\bigotimes_{i=1}^p\CH^{(m_i,s_i)}\)
\end{gather*}
between the space $L^2\(G^{\text{L}_\Gamma}\)$ encoding the gravitational degrees of freedom and the spaces $\CH^{(m_i,s_i)}$
encoding the degrees of freedom of the particles attached to the nodes $\text{N}_\Gamma^{(p)}$.
At this stage, gravity and the particles are not coupled yet. The coupling appears after imposing the constraints.
In particular, imposing the Gauss constraint leads to the kinematical Hilbert space
\begin{gather*}
\Hk(\Gamma;M_1,\dots,M_p) = \text{Cyl}(\Gamma;M_1,\dots,M_p)/G^{\otimes \text{N}_\Gamma} .
\end{gather*}
The part of the symmetry group acting on the nodes with particles
acts simultaneously on cylindrical functions and particles states.
It is clear that the requirement of the invariance under this
simultaneous action is responsible for the coupling between
the gravitational degrees of freedom and the degrees of freedom of the particles.

\subsubsection*{Physical scalar product and transition amplitudes}
\label{subsec_phys3d}

The kinematical Hilbert space $\Hk$ carries an action of the quantum curvature constraint,
which still needs to be imposed in order to obtain the physical Hilbert space~$\Hp$.
For this purpose, we are going to construct a sum over histories (or spin foam) representation of the so-called
``projector'' onto physical states
\begin{gather}\label{projectorF}
\CP_F\equiv\prod_{x\in\Sigma}\delta\big(\hat{F}[\omega]\big)
=\int \pD{\alpha}\,\exp\left(\I\int_\Sigma\tr\big(\alpha\hat{F}[\omega]\big)\),
\end{gather}
where $\alpha\in\su(2)$. For the moment this ``projector'' is mathematically ill-def\/ined, even if its
physical content is clear.
The fact that it is written as an inf\/inite product of delta distributions implies that
one has to introduce a regularization to give it a meaning,
but also that it cannot be a ``projector'' because $\CP_F^2$
does not make sense. We will assume in this section that $\Sigma$
is a~closed Riemann surface of genus~$\text{g}$ with no punctures.

The projector can be def\/ined using the Gelfand--Naimark--Segal (GNS) construction
(see~\cite{Ashtekar:2002sn} for examples of the GNS construction in LQG).
The idea is to f\/ind a positive linear functional $\varpi$ on the space $\text{Cyl}(\Sigma)$ of cylindrical functions,
which is called GNS state and def\/ines the physical scalar product
between two elements $\Psi_1,\Psi_2\in\text{Cyl}(\Sigma)$ as follows:
\begin{gather*}
\langle \Psi_1, \Psi_2 \rangle_{\text{phys}} \equiv \varpi\big(\overline{\Psi_1} \Psi_2\big).
\end{gather*}
This state should enforce the projection onto the physical space, which can be written formally~as
\begin{gather*}
\langle \Psi_1, \hat{F}[\omega] \Psi_2 \rangle_{\text{phys}}=0.
\end{gather*}
Two ingredients are needed in order to make this construction concrete and rigorous:
\begin{itemize}\itemsep=0pt
\item A cellular decomposition of the surface $\Sigma$ which supports the graph $\Gamma$.
This decomposition is in fact arbitrary, and in the end the physical scalar product will not depend on it.

\item A regularization of the curvature operator $\hat{F}[\omega]$. The classical curvature $F[\omega]$ is
not a~cylindrical function and, when promoted to a quantum operator
$\hat{F}[\omega]$, its action by multiplication does not leave the
space of cylindrical functions invariant. For this reason,
we need to introduce a regularization of the quantum curvature constraint
such that its action on $\text{Cyl}(\Sigma)$ is well-def\/ined.
The projection onto the f\/lat connections enforces the condition
that the holonomy of the connection around any cell be trivial.
It is therefore natural to trade the formal def\/inition~(\ref{projectorF})
for the discrete operator
\begin{gather}\label{canonical projector}
\CP \equiv \prod_{c} \delta(g_c),
\end{gather}
where $g_c\in \SU(2)$ is the holonomy around the cell $c$ and the product runs over the f\/inite set of cells
of the cellular decomposition\footnote{Another regularization was recently introduced in~\cite{Bonzom:2011hm}. It
is particularly interesting because it mimics the situation in four dimensions.}.
This regularized ``projector'' onto the physical states is a well-def\/ined distribution because,
in contrast to~(\ref{projectorF}), the product is f\/inite. It is therefore a good candidate for the GNS state.
As a result, the physical scalar product between two cylindrical functions $\Psi_1$ and $\Psi_2$ is def\/ined by
\begin{gather*}
\langle \Psi_1, \Psi_2 \rangle_{\text{phys}} \equiv \varpi(\overline{\Psi_1} \Psi_2)
= \langle \Psi_1, \CP \Psi_2 \rangle_{\text{kin}}.
\end{gather*}
\end{itemize}
Nonetheless, this is not the end of the story because one has to verify
that the product of delta distributions does not introduce divergences and that $\varpi$ is
a positive state. This indeed turns out to be the case. To address the issue of the divergences,
it is useful to use the following bound on the physical scalar product:
\begin{gather*}
\langle \Psi_1, \Psi_2 \rangle_{\text{phys}} \leq C \sum_j (2j+1)^{2-2\text{g}},
\end{gather*}
where $C$ is a constant depending on the states. This bound can easily
be obtained by decompo\-sing the delta distributions into $\SU(2)$ representations.
As a result, the physical scalar product is well-def\/ined when the genus
$\text{g}\leq 2$, but it requires an additional regularization
when $\text{g}=0$ and $\text{g}=1$.
The case of the sphere $\text{g}=0$ is easy because the divergence
is due to a redundancy in the product of delta distributions in~$\CP$,
which is a consequence of the discrete analogue of the Bianchi identity.
One can show that eliminating a single arbitrary cell holonomy from
the pro\-duct in~\eqref{canonical projector} makes the ``projector'' well-def\/ined.
The case of the torus ($\text{g}=1$) is more subtle and needs an explicit description of
the physical Hilbert space in order to be dealt with~\cite{Noui:2004iy}.

As a result, one has a well-def\/ined physical scalar product between cylindrical functions. However, one does
not have an explicit description of the physical Hilbert space yet. To obtain such a description, we need to quotient
the kinematical Hilbert space endowed with the physical scalar product by the null states. The null states
have a vanishing physical norm and form an ideal.
Their presence is a consequence of the fact that the invariance of $\Sigma$ under dif\/feomorphisms
was not taken into account. It is possible to show that the physical scalar product as def\/ined above is invariant
under dif\/feomorphisms and therefore compatible with the quotient of ${\cal H}_{\text{kin}}$
by the ideal of null states. This quotient def\/ines the physical Hilbert space where
a concrete basis can be identif\/ied.

\begin{figure}[t]
\centering
\includegraphics[scale=0.42]{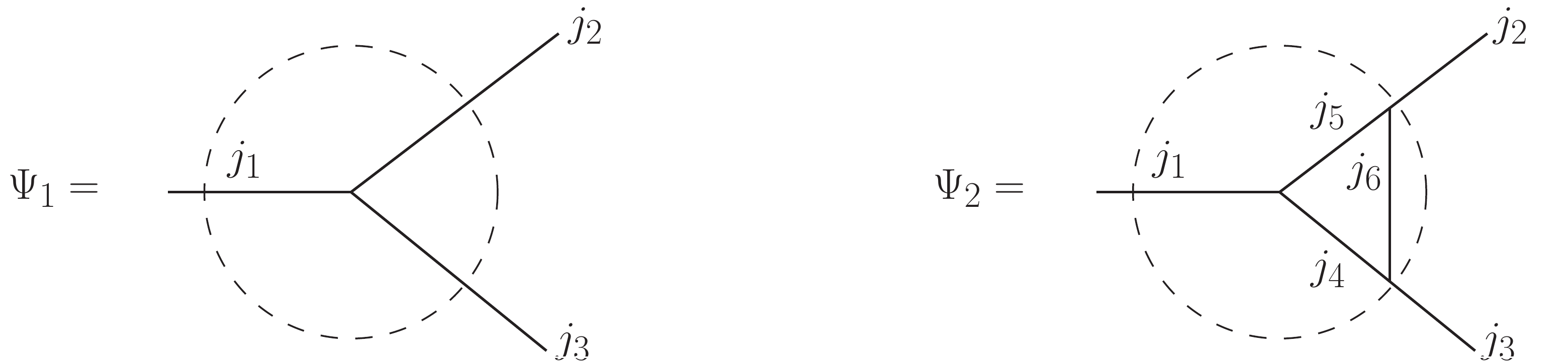}
\caption{Two spin networks states $\Psi_1$ and $\Psi_2$ which dif\/fer only in a small neighborhood.}
\label{physPR}
\end{figure}

It is now straightforward to see the relation between the physical scalar product
and the Ponzano--Regge model~\cite{Noui:2004iy}\footnote{Note that
the relation between the Ponzano--Regge boundary states and spin networks
was realized for the f\/irst time in \cite{Foxon:1994nq}.}.
To illustrate this link, let us consider two spin network states $\Psi_1$ and $\Psi_2$
associated with two graphs $\Gamma_1$ and $\Gamma_2$ which dif\/fer
only in a small region, as illustrated in Fig.~\ref{physPR}.
In this small region, $\Gamma_1$ has one vertex, whereas
$\Gamma_2$ has three vertices. Outside of this region, the spin networks (i.e.\ the graphs and the colorings)
are exactly the same. A direct calculation shows that
\begin{gather}\label{phys scalar prod 3d}
\langle \Psi_1,\Psi_2 \rangle_{\text{phys}} = \sqrt{d_{j_4}d_{j_5}d_{j_6}}
\left\{\begin{array}{ccc}
         j_1 & j_2 & j_3\\
         j_4 & j_5 & j_6\end{array}\right\},
\end{gather}
with the notations of the picture.
This factor precisely agrees with the one from \eqref{3d partition function} predicted by the Ponzano--Regge model!

The generalization of this construction to the case in which particles are present can be done easily
since on both sides we already know how the particles are incorporated.
Indeed, it just amounts to replacing~(\ref{canonical projector}) by a new projector which enforces
that the holonomy of the connection around a particle is in the conjugacy class~$\mathcal{O}_m$
associated to its mass~$m$~\cite{Noui:2004iz}.
At points where there are no particles, the connection remains f\/lat.
Therefore, the computation of the physical scalar product leads to the same expression~(\ref{phys scalar prod 3d}),
where now the dimension~$d_j$ associated to the face crossed by a particle is replaced by the character $\chi_{j}(m)$.

\subsubsection{Positive cosmological constant}
\label{subsec-positive}

According to the previous discussion, three-dimensional Riemannian gravity
with $\Lambda=0$ is an example of a theory in which there is perfect agreement
between LQG and spin foam quantization techniques.
However, the situation is still not quite satisfactory because the Ponzano--Regge model
is in general mathematically ill-def\/ined (for closed spacetimes without boundaries)
and, as a result, the link between the canonical and covariant quantizations stays
at a formal level. For instance, the evaluation of the path integral with
particles associated to closed spacetime loops leads necessarily to divergences due to the
non-compactness of the gauge group.

This provides a strong motivation for the study of the relation between LQG
and spin foam models in three-dimensional Riemannian gravity with a positive cosmological constant,
where the gauge group is compact and no divergences are expected.
However, in this case, the link between spin foam amplitudes and kinematical
states of LQG is far from being trivial. The latter are constructed from representations of the classical group $\SU(2)$,
whereas the former are def\/ined from representations of the quantum group $\Uq(\su(2))$ (when there are no particles).
Understanding how the structure of quantum groups emerges from the implementation of the curvature constraint
can be viewed as a very non-trivial test for LQG.
A f\/irst step towards a resolution of this problem has been done in~\cite{Noui:2011im, Perez:2010pm}
and we would like to present here the main results of this work.

The idea is to generalize what has been done in the case with $\Lambda=0$ and to regularize the curvature
constraint in~(\ref{continuous constraints}) in order to construct a ``projector'' onto
physical states and the physical scalar product. Then, we expect
to recover the Turaev--Viro amplitudes from the physical scalar
product between kinematical states.
When the cosmological constant is non-vanishing, the two constraints
(\ref{continuous constraints}) are equivalent to the requirement that the $\SU(2)$ connections $A_\pm$ def\/ined by
\begin{gather}\label{self anti self connection}
A_\pm = \omega \pm \sqrt{\Lambda} e
\end{gather}
be f\/lat. These two connections are in fact the self-dual and anti self-dual components of the associated
$\Sp(4)$ Chern--Simons connection.

At the kinematical level, the Gauss constraint has already been imposed, and therefore it is enough to impose that only one
of the connections (\ref{self anti self connection}) be f\/lat in order to ``project'' onto the physical states,
as clearly follows from the following relation
\begin{gather*}
F[A_\pm] = (F[\omega] + \Lambda e \wedge e) \pm \sqrt{\Lambda}\,T[e,\omega].
\end{gather*}
We will denote generically $A$ one of these connections.
If one mimics the construction for the f\/lat case ($\Lambda=0$) presented in the previous subsection,
one should impose that the holonomy of one of these connections be trivial around each cell. However, contrary to
what happens in the f\/lat case, the holonomy now depends on the non-commuting
variables $e$ and $\omega$, and therefore one encounters ordering problems
when promoting the classical variables to quantum operators.
These ordering problems are quite dif\/f\/icult to solve, and to our knowledge
no solution giving the right physical scalar product has been obtained.

In~\cite{Noui:2011im}, the quantization of the (one-parameter
family of) classical (kinematical) obser\-vab\-les~$g_\gamma[A]$, def\/ined as the holonomy of the connection $A$
along the path $\gamma$, has been explored. This operator is
the building block for the construction of the ``projector''
onto the physical states. Its action on the vacuum creates a Wilson line excitation,
i.e.\ it acts simply by multiplication by the holonomy of $\omega$ along the
path, namely
\begin{gather*}
g_{\gamma}[A] \triangleright \unit=h_{\gamma}[\omega],
\end{gather*}
where $g_{\gamma}[A]$ and $h_{\gamma}[\omega]$ denote the holonomies of the Chern--Simons connection $A$
and of the spin connection $\omega$, respectively. If however the holonomy
$g_\gamma[A]$ acts on a spin network, one expects a less trivial result.
This is because the operator $e$ in $g_\gamma[A]$ acts as a derivative operator with respect
to the components of the connection.
As a result, one expects additional contributions when the path $\gamma$ in $g_\gamma[A]$
is self-intersecting or the spin-network contains nodes on (or links transversal to)~$\gamma$.

The simplest non-trivial example is the action on a
transversal Wilson loop in the fundamental representation.
To def\/ine the quantization of~$g_\gamma[A]$, we would like to quantize each term in
its series expansion  in powers of $\sqrt{\Lambda}$.
But the problem is that quantization of these terms is potentially ill-def\/ined due
to the non-commutativity of the operators associated to~$e$.
A similar problem has recently been investigated in~\cite{Sahlmann:2011uh}, where the authors provided
a new derivation of the expectation values of holonomies in Chern--Simons theory.
In the analysis of~\cite{Sahlmann:2011uh}, the same sort of ordering ambiguities arises
due to the replacement of holonomy functionals under the path integral
by a complicated functional dif\/ferential operator. It was shown that
the known result can be recovered once a mathematically-preferred ordering,
dictated by the so-called Duf\/lo isomorphism, is adopted.
Therefore, following this example, one can also make use of this mathematical insight.
But in our case the Duf\/lo map does not do all the job. In fact, since the ambiguities in the quantization of
the Chern--Simons holonomy arise due to the presence of non-linear terms in the $e$ f\/ield,
a second piece of information has to be taken into account,
namely the quantum action of f\/lux operators in LQG.
Combining these two elements leads to a well-def\/ined quantization
for each term in the perturbative expansion in~$\sqrt{\Lambda}$.
Moreover, the series can be summed and the result can be expressed
in a closed form, leading to algebraic structures
remarkably similar to those appearing in the skein relation
def\/ining the Jones polynomial.

More precisely, if one concentrates on a single intersection (a {\em crossing})
between the path def\/ining the holonomy~$g_\gamma[A]$
and a transversal spin-network link colored by the fundamental representation $j=1/2$, one obtains
a formula which looks identical to the skein relation~(\ref{skein1})
with  $q=\exp(\I \lp/\ell_c)$ where $\lp$ is the Planck length
and $\ell_c=\Lambda^{-1/2}$ the cosmological length.
Unfortunately, despite the strict resemblance with the Jones polynomial,
this result is not suf\/f\/icient yet to recover these polynomials from LQG\footnote{Note that the quantum parameter $q$
obtained here is slightly dif\/ferent from the one obtained in the Chern--Simons
quantization, where $q=\exp(2\pi \I/(k+2))$. In this last relation, the level $k \propto \lp/\ell_c$ is shifted
by the $\SU(2)$ Coxeter number (which takes the value $N$ when the gauge group is $\SU(N)$).
The shift in the level is due to a one-loop quantum correction to the path integral, and this
correction is exact in the sense that there are no higher order contributions.
However, one does not see the emergence of this shift in the canonical quantization, neither in LQG nor
in the combinatorial quantization, where $q$ is in fact determined only at f\/irst order in~$1/k$
when $k$ goes to inf\/inity. Even in~\cite{Sahlmann:2011uh} where the authors computed Wilson loop
expectation values using the LQG techniques supplied by the Duf\/lo map, one recovers the Jones polynomials
only up to the shift in the def\/inition of the quantum parameter.
Understanding how to recover the shift in the context of the Hamiltonian quantization is certainly
a very interesting question, and some arguments have already been given by \cite{MR1011228}.}
and should be considered only as a preliminary step in this direction.
It suggests that it is not impossible to reproduce Witten's results from the loop quantization,
but the precise way to achieve this still remains unknown. The solution of the problem would require
a precise regularization of the action of the quantum holonomy~$g_\gamma[A]$ on the
holonomy~$h_\gamma[\omega]$ (which is the building block of kinematical states)
associated to the same path $\gamma$, and it is easy to imagine that such a regularization procedure
is full of ordering ambiguities.

\section{Four-dimensional quantum gravity}
\label{sec_4d}

In this section, we focus on the  case of four-dimensional gravity.
We f\/irst start with a brief reminder of the basic elements of LQG followed by the introduction of
the so-called projected spin networks. These states, being an extension of the usual spin network states,
allow to put LQG in a covariant form and provide the habitat for the boundary states
of four-dimensional spin foam models.
Then we recall some facts about the canonical analysis of the Plebanski formulation of gravity
and give an overview of various proposals for four-dimensional spin foam models available in the literature.
These proposals are then compared with the canonical quantization and f\/inally the latter is used to deduce
some generic restrictions on the form of spin foam amplitudes imposed by consistency with the canonical approach.

\subsection{Canonical approach -- spin networks and projected spin networks}
\label{subsec_spinnet}

\subsubsection{Loop quantum gravity in four dimensions}

The canonical loop quantization in $3+1$ spacetime dimensions follows the same ideas which
were presented in Section~\ref{subsec-canq3d}. However, there are several important dif\/ferences
compared to the three-dimensional Riemannian case considered there.
The f\/irst one is that in the Lorentzian signature the gauge group in the tangent space is non-compact.
This gives rise to certain dif\/f\/iculties in rigorously def\/ining various mathematical structures
important for the loop quantization. In particular, a naive generalization of spin networks
to a non-compact group is supplied with serious problems~\cite{Freidel:2002xb}.
The second dif\/ference is that the canonical analysis of the f\/irst order Hilbert--Palatini formulation
of general relativity, or its Plebanski reformulation presented in the next subsection, leads to
second class constraints which cannot be solved in a covariant way.
This fact makes the canonical structure of the theory quite involved since the symplectic
structure relevant for quantization is given now by Dirac brackets~\cite{Alexandrov:2000jw}.
Or equivalently, one can solve explicitly the second class constraints,
which leads to a simple symplectic structure~\cite{Alexandrov:1998cu,BarroseSa:2000vx,Cianfrani:2008zv,Geiller:2011cv},
but the price to pay is a very complicated form of
the f\/irst class constraints and the loss of explicit covariance.

Fortunately, there is a simple way to avoid both of these problems. It involves two ingredients.
First, one changes a bit the starting point. It is now given by the so-called Holst action \cite{Holst:1995pc}
\begin{gather*}
S_{(\im)}[e, \omega]=\frac{1}{2} \int_{\cal M} \eps_{IJKL} e^I \wedge e^J \wedge
\(F^{KL}[\omega]+\frac{1}{\im}\star F^{KL}[\omega]\),
\end{gather*}
which dif\/fers from the usual Hilbert--Palatini action by the presence of the second term.
However, this term does not af\/fect the equations of motion and at the classical level the two actions
are completely equivalent. The coupling constant $\im$ in front of the second term, which is known as the
Immirzi parameter, is therefore classically irrelevant.
The second ingredient, which is responsible for all the simplif\/ications, is to impose a partial gauge-f\/ixing.
Namely, one requires that $e^0_a=0$. This ``time gauge'' f\/ixes the boost part of the gauge freedom in the tangent space,
which therefore reduces to the $\SU(2)$ gauge group thereby solving the non-compactness issue of the Lorentz group.
Since the Lorentz covariance is broken anyway, one can also solve explicitly the second class constraints.
As a result, one f\/inds a canonical formulation very similar to the one we had in the three-dimensional Riemannian case:
the phase space is parametrized by the densitized inverse triad and an $\SU(2)$ Ashtekar--Barbero (AB) connection
\begin{gather}
\tE_i^a=\hf\eps^{abc}\eps_{ijk} e^j_b e^k_c,
\qquad
A^i_a=\Gamma_a^i(\tE)-\im \omega_a^{0i},
\label{ABcon}
\end{gather}
where
$
\Gamma_a^i(\tE)=\hf{\eps^i}_{jk}\omega_a^{jk}
$
is the Christof\/fel connection compatible with the triad.
Note that their commutation relation
\begin{gather}
\label{poisson-bracket-4d}
\big\lbrace A^i_a(x),\tE^b_j(y)\big\rbrace=\im\,\delta^i_j\delta^b_a\,\delta^2(x-y)
\end{gather}
involves the Immirzi parameter and this fact leads to important physical consequences.
On this phase space, one should impose a set of f\/irst class constraints generating the gauge
symmetries: three Gauss constraints~$G_i$ generate~$\SU(2)$ transformations, three vector constraints~$H_a$ are responsible
for spatial dif\/feomorphisms, and the Hamiltonian constraint $H$ encodes the time reparametrizations.

The loop quantization proceeds then as in Section~\ref{subsec-canq3d}. Assuming that the holonomies
of the AB connection give rise to well-def\/ined operators in the Hilbert space of quantum gravity,
one constructs a kinematical Hilbert space $\Hk$ with the scalar product induced by the same formula~\eqref{physscalprod}
def\/ined on cylindrical functions. The orthonormal basis in~$\Hk$ is then provided by
the usual $\SU(2)$ spin networks.

On this Hilbert space one still has to impose the vector and Hamiltonian constraints.
The conceptual meaning of the former is clear -- it amounts to considering the equivalence classes
of embedded spin networks up to three-dimensional spatial dif\/feomorphisms.
As a result, only some of their topological properties become
important\footnote{Nevertheless, even at this level there seem to remain some unsolved issues
which, in particular, lead to the fact that the resulting Hilbert space is still non-separable \cite{Grot:1996kj}.}.
On the other hand, the imposition of the Hamiltonian constraint is much more complicated
and has not been properly implemented so far.
The dif\/f\/iculties start from the form of its classical expression which, in contrast to all other constraints,
is non-polynomial. Due to this, its quantization is very involved and supplied with an inf\/inite set of ambiguities.
Even if one chooses a favorite quantization, f\/inding the kernel of the corresponding constraint operator seems
to be an impossible task.

Despite these problems, at the kinematical level LQG furnishes a nice picture of quantum space
where the states are represented by $\SU(2)$ spin networks. Moreover, it allows for computation of spectra
of various geometric operators such as area~\cite{Ashtekar:1996eg, Rovelli:1994ge},
volume~\cite{Ashtekar:1997fb, Rovelli:1994ge} and length~\cite{Bianchi:2008es, Thiemann:1996at}.
In particular, these results provide a geometric meaning to the coloring of spin networks:
the representations attached to the links give the area of transverse surfaces,
whereas the intertwiners assigned to the nodes encode information about the volume of the surrounding region.

\subsubsection{Projected spin networks}
\label{subsec-prspinnet}

However, the structure provided by LQG is not suf\/f\/icient for our purpose, which is to relate the canonical
quantization to the spin foam approach.
The problem is that, due to the partial gauge-f\/ixing, it operates with an~$\SU(2)$ subgroup of the initial Lorentz
gauge group, whereas none of the existing spin foam models make such a reduction. Therefore, we need
to understand how the results presented above can be reformulated in a Lorentz-covariant way.

This can be done using the so-called projected spin networks~\cite{Livine:2002ak} which provide a bridge
connecting the canonical loop quantization with the spin foam approach.
These objects can be introduced for any group $G$ and its compact
subgroup $H$, and can be viewed as the usual spin networks
for the group $G$ with a special choice of intertwiners depending on an element of the factor space $X=G/H$.
Namely, let $H_x$ be the stationary subgroup of $x\in X$ and $\CH_G^{(\lambda)}$ be the representation space
of an irreducible representation $\lambda$ of $G$.
This representation can be decomposed with respect to the subgroup $H_x$ as the direct sum
\begin{gather}
\CH_G^{(\lambda)}=\bigoplus_j \CH_H^{(j)},
\label{decomposGH}
\end{gather}
and we use~$p$ and~$m_j$ to label basis elements in $\CH_G^{(\lambda)}$ and $\CH_H^{(j)}$, respectively.
With these def\/initions, the class of intertwiners giving rise to projected spin networks is given by
\begin{gather}
\bigotimes_{k=1}^{\text{L}}\CH_G^{(\lambda_k)}\quad\ni\quad
\Int^{(\vec{\lambda},\vec{\jmath},\inter)}_{p_1\dots p_{\text{L}}}(x)=\mathop{\sum}\limits_{m_{j_1}\dots m_{j_{\text{L}}}}
\inter^{(\vec{\jmath})}_{ m_{j_{1}}\dots m_{j_{\text{L}}}} \prod_{k=1}^{\text{L}} \Db^{(\lambda_k)}_{p_k m_{j_{k}}}(g_{x}),
\label{genBC}
\end{gather}
where $g_x$ is a representative of $x\in X$ in $G$, $\vec{\lambda}$
and $\vec{\jmath}$ are a set of representations of $G$ and $H$, and
\begin{gather*}
\inter^{(\vec{\jmath})}\in {\rm Inv}\left(\bigotimes_{k=1}^{\text{L}}\CH_H^{(j_k)}\right)
\end{gather*}
is an invariant $H$-intertwiner. Thus, besides the ${\text{L}}$ representations $\lambda_k$ coupled by the intertwiner,
it is parametrized by ${\text{L}}$ representations $j_k$ of the subgroup $H$ appearing in the decomposition of~$\lambda_k$
and by an invariant $H$-intertwiner $\inter^{(\vec{\jmath})}$ between these~$j_k$.
Note that by itself the intertwiner~$\Int$ is not invariant. Although it does stay invariant
under transformations from the subgroup~$H_x$, under general $G$-transformations it transforms
in a {\it covariant} way
\begin{gather}
\mathop{\sum}\limits_{q_{1}\dots q_{\text{L}}}
\(\prod_{k=1}^{\text{L}} \Db^{(\lambda_k)}_{p_k q_k}(\gx)\)
\Int^{(\vec{\lambda},\vec{\jmath},\inter)}_{q_1 \dots q_{\text{L}}}(x)
=\Int^{(\vec{\lambda},\vec{\jmath},\inter)}_{p_1 \dots p_{\text{L}}}(\gx\cdot x).
\label{invN}
\end{gather}
In other words, the tensor f\/ield $\Int(x)$ is invariant provided the transformation acts not only on the tensor indices,
but also on the argument~$x$. Such an invariance property was called in \cite{Alexandrov:2008da, Alexandrov:2007pq}
{\it relaxed closure} condition.

\begin{figure}[t]
\centering
\includegraphics[scale=0.45]{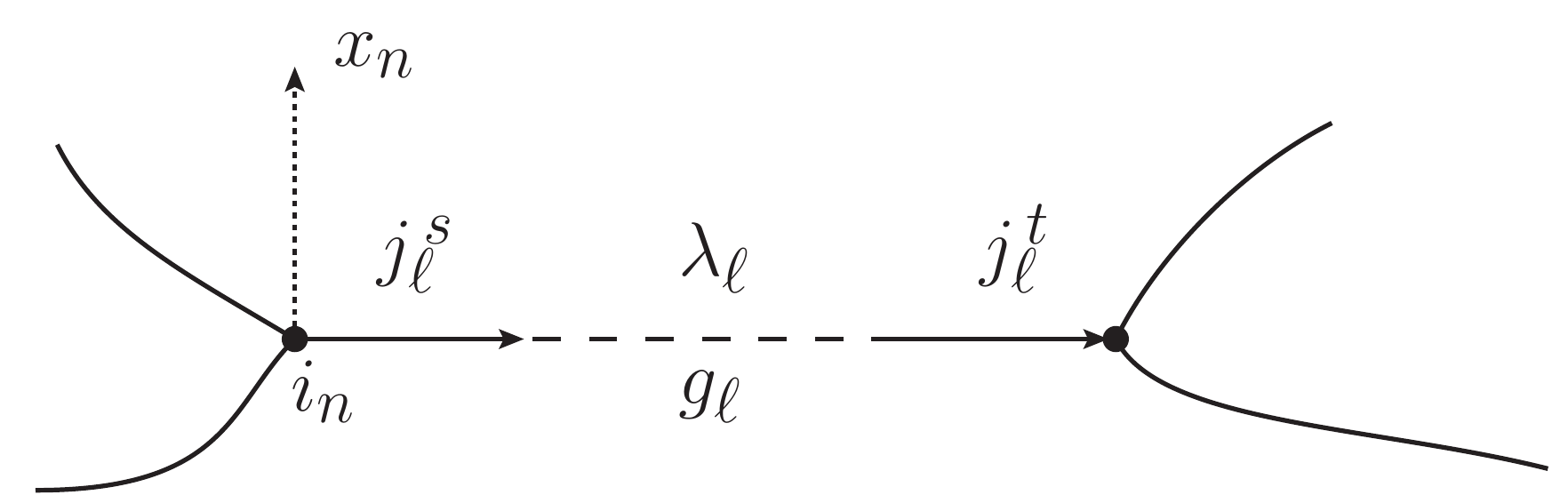}

\caption{Structure of a projected spin network.
The link $\ell$ carries a representation $\lambda_\ell$ of $G$, which is
projected at its source node onto $\CH_H^{(j_{\ell}^{\,s})}$ and at its target node onto $\CH_H^{(j_{\ell}^{\,t})}$.
The node $n$ is labeled by an $H$-invariant intertwiner $i_n$. The projected spin network state depends
on the group elements $g_\ell$ associated to the links and on the normals $x_n$ associated to the nodes.}
\label{projectedSN}
\end{figure}

Combining the intertwiners \eqref{genBC} with the group elements in representations $\lambda_\ell$ as in~\eqref{defSpinNet},
one obtains a projected spin network function
\begin{gather}
\CS_{(\Gamma,\vec{\lambda},\vec\jmath^{\,s},\vec\jmath^{\,t},\vec{\imath})}[g_\ell,x_n]
\equiv
\bigotimes_\ell \Db^{(\lambda_\ell)}(g_\ell)\cdot\bigotimes_n \Int^{(\vec{\lambda}_n,\vec{\jmath}_n,\inter_n)}(x_n).
\label{defPojSpinNet}
\end{gather}
It should be clear that the projected spin networks carry the following coloring:
each link is colored by a representation $\lambda_\ell$ of $G$ and by two representations of $H$,
$j_{\ell}^{\,s}$ and $j_{\ell}^{\,t}$, associated with the source and target nodes of the link,
and each node comes with an $H$-invariant intertwiner~$\inter_n$. The idea is that the matrix element
$\Db^{(\lambda_\ell)}(g_\ell)$ is ``projected'' from the left to $\CH_H^{(j_{\ell}^{\,s})}$
and from the right to $\CH_H^{(j_{\ell}^{\,t})}$ (see Fig.~\ref{projectedSN}).
Since the subgroup $H$ is supposed to be compact,
this projection selects a f\/inite matrix block in the original
(inf\/inite-dimensional if $G$ is non-compact) matrix.
Such a construction allows to avoid the problems arising for the usual spin networks
def\/ined for a non-compact gauge group~\cite{Freidel:2002xb} since all traces become ef\/fectively f\/inite
dimensional. In particular, the projected spin networks are well-def\/ined on the trivial
group elements $g_\ell=\unit$.

It is important to remember that the projected spin networks depend not only on the group
elements $g_\ell\in G$, but also on the elements of the factor space $x_n\in X$.
Due to this additional dependence, and despite the projection to a subgroup, the functions~\eqref{defPojSpinNet}
are $G$-invariant
\begin{gather*}
\CS_{(\Gamma,\vec{\lambda},\vec\jmath^{\,s},\vec\jmath^{\,t},\vec{\imath})}
[\gx_{\ell_s}\cdot g_\ell\cdot \gx_{\ell_t}^{-1},\gx_n \cdot x_n]
=\CS_{(\Gamma,\vec{\lambda},\vec\jmath^{\,s},\vec\jmath^{\,t},\vec{\imath})}[g_\ell,x_n],
\end{gather*}
where $\ell_s$ and $\ell_t$ denote the source and target nodes of the link $\ell$. This equality
can be easily verif\/ied using \eqref{invN}. In fact, the projected spin networks
provide an orthonormal basis in the space of all gauge invariant functions of a $G$-connection
and an $X$-valued scalar f\/ield~\cite{Alexandrov:2005ar}: the connection appears through its holonomies
along the links of the graph $\Gamma$ and the scalar f\/ield through its values at the nodes.
The corresponding Hilbert space structure is induced by the scalar product def\/ined as usual by~\cite{Livine:2002ak}
\begin{gather}
\langle\Psi_{\Gamma,f_1}|\Psi_{\Gamma,f_2}\rangle=
\int_{G^{\text{L}_\Gamma}}\prod_{\ell=1}^{\text{L}_\Gamma}\de g_\ell\,
\overline{f_1(g_1,\dots,g_{\text{L}_\Gamma},x_1,\dots,x_{\text{N}_\Gamma})}
f_2(g_1,\dots,g_{\text{L}_\Gamma},x_1,\dots,x_{\text{N}_\Gamma}),
\label{scalarproduct-pr}
\end{gather}
where $\Psi_{\Gamma,f}$ is a (generalized) cylindrical function def\/ined by a graph $\Gamma$ and a function
$f :  G^{\text{L}_\Gamma}\times X^{\text{N}_\Gamma} \longrightarrow \Cmat$, and the scalar product~\eqref{scalarproduct-pr}
can be checked to be independent on the choice of $x_n\in X$ due to the left and right invariance of the
Haar measure.

In the case of four-dimensional gravity we are interested in, $G$ should be taken to be either
the Lorentz group or $\Sp(4)$, depending on the signature, and $H={\rm SU(2)}$. The projected spin networks
appear then as a habitat for kinematical states in the Lorentz-covariant approach to the loop quantization
(see, for example,~\cite{Alexandrov:2010un}).
Moreover, we will see that they also represent the boundary states of all spin foam models
of gravity which have been introduced up to now. Thus, they provide the most direct link
between the canonical and spin foam quantizations.

\subsubsection{Lorentz-covariant form of loop quantum gravity}
\label{subsubsec-covLQG}

The advantage of projected spin networks can be seen immediately from the fact that
they allow to formulate LQG in a Lorentz-covariant way.
To see how this comes about, we should better understand the origin of their arguments, the connection
and the $X$-valued scalar f\/ield. The latter has a very clear geometric meaning: it def\/ines
the normal to the foliation used to perform the $3+1$ decomposition of spacetime. In particular, choosing
this normal to be time-directed, one reproduces the time gauge used in the canonical formulation of
the Holst action leading to LQG.

The choice of the connection def\/ining the holonomy in the f\/irst argument of \eqref{defPojSpinNet}
is a more subtle issue. The most evident possibility would be to take the usual spin connection $\omega^{IJ}$.
However, due to the second class constraints mentioned in the beginning of this section,
it has very non-trivial commutation relations.
As a result, the holonomies constructed from $\omega^{IJ}$
do not diagonalize the action of the area operator \cite{Alexandrov:2001pa},
which is in drastic contrast with the situation in LQG. Due to this, the spin connection does not seem to be suitable
for the loop quantization.

Fortunately, one can construct other candidates for the connection to be used in holonomy operators and
in projected spin networks. In particular, in \cite{Alexandrov:2001wt} using the approach based on the Dirac brackets
and in \cite{Geiller:2011cv} by explicitly solving the second class constraint (see also \cite{Cianfrani:2008zv}),
it was found that there exists a {\it Lorentz} connection $\SSA^{IJ}$ which provides a generalization of
the $\SU(2)$ AB connection \eqref{ABcon} in the following sense:
\begin{itemize}\itemsep=0pt
\item
In the time gauge $\chi=0$, it reduces to the AB connection
\begin{gather*}
\SSA_a^{IJ}\mathop{=}\limits_{\rm time\ gauge}
\begin{cases}
0, &\ \ I=0,\ J=i,\\
{\eps^{ij}}_{k} A_{a}^{k},
&\ \ I=i,\ J=j.
\end{cases}
\end{gather*}
\item
Its commutation relations generalize the symplectic structure \eqref{poisson-bracket-4d} in a natural way:
\begin{gather}
\{ \SSA_a^{IJ}, \SSA_b^{KL}\}_D  = 0.
\qquad
\{ \SSA_a^{IJ},\eps^{bcd}e^K_c e^L_d\}_D
=\im\, \delta_a^b\, I^{IJ,KL}(x)\,\delta^3(x-y),
\label{comSSA}
\end{gather}
where  $I_{KL}^{IJ}(x)=\delta^{I}_{[K}\delta_{L]}^{J}
-2\sigma\, x^{[J}_{\mathstrut}\delta^{I]}_{[K} x_{L]}^{\mathstrut}$ is the projector on the $\SU(2)_{x}$
part of the Lorentz group, so that in the time gauge $x^I=\delta^I_0$ one recovers again~\eqref{poisson-bracket-4d}.
In particular, the area operator evaluated on holonomies of this connection leads exactly to the same spectrum,
given by the Casimir of $\SU(2)$ \cite{Alexandrov:2001wt,Geiller:2011bh}, as in LQG.
\end{itemize}
The explicit expression of $\SSA$ through the spin connection and the normal to spacelike slices is given by
\begin{gather}
\SSA_a^{IJ}\equiv
I_{KL}^{IJ}(1-{\im}\star) \omega_a^{KL}
+2(1+\im\star)x^{[J}\p_a x^{I]}.
\label{conSU2}
\end{gather}

{\sloppy Moreover, in \cite{Alexandrov:2002br} it was shown that the loop quantization of the symplectic structure \eqref{comSSA}
using projected spin networks gives a quantum theory equivalent to LQG but formulated in a~Lorentz-covariant form.
The clue to this result is a constraint satisf\/ied by the connection $\SSA^{IJ}$:
\begin{gather}
\Big(x^{[J}_{\mathstrut}\delta^{I]}_{[K} x_{L]}^{\mathstrut}\Big) \SSA_a^{KL}=\sigma x^{[J}\p_a x^{I]},
\label{su2con}
\end{gather}
as can be easily verif\/ied using \eqref{conSU2}.
The operator acting on the connection on the left-hand side is the projector on the boost part
of the Lorentz algebra orthogonal to $\su(2)_x$.
Therefore, for a~constant~$x^I$, this relation implies that only the $\su(2)_x$ part of the connection
is non-trivial and its holonomy $g_\ell(\SSA)$ belongs to the subgroup $\SU(2)_x$. Since by def\/inition such operators
do not mix various terms in the decomposition~\eqref{decomposGH}, one obtains the following result\footnote{We will use $h$
to denote holonomies of the subgroup $H\subset G$.}
\begin{gather}
\CS_{(\Gamma,\vec{\lambda},\vec\jmath^{\,s},\vec\jmath^{\,t},\vec{\imath})}[g_\ell(\SSA),x_n]
= \delta_{\vec{\jmath}^{\,s},\vec{\jmath}^{\,t}}\CS_{(\Gamma,\vec{\jmath},\vec{\imath})}^{\mathstrut} [h_\ell(A)].
\label{wl}
\end{gather}
Thus, the projected spin networks evaluated on the connection $\SSA$ are non-vanishing only for those
$\SU(2)$ spin labels, which are the same on both ends of every link, $j_{\ell}^{\,s}=j_{\ell}^{\,t}\equiv j_\ell$,
do not depend on the representation labels $\lambda_\ell$, and coincide with the usual $\SU(2)$ spin networks
evaluated on the AB connection.
Furthermore, the constraints \eqref{su2con} also af\/fect the scalar product, which is not given by
the simple formula \eqref{scalarproduct-pr}, but should involve the $\delta$-functions of constraints in the measure.
The correct kinematical scalar product is then given by
\begin{gather}
\langle\Psi_{\Gamma,f_1}|\Psi_{\Gamma,f_2}\rangle_{\rm kin}=
\int_{G^{\text{L}_\Gamma}}\prod_{\ell=1}^{\text{L}_\Gamma}
\big[\de g_\ell\,\delta_{H}\bigl(g_{x_{\ell_s}}^{-1}\,g_\ell \,g_{x_{\ell_t}}^{\mathstrut}\bigr)\big]
\overline{f_1(g_\ell,x_n)}f_2(g_\ell,x_n),
\label{scalarproduct-kin}
\end{gather}
where $\delta_H(g)$ is the $\delta$-function on the group with support on the subgroup $H$ such that
$\int_G \de g \delta_H(g)=\int_H\de g$. It is clear that for the states~\eqref{wl}, the formula~\eqref{scalarproduct-kin} reduces to the usual scalar product on the $\SU(2)$ spin networks.

}

This construction identif\/ies the kinematical Hilbert spaces of LQG
and Lorentz-covariant quantization. Besides, it provides an important lesson:
the projected spin networks form an extended state space which should be reduced to the kinematical Hilbert space
by imposing certain constraints on their arguments, i.e.\ on the {\it connection} entering the def\/inition
of the quantum holonomy operators. In fact, these constraints (in particular~\eqref{su2con}) are nothing else but
the {\it secondary second class} constraints conjugated to the simplicity constraints, playing the prominent role
in the spin foam approach.
We derive them in the next subsection by studying the Plebanski action.

\subsection{Plebanski formulation and its canonical analysis}
\label{subsec_Pleb}

Most approaches to four-dimensional spin foam models take as a starting point the
formulation of general relativity as a constrained topological f\/ield theory.
As we already mentioned in Subsection~\ref{subsubsec_strategy}, this can be
done by considering the Plebanski modif\/ication of the BF action~(\ref{BF action}).
The Plebanski action with non-vanishing cosmological constant is given by
\begin{gather}\label{plebanski action}
S_\text{Pl}[\omega,B,\lmul]=\int_\CM\left(\Tr(B\wedge F)
+\f{\Lambda}{2}\,\Tr(\star B\wedge B)+\frac{1}{4}\,\Tr(\lmul\cdot B\wedge B)\right).
\end{gather}
In this expression, the density tensor $\lmul^{IJKL}=\lmul^{[IJ][KL]}$
is symmetric under the exchange of the pairs $[IJ]$ and $[KL]$, and satisf\/ies
the tracelessness condition $\eps_{IJKL}\lmul^{IJKL}=0$.
It generates the 20 simplicity constraints ensuring that the 36 components of $B^{IJ}_{\mu\nu}$
reduce to 16 components associated to a tetrad f\/ield $e^I_\mu$. To understand how this comes about,
let us look at the equations of motion derived from (\ref{plebanski action}). They are given by
\begin{gather*}
DB=\de B+[\omega,B]=0,
\\
F^{IJ}+\frac{1}{2}\(\Lambda\, \eps^{IJKL}+\lmul^{IJKL}\)B_{KL}=0,
\\
B^{IJ}\wedge B^{KL}=\sigma\CV\eps^{IJKL},
\end{gather*}
where
\begin{gather*}
\CV=\frac{1}{4!} \eps_{IJKL}B^{IJ}\wedge B^{KL},
\end{gather*}
and $\sigma$ is as usual the sign distinguishing the Riemannian and Lorentzian signatures.
The last equation of motion is known as the simplicity constraint.
In the non-degenerate case, i.e.\ when $\CV\neq0$,
one can show that this condition implies that the $B$ f\/ield can be written as
\begin{itemize}\itemsep=0pt
\item topological sector (I${}^\pm$): $B^{IJ}=\pm e^I\wedge e^J$,
\item gravitational sector (II${}^\pm$): $B^{IJ}=\pm\frac{1}{2}\,\eps^{IJ}_{~~KL}e^K\wedge e^L$.
\end{itemize}
Plugging the solution for the gravitational sector back into the Plebanski action,
one recovers the usual Hilbert--Palatini formulation.
This demonstrates that, when Plebanski theory is restricted to this sector, the two formulations are equivalent
at the classical level.

It is useful to note that, again for $\CV\neq0$, the simplicity constraints
can be equivalently rewritten as
\begin{gather}
\label{simplicity constraint}
B^{IJ}\wedge B^{KL}=\sigma\CV\eps^{IJKL}
\quad\Longleftrightarrow\quad
\eps_{IJKL}B^{IJ}_{\mu\nu}B^{KL}_{\rho\sigma}=\sigma{\cal V}   \eps_{\mu\nu\rho\sigma}.
\end{gather}
Of these two equivalent forms, it is the second one that will be important for our purposes.
It can be obtained from an action similar to~(\ref{plebanski action}) which takes the form
\begin{gather}
S_\text{Pl}[\omega,B,\lmul]\nonumber\\
\qquad{}=\int_\CM\de^4x\left[\eps^{\mu\nu\rho\sigma}\left(\Tr(B_{\mu\nu}F_{\rho\sigma})
+\f{\Lambda}{2}\,\Tr(\star B_{\mu\nu}B_{\rho\sigma})\right)
+\frac{1}{4} \lmul^{\mu\nu\rho\sigma}\Tr(\star B_{\mu\nu}B_{\rho\sigma})\right].\label{plebanski action2}
\end{gather}
The dif\/ference is the nature of the Lagrange multiplier f\/ield $\lmul$. Whereas before it lived in the tangent space,
here it is a spacetime tensor, satisfying the same symmetry properties as $\lmul^{IJKL}$.
The canonical analysis of this theory was carried out for the f\/irst time in~\cite{Buffenoir:2004vx},
and further developed in \cite{Alexandrov:2006wt}, whereas the canonical structure of the action
(\ref{plebanski action}) has been elucidated in~\cite{Alexandrov:2008fs}.
The canonical analysis of~\eqref{plebanski action} turns out to be simpler than that of~\eqref{plebanski action2}, and not surprisingly can be obtained from the latter by
solving some of the auxiliary constraints.
Here we present the results of the original analysis of~\cite{Buffenoir:2004vx}.

To perform the canonical analysis of the action \eqref{plebanski action2}, it is convenient to
add new non-dynamical variables $\mu_a$ and $\pi^a$ in order to enforce the vanishing of
the momenta conjugated to~$B_{0a}$. This can be done by adding to the action the following term
\begin{gather*}
\int_\CM\de^4x\Big(\Tr(\pi^a\partial_0B_{0a})+\Tr(\mu_a\pi^a)\Big).
\end{gather*}
The variables $\omega_0$, $\lmul$ and $B_{0a}$ are non-dynamical
(they appear without their time derivatives), but the fact that $B_{0a}$ appears
quadratically in the action prevents us from treating it as a true Lagrange multiplier, and this
is the reason for which we have to use this additional term.
The action can then be written in its canonical form as
\begin{gather}
S_\text{Pl}[\omega,B,\lmul,\pi,\mu]=
\int_\Rmat\de t\int_\Sigma\de^3x\Big(\Tr(\tP^a\partial_0\omega_a)+\Tr(\pi^a\partial_0B_{0a})-H_T\Big),
\label{canactPleb}
\end{gather}
where we have introduced $\tP^a=\eps^{abc}B_{bc}$ and $H_T$ is the total Hamiltonian given by a linear
combination of primary constraints.
The results of the Hamiltonian analysis can be summarized as follows.
The total list of f\/irst class constraints is given by
\begin{gather*}
\CG = D_a\tP^a-[\tP^a,B_{0a}]\approx0,\\
\CK_a = \eps_{abc}\Tr(\pi^b\tP^c)\approx0,\\
\CK_0 = \Tr(\pi^aB_{0a})\approx0,\\
\CD_a = \Tr(\tP^b\partial_a\omega_b)-\partial_b\Tr(\omega_a\tP^b)
+\Tr(\pi^b\partial_aB_{0b})-\partial_b\Tr(\pi^bB_{0a})\approx 0,
\\
\CD_0 = \eps^{abc}\Tr(B_{0a}F_{bc})+\Lambda\Tr(\star \tP^aB_{0a})\approx0.
\end{gather*}
Among these f\/irst class constraints, the f\/irst three are primary and the last
two are secondary arising from the requirement that $\CK_a$ and $\CK_0$ be preserved in time.
The constraint $\CG$ is the generators
of the Lorentz gauge transformations, while the remaining constraints
are related to the generators of the dif\/feomorphisms.

Apart from the f\/irst class constraints, the phase space variables of
the Plebanski theory are also subject to second class constraints. These are given by
\begin{subequations}\label{seccl-cosntr}
\begin{gather}
\Phi(B,\pi)^a_{~b} = \Tr(B_{0b}\pi^a)-\f{1}{3}\,\delta^a_{~b}\CK_0=0,\\
\Phi(\tP,\pi)^{ab} = \Tr(\tP^{(a}\pi^{b)})=0,\\
\Phi(B,B)_{ab} = \Tr(\star B_{0a}B_{0b})=0,
\\
\Phi(B,\tP)^a_{~b} = \Tr(\star B_{0b}\tP^a)-\delta^a_{~b}\CV=0,\\
\Phi(\tP,\tP)^{ab} = \Tr(\star \tP^a\tP^b)=0,\\
\Psi^{ab} = \eps^{cd(a}\Tr(\star D_c\tP^{b)}B_{0d})=0.
\end{gather}
\end{subequations}
One can recognize that the second class constraints $\Phi(B,B)$, $\Phi(B,\tP)$ and
$\Phi(\tP,\tP)$ are the various components of the simplicity constraints~\eqref{simplicity constraint}.
Moreover, one can solve explicitly the f\/irst four constraints,
which removes 14 out of 18 components of $B_{0a}^{IJ}$ and $\pi^a_{IJ}$.
The remaining four components represent the usual
lapse $\tNn$ and shift $\nd^a$ and their conjugated momenta, which are treated
in this formulation as dynamical variables. In particular, the f\/irst class constraints~$\CK_a$ and~$\CK_0$
are responsible for their transformations under dif\/feomorphisms. Reinterpreting~$\tNn$ and~$\nd^a$
as Lagrange multipliers, one establishes a direct
contact with the canonical formulation of the Hilbert--Palatini
action~\cite{Alexandrov:2006wt}\footnote{Note that these results apply only in the gravitational sector.
In the topological sector the canonical structure changes drastically.
But the two sectors stay disconnected from each other.}.

In contrast, the remaining second class constraints, $\Phi(\tP,\tP)^{ab}$ and $\Psi^{ab}$,
cannot be solved in a Lorentz-covariant way. The meaning of the former is to say that
$\tP^a_{IJ}$ is expressed in terms of the triad $\tE^a_i$ and the
normal $x^I$ introduced in the previous subsection.
On the other hand, the latter is a restriction on the connection $\omega_a^{IJ}$
which allows to express six of its components in terms of the triad and the normal.
In fact, it is a secondary constraint obtained by stabilization procedure
from $\Phi(\tP,\tP)^{ab}$. They are conjugated to each other and together form a second class system.
Thus, when it is said that the simplicity constraints are of second class, it is important to remember
that this is due to the presence of these secondary constraints which do not commute with
the primary ones\footnote{For example, in the topological sector there are also secondary simplicity constraints,
which however commute with the original simplicity. As a result, all constraints appear as f\/irst class~\cite{Liu:2009em}.}.

The presence of the second class constraints modif\/ies the symplectic structure of the theory:
the usual Poisson bracket on the phase space has to be replaced by the appropriate Dirac bracket
which is simply the symplectic structure induced on the constraint surface.
The Dirac brackets of the variables parametrizing the phase space have been computed in~\cite{Alexandrov:2006wt}.
These results in particular imply that $\tP^a$ and $\omega_a$ are not canonically conjugated anymore and
the spin connection becomes non-commutative. As we will see in the next subsection,
instead of quantizing the resulting complicated symplectic structure,
in the spin foam approach one follows a dif\/ferent strategy which ignores the presence
of the secondary constraints. We will discuss implications of this approach in Section~\ref{subsec-EPRLLQG},
whereas in Section~\ref{subsec_lessons} we will try to formulate a~spin foam quantization
taking into account the canonical structure we have just described.

\subsection{Four-dimensional spin foam models}
\label{subsec_4dmodels}

In this subsection we present the spin foam models of four-dimensional gravity which have received
the most attention. These are the Barrett--Crane (BC) model \cite{Barrett:1999qw, Barrett:1997gw}, the
Engle--Pereira--Rovelli--Livine (EPRL) model \cite{Engle:2007wy}, and the Freidel--Krasnov (FK) model \cite{Freidel:2007py}.
We outline their derivations, give the resulting constructions, and discuss their relations to the canonical quantization.
In particular, we express their boundary states in terms of
the projected spin networks introduced in Section~\ref{subsec-prspinnet}.

All these models are based on the strategy outlined in Section~\ref{subsubsec_strategy}.
Namely, they are obtained by implementing a quantum version of discretized simplicity constraints
of the Plebanski formulation in the partition function of four-dimensional BF theory.
The latter can be found in the same way as its three-dimensional analogue in Section~\ref{sec_3dSF}.
The only dif\/ference is that instead of discretizing the triad one-form as in three dimensions,
one has to discretize the bivector f\/ields~$B^{IJ}_{\mu\nu}$.
This can be done by integrating the bivectors along the two-cells (triangles) of $\Delta$ dual
to the faces~$f\in\Delta^*$, in order to obtain Lie algebra elements~$B_f^{IJ}$.
Then one can proceed as in the three-dimensional case and integrate over the discretized $B$ f\/ield in the
partition function to obtain the analogue of~\eqref{ZBF3delta}:
\begin{gather}
\mathcal{Z}_{\rm BF}(\Delta^*)=\left(\prod_{\vphantom{f}e\in\Delta^*}\int_{G}\de g_e\right)\prod_{f\in\Delta^*}\delta(g_f),
\label{ZBF4delta}
\end{gather}
where the group $G$ is now either ${\rm SL}(2,\Cmat)$ or $\Sp(4)$. Using the Peter--Weyl decomposition
and integrating out the group elements, one arrives at a representation of the partition function similar to
\eqref{ZBF3final}, where the vertex amplitude is given by the evaluation on a f\/lat connection
of the spin network dual to the boundary of a four-simplex and represented by the pentagon graph
\begin{gather}\label{vampBF}
\begin{split}
\includegraphics{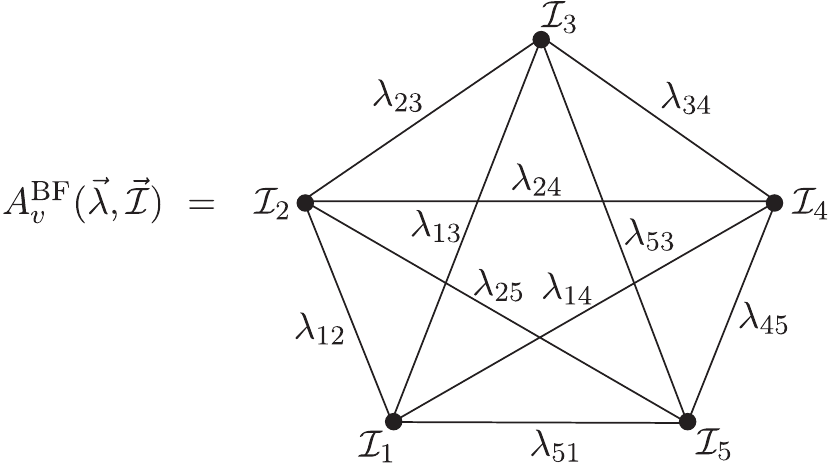}
\end{split}
\end{gather}
This result also tells us that the space of the boundary states in quantum BF theory
is spanned by the usual spin networks of the gauge group~$G$.

The next and the main step is to properly discretize, quantize and incorporate the simplicity constraints.
This is where dif\/ferent spin foam models put forward dif\/ferent proposals and start to diverge from each other.
The problem is that in general there is no unique, well-def\/ined procedure to make such a ``quantum reduction"
because from the very beginning it goes against the principles of Dirac's quantization, which state
that the second class constraints should be implemented prior to quantizing the initial symplectic structure.
Nevertheless, keeping this dif\/f\/iculty in mind, but proceeding with the suggested strategy, one can build
interesting models and some of them turn out to be closely related to the results of the canonical approach.

\subsubsection{BC model}
\label{subsubsec-BC}

The BC model was the f\/irst spin foam model of four-dimensional gravity.
The f\/irst step towards its def\/inition is the discretization of
the quadratic simplicity constraints~\eqref{simplicity constraint}.
In the discrete setting, they can be split into three family of constraints, depending upon the
position of the triangles def\/ining the discretized $B$ f\/ield. The various constraints are given by
\begin{itemize}\itemsep=0pt
\item diagonal simplicity: $\eps_{IJKL}B^{IJ}_fB^{KL}_f=0$, if two triangles are the same,
\item cross simplicity: $\eps_{IJKL}B^{IJ}_fB^{KL}_{f'}=0$, if two triangles share an edge,
\item volume simplicity: $\eps_{IJKL}B^{IJ}_fB^{KL}_{f'}=\pm\CV_v$, if two triangles meet only at a vertex,
\end{itemize}
where $\CV_v$ is the volume of the dual four-simplex.
It is convenient to distinguish between these three types of constraints because they play
rather dif\/ferent roles in the spin foam quantization.
Besides, note that the bivectors~$B_f$ possess a clear geometric interpretation: their norm is equal
to the area of the triangle dual to the face~$f$, and their tensor structure encodes its orientation in spacetime.
Given this interpretation, the bivectors should satisfy an additional constraint,
which originates from the fact that the triangles forming the boundary of a tetrahedron $\tau$
cannot be arbitrary. They are restricted to fulf\/ill the following condition:
\begin{gather}
\sum_{f\subset\tau}B_f^{IJ}=0,
\label{closure}
\end{gather}
which is known as the so-called closure constraint. One can show~\cite{Livine:2007ya} that
this constraint together with the diagonal and cross simplicity implies the volume simplicity.
Due to this, the volume simplicity is often ignored and we remain with the problem of implementing
the other constraints in the boundary states and in the partition function of the topological BF theory.

To translate the constraints to the quantum level, one identif\/ies the classical bivectors $B^{IJ}_f$ with the
generators $T^{IJ}$ of the Lie algebra of the gauge group $G$.
This identif\/ication is justif\/ied by the observation that in BF theory the $B$ f\/ield is canonically conjugated
to the spin connection. As a result, in the boundary Hilbert space it is represented by a derivative operator
whose action on the spin networks is equivalent to an insertion of the generator of the Lie algebra
in the representation associated with the given face. This can be clearly seen from the action of the smeared
$B$ f\/ield on the holonomy in representation $\lambda$. This action can be schematically represented as
\begin{gather*}
\int_\Sigma  \hat{B}
=-\I \hbar\int_\Sigma
\frac{\delta}{\delta\omega}
\   :\qquad
g_{\ell}^{(\lambda)}[\omega]
\ \ \mapsto\ \
-\I \hbar\,g_{\ell_1}^{(\lambda)}[\omega]\cdot \hT^{(\lambda)}\cdot g_{\ell_2}^{(\lambda)}[\omega],
\end{gather*}
where $\Sigma$ is a two-dimensional surface transverse to the link $\ell=\ell_1\cup \ell_2$.
Thus, the quantization map
\begin{gather}
B_f \ \ \mapsto\ \ T_f\equiv T^{(\lambda_f)},
\label{quantmapBC}
\end{gather}
used to quantize the simplicity constraints, is a simple consequence of the symplectic structure
of BF theory, in accordance with our strategy which implies the ignorance of the second class constraints
in the process of quantization.

Once this identif\/ication has been made, the simplicity constraints are imposed as conditions
\begin{gather}
\hat \Phi \cdot \CS_{(\Gamma,\vec{\lambda},\vec{\Int})}=0
\label{condBC}
\end{gather}
on the boundary states of the BF theory, which has the ef\/fect of f\/ixing the representations and the intertwiners coloring the graphs.
The easiest constraint to impose is the diagonal simplicity, which involves only the bivectors referring to one face.
As a result, it restricts the allowed representations~$\lambda_f$.
More precisely, using the map~\eqref{quantmapBC}, one f\/inds that it leads to the condition that
the second quadratic Casimir operator of $G$ should vanish:
\begin{gather}\label{diagonal condition}
C^{(2)}_G(\lambda_f)=T_f\star T_f=0.
\end{gather}
The representations satisfying this condition are commonly called simple
and will be denoted by $\lambda^s$.
To display more explicitly what the simple representations are,
we distinguish the cases in which $G$ is either $\Sp(4)$ or ${\rm SL}(2,\Cmat)$.
For the former, the unitary irreducible representations are labeled by two half-integers, $\lambda=(j^+,j^-)$,
and for the Lorentz group, representations of the principal series
are given by $\lambda=(n,\rho)$ with $n\in \Zmat/2$, $\rho\in\Rmat$.
We use the principal series only because they are the only ones which appear in the
Plancherel formula for the Lorentz group.
The two quadratic Casimir operators are given, respectively, by\footnote{Note that
the Riemannian and Lorentzian cases are related by the formal identif\/ication $2j^\pm +1=n\pm \I \rho$.}
\begin{alignat*}{3}
& C^{(1)}_{\Sp(4)}(j^+,j^-)=2j^+(j^++1)+2j^-(j^-+1),
\qquad &&
C^{(1)}_{{\rm SL}(2,\Cmat)}(n,\rho)=n^2-\rho^2-1, &
\\
&
C^{(2)}_{\Sp(4)}(j^+,j^-)=2j^+(j^++1)-2j^-(j^-+1),
\qquad &&
C^{(2)}_{{\rm SL}(2,\Cmat)}(n,\rho)=2n\rho. &
\end{alignat*}
Thus, the solution to the condition~\eqref{diagonal condition} is provided by the representations with
$j^+=j^-$ in the Riemannian case and $n=0$ (or $\rho=0$) in the Lorentzian case.

The next step is to impose the cross simplicity constraint.
It involves dif\/ferent faces dual to triangles sharing an edge,
and these triangles are in the boundary of a tetrahedron $\tau$.
Since a tetrahedron is represented in the quantum theory by an intertwining operator,
the imposition of the cross simplicity requires that
we look for intertwiners that are compatible with the coupling of simple representations.
More precisely, since it is always possible to decompose an intertwiner
with arbitrary valency into the sum of three-valent intertwiners, the cross simplicity is equivalent
to the condition that the intermediate representations of the decomposition be simple as well.
A~solution to this condition, called BC intertwiner, was proposed in~\cite{Barrett:1997gw}
and was shown to be unique in~\cite{Reisenberger:1998bn}.

Finally, it remains to implement the closure constraint, which simply amounts to
imposing the invariance of the intertwiners with respect to the group $G$. Since however
the original boundary states of BF theory represented by $G$ spin networks
already possess this invariance, there is actually nothing to impose.
As a result, we arrive at the boundary state space of the BC model, consisting
of spin networks for the group $G$ with representations restricted to be simple and intertwiners f\/ixed to
coincide with the BC intertwiner. We can denote such states as $\CS^{\rm BC}_{(\Gamma,\vec\lambda^s)}$.

The restrictions on representations and intertwiners obtained by imposing the simplicity constraints
should now be inserted into the spin foam representation of the partition function for BF theory.
The result is supposed to give a model for four-dimensional quantum gravity.
Since the constraints implemented in this way do not change the summand, but only the set
of group theoretic data which one sums over, the vertex amplitude is given by the same formula~\eqref{vampBF}
as in BF theory, but with arguments satisfying the above requirements.
Denoting by~$\Gamma_\sigma$ this pentagon graph (let us remind that it is dual to the boundary
of a four-simplex), we therefore can write
\begin{gather}
A_v^{\rm BC}(\vec\lambda^s)=\CS^{\rm BC}_{(\Gamma_\sigma,\vec\lambda^s)}\[\unit\].
\label{BCamplit_simplex-one}
\end{gather}
This def\/ines the BC spin foam model. Strictly speaking, to complete its def\/inition one should also provide
the face and edge amplitudes. From the procedure outlined above, one f\/inds that $A_e=1$ and
$A_f=d_{\lambda_f}$. However, one can expect on a general ground that this choice is modif\/ied by
the implementation of the simplicity constraints. See in particular the discussion in Section~\ref{subsubsec-measure}
where this choice is related to a non-trivial measure in the canonical path integral~\cite{Bojowald:2009im}.
Although several proposals for these lower dimensional amplitudes have been put forward,
none of them was suf\/f\/iciently convincing. Thus, it is still considered as an important open issue,
especially because the convergence properties of spin foam transition functions strongly depend
on the choice of~$A_f$ and~$A_e$~\cite{Crane:2001as, Perez:2000fs}.

The BC model has been extensively studied in the literature.
In particular, for some special choice of the face and edge amplitudes
it has been shown to be f\/inite~\cite{Crane:2001as, Perez:2000bf}
and to possess an interpretation in terms of a group f\/ield theory \cite{Perez:2000ec}.
It also was generalized in order to allow for the presence of timelike bivectors \cite{Perez:2000ep}
where however the f\/initeness properties get lost.

At the same time, no relation with the canonical quantization has been established, except
one remarkable fact. It has been noticed in~\cite{Alexandrov:2002br, Livine:2002ak}
that the boundary states of the BC model can be expressed in terms of the projected spin networks.
Indeed, using the representation for the BC intertwiner found in~\cite{Freidel:1999rr}, it is easy to show that
\begin{gather}
\CS^{\rm BC}_{(\Gamma,\vec\lambda^s)}[g_\ell]=\( \prod_n \int_X \de x_n\)
\CS_{(\Gamma,\vec\lambda^s,0,0,\unit)}[g_\ell,x_n],
\label{bs-prjBC}
\end{gather}
i.e.\ the boundary states are obtained from general projected spin networks by f\/ixing
the $\SU(2)$ representations one projects onto to be the trivial ones $j^s_\ell=j^t_\ell=0$,
which also forces the $\SU(2)$ intertwiners to be trivial $\inter_n=\unit$, and by integrating over
all factor space elements living at the nodes\footnote{In fact, in \cite{Freidel:1999rr} the notion of simple spin networks
representing the boundary states of the BC model has been generalized to any dimension $d\ge 4$.
In particular, for a general gauge group~$G$, a representation~$\lambda$ is called simple
if its decomposition~\eqref{decomposGH} into irreducible representations of
the maximal compact subgroup $H$ contains the trivial representation.
Remarkably, such simple spin networks are given by the same formula~\eqref{bs-prjBC}
in all dimensions, provided $G$ is the corresponding group Spin($d$).}.
The integral is necessary in order to ensure the invariance of the intertwiners, which should be contrasted
with the covariance property~\eqref{invN} valid without integration.
Due to this, one can rewrite the BC vertex amplitude~\eqref{BCamplit_simplex-one}
in the following way:
\begin{gather*}
A_v^{\rm BC}(\vec\lambda^s)=\int \prod_n\de\gl_n\,
\CS_{(\Gamma,\vec\lambda^s,0,0,\unit)}\big[\gl_{\ell_s}^{\mathstrut}\gl_{\ell_t}^{-1},x_0 \big].
\end{gather*}
This integral representation was very useful in the asymptotic analysis of the vertex amplitude
and plays an important role in attempts to overcome the shortcomings of the BC model.
Although the relation~\eqref{bs-prjBC} suggests a possible connection with the canonical quantization,
where the projected spin networks have been introduced for the f\/irst time,
it does not explain how to reproduce the restrictions on the labels on the canonical side.
As a result, the lack of a clear connection with the loop approach called for modif\/ications
of the BC model~\cite{Alexandrov:2007pq, Engle:2007uq}.

In fact, this was not the only reason.
A lot of criticism has been raised concerning the ability of the BC model
to reproduce correctly quantum simplicial geometry. It seems that the asymptotic (i.e.\ the large
spin limit) behavior of the vertex amplitude \cite{Baez:2002rx,Barrett:2002ur,Freidel:2002mj}
does not lead to the exponential of the Regge action, but rather to crippled geometric conf\/igurations.
Another serious issue was the inability of the BC model to reproduce the graviton
propagator~\cite{Alesci:2007tx}. Most of these problems seem to be related to the
so-called ultralocality property of the BC model, which originates in the uniqueness of the BC intertwiner
and results in the fact that the simplices do not talk to each other.
This feature can be traced back to the way in which the simplicity constraints have been implemented.
Thus, if one wishes to def\/ine a new model which is not
plagued by the same problems, the constraints have to be imposed in an alternative
way\footnote{One should however mention some recent developments in the non-commutative f\/lux
formulation of group f\/ield theory pointing towards a possible reconsideration of the BC model \cite{Baratin:2011tx}.
We refer the reader to this last paper for a discussion of various arguments
that have been put forward in this context.}.
Such alternatives have been suggested by the EPRL and FK models, which are presented in the rest of this subsection.

\subsubsection{EPRL model}
\label{subsubsection-EPRL}

The EPRL model is a straightforward generalization of the BC model based on three new inputs:
\begin{itemize}\itemsep=0pt
\item
linearization of the simplicity constraints,
\item
inclusion of the Immirzi parameter,
\item
imposition of the simplicity constraints in a weak sense.
\end{itemize}
Let us present these new ideas and their ef\/fects on the spin foam quantization.

As we know from the classical analysis, the Plebanski formulation is equivalent
to four-dimensional gravity only in one of the sectors of solutions to the simplicity constraints.
Besides the gravitational sector, there is a topological sector which may ``interfere''
with the physical one at the quantum level.
The BC model does not distinguish between the two sectors and therefore
one cannot be sure that one describes pure gravity.
On the other hand, there is a simple way to resolve this problem.
It relies on the geometric meaning of the diagonal and cross simplicity constraints.
What they simply mean is that the four triangles described by bivectors $B_f$,
which belong to the same tetrahedron $\tau \supset f$,
lie in the same hyperplane. This is equivalent to the condition that there exists a vector $x_\tau$,
normal to the tetrahedron, such that~\cite{Alexandrov:2007pq, Engle:2007uq}
\begin{gather}
(\star B_f)^{IJ}(x_\tau)_J=0.
\label{linearsim}
\end{gather}
This linear version of the simplicity constraints turns out to be stronger than the original one~\eqref{simplicity constraint}.
It is satisf\/ied only by the gravitational sector $\text{II}^\pm$ of solutions for the $B$ f\/ield.
The topological sector $\text{I}^\pm$ can also be described in a similar way, but requires a dif\/ferent linear
constraint which can be obtained from~\eqref{linearsim} by removing the star operator.
The new models, EPRL and FK, are both based on the linearization \eqref{linearsim}.
One of the consequences of this is that from the very beginning, the formulation features the normals $x_\tau$.
They turn out to coincide with the elements $x_n\in X$ appearing as arguments of the projected spin networks
(recall that the tetrahedra $\tau\in \Delta$ are dual to the spin foam edges $e\in\Delta^*$ which in turn
correspond to the nodes $n\in \Gamma$ of the boundary spin networks).
This fact facilitates the comparison with the canonical framework.

The second point is to incorporate the Immirzi parameter into the quantization scheme.
This is, of course, motivated by the hope to f\/ind a spin foam model
consistent with LQG, whose results depend on this (classically irrelevant) parameter in a non-trivial way.
This goal can be easily achieved by replacing the $B$ f\/ield in the BF part of the action
by its  Holst generalization $B+\gamma^{-1}\star B$.
Although this replacement does not af\/fect the partition function
of the unconstrained theory, it does change the quantization of the bivectors $B_f$.
As a result, the previous quantization map~\eqref{quantmapBC} takes now a $\im$-dependent form,
\begin{gather}\label{B generator}
B_f+\f{1}{\gamma}\star B_f \ \mapsto \ T_f
\qquad\Leftrightarrow\qquad
B_f\ \mapsto\ \f{\gamma^2}{\gamma^2-\sigma}\left(T_f-\f{1}{\gamma}\star T_f\right),
\end{gather}
provided that $\gamma^2\neq\sigma$. Using this map in the simplicity constraints,
one obtains a quantum theory which also depends non-trivially on the Immirzi parameter.
What remains to be understood is how the constraints are actually imposed.

We saw that the problems of the BC model were related to an over-imposition of the simplicity constraints.
Mathematically this is explained by the fact that, after quantization \eqref{quantmapBC}, they become
{\it non-commuting} operators. In other words, they represent a second class system,
and imposing them in a strong way as in~\eqref{condBC}, one actually requires that all their commutators be vanishing as well.
To avoid this problem, it has been suggested~\cite{Engle:2007uq} to replace~\eqref{condBC} by a~weaker condition
where one requires only the vanishing of their matrix elements:
\begin{gather*}
\langle\Psi'|\hat \Phi|\Psi\rangle=0.
\end{gather*}
Moreover, the initial quadratic simplicity constraints are supposed to be replaced by their li\-neari\-zed version
\eqref{linearsim} and quantized using the $\im$-dependent map \eqref{B generator}.
In fact, two other methods have also been suggested: using the so-called master constraint~\cite{Engle:2007wy}
and using the coherent states \cite{Conrady:2010kc} playing an important role in the FK model.
All three methods lead to the same construction and we present only the f\/inal result
encoded in the following two equations:
\begin{subequations}
\begin{gather}
\left(1+\f{\sigma}{\gamma^2}\right)C^{(2)}_G(\lambda_f)-\f{2\sigma}{\gamma}\,C^{(1)}_G(\lambda_f)  =  0,
\label{constraint 1}
\\
C^{(2)}_G(\lambda_f)-2\gamma\, C_H(j_{\tau f})  =  0.
\label{constr_cross}
\end{gather}
\label{constrEPRL}
\end{subequations}
Several comments concerning this result are in order:
\begin{itemize}\itemsep=0pt
\item
In all three methods, not all the simplicity constraints are treated on an equal footing.
In fact, the diagonal simplicity is still imposed in a strong way as in \eqref{condBC}.
The distinguishing role of this constraint is justif\/ied by the fact that it lies in the center of the algebra
formed by the quantum constraint operators, i.e.\ it commutes with all the other constraints, and can therefore be considered
as f\/irst class. It is this constraint that, after using~\eqref{B generator},
leads to the f\/irst equation~\eqref{constraint 1}.
\item
The second equation~\eqref{constr_cross} is a consequence of the cross simplicity.
The linearized version of this constraint~\eqref{linearsim} depends explicitly on the normal to the tetrahedron $\tau$
which can be considered as an element~$x_\tau\in X$. In its quantum version, this dependence appears through
the Casimir operator $C_H$, since the subgroup $H=\SU(2)$ appearing in this equation is the stabilizer of $x_\tau$.
(This is why we put two indices on~$j_{\tau f}$.)
\item
The presence of the $\SU(2)$ subgroup and the normal $x_\tau$ immediately implies that
the solution to the quantum simplicity constraints will lie in the class of projected spin networks.
The $\SU(2)$ representations $j_{\tau f}$ are then identif\/ied with the representations
$j^s_{\ell}$ and $j^t_{\ell}$ attached to the ends of links of the boundary graph and
appearing in the construction of Section~\ref{subsec-prspinnet}. Equation \eqref{constr_cross}
f\/ixes these representations in terms of $\lambda_f$ and thereby constrains the intertwiner
associated with the tetrahedron $\tau$.
\item
Strictly speaking, the two equations \eqref{constrEPRL} do not have common solutions neither
in the Riemannian nor in the Lorentzian case. It is assumed however that they may acquire quantum corrections
due to ordering ambiguities in the Casimir operators. Moreover, these corrections are assumed to be such that
the equations do have non-trivial solutions. What is presented below corresponds to one particular
choice of such corrections, but it should be mentioned that there are also other interesting possibilities~\cite{Alexandrov:2010pg,Ding:2010ye}.
\end{itemize}

One can show that it is possible to adjust the quantum corrections to the Casimir operators
such that the equations~\eqref{constrEPRL} have the following solution
in the Riemannian and Lorentzian case, respectively,
\begin{gather}
\lambda(j)=\(\hf\, (1+\im)j,\hf\, |1-\im| j\),
\qquad
\lambda(j)=(j,\im j),
\label{jmap}
\end{gather}
We see that the representations $\lambda_f$ of the group $G$ are completely determined
by the $\SU(2)$ representations $j_{\tau f}$.
An important consequence of this fact is that even in the Lorentzian case
the spectrum of allowed representations becomes discrete, despite the non-compactness of $G=\SL(2,\mathbb{C})$.
As a result, the same pattern holds for various geometric operators \cite{Ding:2010ye}.
On the other hand, in the Riemannian case there is a clash between the discrete nature of representation labels
and the continuous Immirzi parameter. Indeed, for $\lambda(j)$ def\/ined in \eqref{jmap}, it belongs to~$[\mathbb{N}/2]^2$
only if the Immirzi parameter is a rational number. This is a very unexpected result
which has no clear justif\/ication coming from the canonical quantization\footnote{Note, however, that a similar phenomenon occurs in
three-dimensional Riemannian gravity, where the level~$k$ (related to the cosmological constant) of the Chern--Simons theory is
discrete at the quantum level. It is clear from the covariant point of view that $k$ must be discrete but, to our knowledge,
no explanation coming from the canonical point of view exists.
In the Lorentzian regime, there are no restrictions on the values of~$k$.}.

The restrictions on representations \eqref{jmap} imply that
the boundary states of the EPRL model coincide with the following particular class of
projected spin networks:
\begin{gather}
\CS^{\rm EPRL}_{(\Gamma,\vec{\jmath},\vec{\imath})}[g_\ell,x_n]=
\CS_{(\Gamma,\vec{\lambda}(j),\vec{\jmath},\vec{\jmath},\vec{\imath})}[g_\ell,x_n].
\label{bsEPRL}
\end{gather}
This formula represents the kinematical setup of the EPRL model.
In principle, one should also impose the closure constraint \eqref{closure}, which is not satisf\/ied automatically anymore
because of the explicit dependence on $x_\tau$.
This can be easily achieved by integrating over the normals as in~\eqref{bs-prjBC}. This is what
has been done in the original version of this model~\cite{Engle:2007wy}.
However, as was argued in~\cite{Alexandrov:2008da, Alexandrov:2007pq}
and will be justif\/ied below in Section~\ref{subsubsec-quantconstr},
the closure constraint must be relaxed and the integration over $x_\tau$ should be omitted.
And it seems that now there is a~convergence towards this viewpoint, see for instance \cite{Baratin:2010wi,Bonzom:2009wm,Rovelli:2010ed}.

The dynamics of the model is encoded in the EPRL vertex amplitude which is as usual obtained
by restricting the labels in the BF vertex~\eqref{vampBF} according to~\eqref{jmap}.
Thus, it is completely determined by our kinematical result and appears
as the simplex boundary state (with the closure constraint implemented) evaluated on a f\/lat connection.
Equivalently, it can be written as the following integral:
\begin{gather}
A_v^{\rm EPRL}(\vec \jmath,\vec\imath)=\int \prod_n\de\gl_n\,
\CS_{(\Gamma,\vec{\lambda}(j),\vec{\jmath},\vec{\jmath},\vec{\imath})}\big[\gl_{\ell_s}^{\mathstrut}\gl_{\ell_t}^{-1},x_0 \big].
\label{EPRLamplit_simplex}
\end{gather}
We postpone a more detailed discussion of the relations between this model and the canonical quantization
to Section~\ref{subsec-EPRLLQG}.

\subsubsection{FK model}
\label{subsubsection-FK}

Finally, we present the FK spin foam model. It is based on the same set of ideas which underlie the EPRL model,
namely, the linearization of the simplicity constraints, the inclusion of~$\im$ and the imposition of the constraints
in a weak sense, but this last step is realized in a dif\/ferent way.
Instead of imposing the simplicity on the boundary states, the FK model
uses a coherent state decomposition in the path integral of BF theory
and imposes the constraints on the semi-classical bivectors representing the coherent states.
Let us explain the main steps of this construction by focusing on the Euclidean theory.

We start by recalling some facts about coherent states on a group.
For the $\SU(2)$ group, these states are def\/ined by acting with an $\SU(2)$ element
on the highest weight state $|j,j\rangle$ in a~representation of spin $j$
\begin{gather*}
|h,j\rangle\equiv\Db^{(j)}(h)|j,j\rangle.
\end{gather*}
They form an overcomplete basis and can be used to write a decomposition of the identity as
\begin{gather}
\unit_j=d_j\int_{\mathbb{S}^2}\de n\,|j,n\rangle\langle j,n|,
\label{decompos-one}
\end{gather}
where we took into account the possibility to reduce the integral to the quotient space $\mathbb{S}^2=\SU(2)/\text{U}(1)$.
The advantage of these states is that they possess a geometric interpretation: the state $|j,n\rangle$
describes a vector in $\Rmat^3$ of length $j$ and of direction $\vec n$,
given by the action of $h_n$ on a unit reference vector,
\begin{gather}
\langle j,n|\vec J|j,n\rangle=j\vec n.
\label{ecal-J}
\end{gather}
Furthermore, they minimize the uncertainty of the quadratic operator $J^2$ and therefore
can be considered as semi-classical states.
To get the coherent states for the $\Sp(4)$ group, which we are really interested in,
it is suf\/f\/icient to pair two $\SU(2)$ coherent states
\begin{gather*}
|\lambda,\mathbf{n}\rangle\equiv|j^+,n^+\rangle\otimes|j^+,n^+\rangle,
\end{gather*}
where $\lambda=(j^+,j^-)$ and $\mathbf{n}=(n^+,n^-)$. It is clear that they satisfy the decomposition
of the identity analogous to~\eqref{decompos-one}.

Let us now consider the partition function of BF theory obtained by
expanding~\eqref{ZBF4delta} in the sum over representations, but {\it before} performing
the integration over the holonomies $g_e$. Since an edge $e\in\Delta^*$ connects two four-simplices
sharing the tetrahedron $\tau\in\Delta$ dual to the edge, the holonomies can be represented as a product
of two ``half-holonomies'', $g_e=g_{\sigma\tau}g_{\sigma'\tau}^{-1}$,
where $g_{\sigma\tau}$ goes from the center of the four-simplex $\sigma$ to the
center of the tetrahedron $\tau$.
Inserting the decomposition of the identity in terms of the $\Sp(4)$ coherent states
between the two group elements of this representation
for all $g_e$, one f\/inds that the partition function is given by
\begin{gather}
\mathcal{Z}_{\rm BF}(\Delta^*)=\sum_{\lambda_f}\prod_{f}d_{\lambda_f}\int\prod_{(\tau,\sigma)}
\de g_{\sigma\tau}\int\prod_{(\tau,f)}d_{\lambda_f}\de\mathbf{n}_{\tau f}\prod_{(\sigma,f)}
\langle\lambda_f,\mathbf{n}_{\tau f}|g_{\sigma\tau}^{-1}g_{\sigma\tau'}|\lambda_f,\mathbf{n}_{\tau' f}\rangle.
\label{cohstates-BFpf}
\end{gather}
The idea implemented in the FK model is that the simplicity constraints have to
be imposed directly in \eqref{cohstates-BFpf} as certain restrictions
on the representations $\lambda_f$ and the normals $\mathbf{n}_{\tau f}$ labeling the $\Sp(4)$ coherent states.

In order to translate the simplicity conditions in terms of these data,
one uses the semi-classical nature of the coherent states.
The idea is that the state $|\lambda_f,\mathbf{n}_{\tau f}\rangle$ encodes
the geometry of a semi-classical tetrahedron in the sense that $\lambda_f$ represents the area of the triangle dual to
the face $f$, whereas $\mathbf{n}_{\tau f}$ is the normal to the triangle as viewed from the tetrahedron $\tau$.
To extract this information, one should evaluate the expectation value of the appropriate quantum operator.
Since classically the tetrahedron is described by the discretized $B$ f\/ield, it is natural to expect
that one should consider its quantization $\hat B_f$ which is provided by the map \eqref{B generator}.
As a result, we have to consider the following semi-classical bivectors:
\begin{gather*}
\mathbf{X}_{(\lambda_f,\mathbf{n}_{\tau f})}^{IJ}\equiv\langle\lambda_f,
\mathbf{n}_{\tau f}|\(T^{IJ}-\f{1}{\gamma}\star T^{IJ}\)|\lambda_f,\mathbf{n}_{\tau f}\rangle.
\end{gather*}
The simplicity constraints are then imposed by requiring that this bivector is simple, i.e.\ it can be represented
as (the Hodge dual of) wedge product of two one-forms, and therefore it should satisfy
$(\star \mathbf{X}_{(\lambda_f,\mathbf{n}_{\tau f})})^{IJ}(x_\tau)_J=0$ for all $f\subset\tau$ and some normal vector $x_\tau$.
Using \eqref{ecal-J}, this condition can be easily solved and leads to a restriction on $\lambda_f$,
\begin{gather}
j^+_f=\left|\frac{\im+1}{\im-1}\right|j^-_f
\label{constr_repr}
\end{gather}
and to a restriction on $\mathbf{n}_{\tau f}$,
\begin{gather}
\mathbf{n}_{\tau f}=\Big(n_{\tau f}h_{\phi_{\tau f}}^{\left| 1-\frac{1}{\im}\right|},
u_{x_\tau}n_{\tau f}h_{\phi_{\tau f}}^{-\left( 1+\frac{1}{\im}\right)}\epsilon\Big),
\qquad\mbox{with}\quad u_{x_\tau}=g_{x_\tau}^+(g_{x_\tau}^-)^{-1},
\label{cond_nn}
\end{gather}
where $g_x$ is as usual a representative of $x\in X$ in $\Sp(4)$,
the $\text{U}(1)$ elements $h_{\phi}=e^{i \phi \sigma_3}$ take care of the fact that the $\npm$ are def\/ined up to a phase,
and $\epsilon$ is either 1 or the matrix ${\scriptsize\begin{pmatrix} 0 &\!\! \!\!  1\\ -1 & \!\!\! 0\end{pmatrix}}$
such that $h\epsilon=\epsilon\bar h$ for all $h\in \SU(2)$,
depending on whether $\im$ is smaller or larger than 1.
These restrictions are then substituted in the partition function \eqref{cohstates-BFpf} so that instead of
a four-dimensional integral over $\mathbf{n}_{\tau f}$ one is left with a two-dimensional integral over $n_{\tau f}$ only.

Notice that the f\/irst condition \eqref{constr_repr} agrees with
the restriction on representations \eqref{jmap} appearing in the EPRL model
and can be considered as a consequence of~\eqref{constraint 1}.
At the same time, the second condition~\eqref{cond_nn}
clearly distinguishes the two cases $\im<1$ and $\im>1$.
It turns out that the model behaves very dif\/ferently in these two ranges of the Immirzi parameter \cite{Freidel:2007py}.
In the f\/irst case, it literally coincides with the EPRL model.
In particular, the boundary states and the vertex amplitude are given by
\eqref{bsEPRL} and \eqref{EPRLamplit_simplex}, respectively.
In the case $\im>1$, the presence of the matrix $\epsilon$ in \eqref{cond_nn} has drastic consequences.
The boundary states are spanned by all projected spin networks with representations $\lambda_f$
satisfying the constraint \eqref{constr_repr}. In other words, the simplicity constraints
do not lead to any restrictions on the intertwiners. What they f\/ix is the weight of each intertwiner
in the partition function. More precisely, the partition function of the FK model can be represented as the following sum:
\begin{gather*}
\CZ_{\rm FK_{\im>1}}=\sum_{j_f}\sum_{\inter_\tau}\sum_{j_{\tau f}=j_f}^{\im j_f}
\prod_f A_f(j_f)\prod_\tau A_e(\inter_\tau) \prod_{\tau f}\[d_{j_{\tau f}}^2 C_{j_{\tau f}}^{j_f}\]
\prod_v A_v(j_f,j_{\tau f},\inter_\tau),
\end{gather*}
where $A_f$ and $A_e$ are the face and edge amplitudes given by very simple factors in terms of dimensions
of the corresponding representations, $A_v$ is the vertex amplitude given by the formula similar to \eqref{EPRLamplit_simplex}
but with the arbitrary representations $j_{\tau f}$ one projects onto, and f\/inally
\begin{gather*}
C_k^j=\(C^{jjk}_{j{-j}0}\)^2=\frac{(2j)!}{(2j-k)!} \frac{(2j)!}{(2j+k+1)!}
\end{gather*}
is the factor f\/ixing the weight of the projected spin network intertwiners.

Since the FK model for $\im<1$ is identical to the EPRL model, it does not require a separate discussion.
In contrast, the FK construction for $\im>1$ suggests an alternative spin foam model, which however
has not been investigated as deeply as its EPRL cousin. In particular, its relation to the canonical quantization
remains unclear. On the other hand, there are several evidences in favor of this model
compared to EPRL \cite{Alexandrov:2010pg} one. In particular, in the limit $\im \to \infty$, the EPRL model reduces
to the original BC model, whereas the FK model is truly dif\/ferent. But as is well known, the BC model
suf\/fers from various problems. Furthermore, in \cite{Freidel:2007py} it was shown that this model
can also be reproduced in the FK framework, but involves an erroneous step, which makes its results more than suspicious.
Due to these reasons, for an Immirzi parameter $\im>1$, the FK model might be considered as most promising one.

\subsection{EPRL vs. LQG}
\label{subsec-EPRLLQG}

\subsubsection{Identif\/ication of kinematical states}

Let us take a closer look at the results of the EPRL model.
One observes that in both Euclidean and Lorentzian signature, the boundary states, represented by
the particular class of projected spin networks \eqref{bsEPRL}, are labeled by graphs colored
by one spin per link and one invariant $\SU(2)$ intertwiner per node. But this is precisely the coloring
carried by the usual $\SU(2)$ spin networks! This fact suggests to identify the space of boundary states
with the kinematical Hilbert space of LQG \cite{Engle:2007wy}.

This identif\/ication can be mathematically formalized as a certain bijection map $\tau$ from $\Hk$ to the space $\CK$
spanned by the states \eqref{bsEPRL}. It is given by the following integral
formula \cite{Dupuis:2010jn}:
\begin{gather}
\tau: \ \CS_{(\Gamma,\vec{\jmath},\vec{\imath})}\ \mapsto\
\CS^{(\rm EPRL)}_{(\Gamma,\vec{\jmath},\vec{\imath})}[g_\ell,x_n]=
\int_{\rm SU(2)} \prod_\ell\(\de h_\ell\, K(g^{-1}_{x_{\ell_s}}\cdot g_\ell\cdot g^{\mathstrut}_{x_{\ell_t}},h_\ell)\)
\CS_{(\Gamma,\vec{\jmath},\vec{\imath})}[h_\ell],
\label{ident-map}
\end{gather}
where the kernel is
\begin{gather*}
K(g,h)=\sum_j d_j^2 \int_{\rm SU(2)} \de k \,\chi_{{\rm SL}(2,\Cmat)}^{\lambda(j)}(gk)\, \chi_{\rm SU(2)}^j(kh),
\end{gather*}
and $\chi_G^\lambda(g)=\Tr\Db^{(\lambda)}(g)$ denotes the character of representation $\lambda$.
This map lifts $\SU(2)$-invariant states to Lorentz-invariant ones. However, these Lorentz-invariant states are not arbitrary.
The image of $\tau$ can be characterized by the property
that the reduction of $\tau [\CS]$ to the $\SU(2)$ subgroup coincides with
the original $\SU(2)$ state $\CS$. In other words, the inverse map is given the following simple restriction
\begin{gather}
\tau^{-1} :\ \CS^{(\rm EPRL)}_{(\Gamma,\vec{\jmath},\vec{\imath})}\ \mapsto\
\CS_{(\Gamma,\vec{\jmath},\vec{\imath})}[h_\ell]=\CS^{(\rm EPRL)}_{(\Gamma,\vec{\jmath},\vec{\imath})}[h_\ell,x_{(0)}],
\label{invmap}
\end{gather}
where $x_{(0)}=\delta^I_0$ is the normal vector in the time gauge.

This construction however raises some issues.
First of all, in the Euclidean case the map is def\/ined on the whole $\Hk$ only for very special values of
the Immirzi parameter. Indeed, let
\begin{gather*}
\frac{1+\im}{|1-\im|}=\frac{p}{q},
\end{gather*}
where $p,q$ are two coprime integers.
Then \eqref{jmap} implies that $\lambda \in [\Nint/2]^2$ and therefore the map~\eqref{ident-map} is well-def\/ined
only for~\cite{Alexandrov:2010un}
\begin{gather*}
j=  \begin{cases}
\displaystyle\hf_{\vphantom{\hf}} m(p+q), & \quad  \im<1,
\\
\displaystyle\hf m(p-q), & \quad \im>1,
\end{cases} \qquad m\in \Nint.
\end{gather*}
These values run over all half-integers only for $\im=2n+1$, $n\in \Nint$.
From the canonical point of view, it is not clear at all
why the Immirzi parameter should be restricted to one of these values.
Thus, in Euclidean signature the identif\/ication seems to be problematic.

Fortunately, in the physically interesting case of Lorentzian signature, there are no restrictions on the Immirzi
parameter and the map \eqref{ident-map} is always well-def\/ined. But here one encounters another problem: the norm
of the states on $\CK$ induced from the scalar product on the space of all projected spin networks is divergent.
This divergence arises due to the delta function $\delta(\rho-\rho')$
of continuous representation labels and can be traced back
to the non-compactness of the Lorentz group. It has been suggested in~\cite{Rovelli:2010ed}
that the relevant scalar product on the constrained state space $\CK$ is obtained by replacing
$\delta_{\lambda\lambda'}\equiv\delta_{nn'}\delta(\rho-\rho')$ by $\delta_{jj'}$ since $\lambda$ is
determined by the discrete variable~$j$. After such a replacement, $\tau$ becomes a unitary map.
One immediately recognizes that this change of the scalar product is exactly the same as the passage
from~\eqref{scalarproduct-pr} to~\eqref{scalarproduct-kin} in the canonical approach and therefore
it is properly justif\/ied. However, it is important to note that it is equivalent to a certain change
in the integration measure over holonomies which acquires a~$\delta$-function of the secondary second class constraints.
As we will see in the next subsection, if this modif\/ication is done not only in the scalar product,
but also in all other places in a consistent way, it gives rise to interesting new results for the spin foam quantization.

Thus, it seems that, at least in the Lorentzian signature, the EPRL model reproduces
the kinematical structure of LQG. In fact, not only the states are identif\/ied, but there is also
a~way of def\/ining quantum operators of area and volume
in the spin foam context such that their spectra coincide with those of the canonical quantization~\cite{Ding:2010ye}.

However, at the dynamical level the situation is far from being clear.
What has been extensively studied is the asymptotics of the EPRL vertex amplitude at large spins
\cite{Barrett:2009gg,Barrett:2009mw,Conrady:2008mk,Mikovic:2011zv}.
It has been shown that it is dominated by non-degenerate simplicial geometries
whose contribution scales as the exponential of the Regge action. Although this fact is promising, it is not suf\/f\/icient
to claim that the model has the correct quasiclassical limit.
There were also some attempts to relate this vertex amplitude to a Hamiltonian operator in LQG~\cite{Alesci:2008yf,Bonzom:2011tf},
but this issue remains largely unexplored.

\subsubsection{Problems}
\label{subsubsec-problems}

It is natural to compare the map $\tau$ \eqref{ident-map} to the construction
of the covariant representation of LQG of Section~\ref{subsubsec-covLQG}.
More precisely, the inverse map~\eqref{invmap} seems to be essentially the same as the equality \eqref{wl}
reducing the projected spin networks to the usual spin network states. However, between these two there
is a crucial dif\/ference: the map $\tau$ is a formal identif\/ication, whereas \eqref{wl} is a consequence
of special properties of the connection on which these states are evaluated.
This observation raises the question of why the holonomies
entering in the def\/inition of the boundary states are restricted
to belong to an $\SU(2)$ subgroup. In the case of covariant LQG,
this follows from the property \eqref{su2con}.
What is a similar reason in the EPRL model?

First of all, it is impossible that this is the same reason, namely that the holonomies
are def\/ined with respect to a connection satisfying the same constraints. The EPRL model
is derived such that its boundary states are functionals of the usual spin connection $\omega$
whose holonomies do not satisfy any restrictions and span the whole Lorentz group.
Another possibility would be that it is the projection entering the def\/inition of projected spin networks
that is responsible for the restriction to the subgroup. Indeed, the spin connection satisf\/ies
a very appealing property~\cite{Alexandrov:2010pg}:
\begin{gather}
\pr{j}\cdot\omega_a^{(\lambda)}\cdot \pr{j}=A_a^{(j)},
\label{projcon}
\end{gather}
where $A_a$ is the $\su(2)$-valued Ashtekar--Barbero connection \eqref{ABcon},
the upper index in parenthesis indicates an irreducible representation,
$\pr{j}$ denotes the projector to $\CH_{\rm SU(2)}^{(j)}$ in the decomposition~\eqref{decomposGH},
and $\lambda$ is supposed to be related to~$j$ (in the Lorentzian case) as $\lambda(j)=(j,\im(j+1))$.
This is a slight modif\/ication of the condition~\eqref{jmap}.
It also provides a solution to the constraints~\eqref{constrEPRL}.
Moreover it is an {\it exact} solution for the second equation,
i.e.\ it does not require any additional large-spin approximation
or ordering f\/itting in~\eqref{constr_cross}~\cite{Ding:2010ye}.
However, the pro\-per\-ty~\eqref{projcon} is not suf\/f\/icient for our purpose since we need its exponentiated version
\begin{gather}
\pr{j}\cdot \Db^{(\lambda)}(g_\ell(\omega))\cdot \pr{j}\,\mathop{=}^?\,
\Db^{(j)}(h_\ell(A)).
\label{twoproj}
\end{gather}
In other words, the EPRL model projects at nodes only, while LQG requires to project continuously at all points
of the graph. It is clear that these two projections do nto coincide and the equality \eqref{twoproj} is generally not true.
In the absence of a justif\/ication for the reduction to the $\SU(2)$ subgroup in the spin foam context,
it is impossible to claim that at the kinematical level the complete agreement with the canonical quantization
has been already achieved\footnote{Note also the following dif\/ference: the covariant form of LQG can be formulated
using the projected spin networks with arbitrary~$\lambda_\ell$,
whereas the EPRL model requires a specif\/ic map $\lambda(j)$. See p.~\pageref{page-dislabels}
for further discussions on this issue and on how it can be resolved in the spin foam context.}.

Furthermore, going beyond the kinematical level, one reveals another, more serious problem of the EPRL model.
To see this, it is again indispensable to make a comparison with the canonical approach.
Namely, let us compare the imposition of the simplicity constraints in the EPRL model and in the canonical quantization.
The main dif\/ference between the two quantization procedures is the one which is at the heart of the spin foam
approach: whereas the canonical quantization implies that
the second class constraints should be imposed at the classical level,
either by solving them explicitly or by introducing
the Dirac bracket, in spin foam models the constraints
are implemented at the quantum level via certain conditions on an auxiliary state space.
As a consequence, in the two approaches one quantizes
completely dif\/ferent symplectic structures:
in the canonical quantization it is the symplectic structure
induced on the constraint surface, and in the spin foam approach
it is the symplectic structure of BF theory.
As a result, the simplicity constraints are realized
in completely dif\/ferent ways: either as operators identically
vanishing on {\it all} states, or as some non-trivial
operators annihilating only a certain ``physical'' subspace.

As has been shown in \cite{Alexandrov:2010pg}, the second quantization procedure, employed
in most spin foam models, leads to internal inconsistencies. For example, the Hamiltonian turns out to be
not well-def\/ined on the ``physical" subspace given by the constraints and its eigenstates
cannot belong to this subspace. This last fact, in particular, implies that the kernel of the Hamiltonian,
which provides the real physical states of the theory, has a vanishing intersection with the boundary state space
determined in this way.

Moreover, the spin foam quantization completely ignores the presence of the secondary constraints
which are responsible for the fact that the simplicity constraints are of second class. Although they do follow from
the primary simplicity constraints imposed at all times, this holds only in the quasiclassical approximation.
At the full quantum level, they are allowed to have non-vanishing f\/luctuations. At the same time,
in the canonical approach they are completely frozen, which is ensured by inserting their delta function
in the path integral measure.
In particular, this is the reason why it appears in the scalar product \eqref{scalarproduct-kin} and,
as will be shown in the next subsection, this insertion might have important ef\/fects on the quantum dynamics.

Thus, some of the quantization steps leading to the EPRL model
or to the natural identif\/ications with the LQG structures, when critically examined, show
a deep dif\/ference with the canonical quantization, which can be traced back to
the standard spin foam strategy. This strategy can be summarized by the recipe ``f\/irst quantize, then constrain"
meaning that one f\/irst quantizes the unconstrained theory, while the constraints are incorporated only at the quantum level.
Given the above observations, the results of the EPRL model, and any other model derived using
this strategy, should be taken with great care.

\subsection{Lessons from the canonical quantization}
\label{subsec_lessons}

Let us now try to understand what a spin foam model would look like if it was derived using
the standard rules of Dirac's quantization.
In other words, we would like to address the following question:
which properties can we expect to be satisf\/ied by spin foam amplitudes and their boundary states
on the basis of the canonical analysis?

\subsubsection{Measure}
\label{subsubsec-measure}

We start our discussion with the problem of the measure in the (discretized) path integral.
Deriving spin foam models, one usually chooses a trivial measure which, after discretization,
is given either by the Haar measure for variables taking values in a Lie group, or the Lebesgue measure
for algebra-valued f\/ields. On the other hand, the canonical path integral is known to have
a non-trivial measure which might strongly af\/fect the results of quantization.

Let us recall the general form of the canonical path integral representing the partition function
of a dynamical system with coordinates $q$, momenta $p$, primary and secondary second class constraints,
$\phi$ and $\psi$, respectively, and a Hamiltonian $H$. It is given by
\begin{gather}
\CZ = \int \CD q\,\CD p \, \sqrt{|\det\Delta|}\,\delta(\phi)\delta(\psi)\exp\(\I\int\de t\(p_a \dot q^a-H\)\),
\label{phspint}
\end{gather}
where $\Delta$ is the Dirac matrix of second class constraints.
We assume that there are no f\/irst class constraints since they are not essential for our discussion.
If they are present, the Hamiltonian should be replaced by $H_{\rm ef\/f}=H_\Omega-\{\Theta, \Omega\}_+$ where
$\Omega$ is the BRST charge, $H_\Omega$ is an extension of~$H$ commuting with~$\Omega$, and
$\Theta$ is a gauge-f\/ixing fermion. Besides, one should add the Lagrange multipliers and ghosts
to the kinetic term and the integration measure. If the gauge-f\/ixing fermion is appropriately chosen
(see, for example, \cite{Alexandrov:1998cu} for the case of the Hilbert--Palatini formulation),
the ghosts can be integrated out and one obtains the usual Faddeev--Popov path integral.

One can distinguish two types of contributions to the measure in the path integral \eqref{phspint}:
determinants of Faddeev--Popov type and $\delta$-functions of constraints.
We consider the implications of these contributions separately.
The reason for this is that there is a trick \cite{Henneaux:1994jf}
which allows to rewrite the partition function in the Lagrangian form
\begin{gather}
\CZ=\int \CD q\,\CD p \, \CD\lambda\, \mu(q,p)\exp\(\I\int\de t\(p_a \dot q^a-H-\lambda\phi\)\),
\label{confspint}
\end{gather}
where $\lambda$ is the Lagrange multiplier for the primary constraint, $\mu(q,p)$ is a regular local measure,
and the expression in the exponential is the Lagrangian in its f\/irst order formulation.
In passing from \eqref{phspint} to \eqref{confspint}, the $\delta$-function of the secondary constraint has been traded
for an additional local regular contribution which together with $\sqrt{|\det\Delta|}$ gives the measure $\mu(q,p)$.
Thus, this allows to avoid the contributions of the second type altogether.

The local measure $\mu(q,p)$ has been computed for the Plebanski formulation in \cite{Buffenoir:2004vx,Engle:2009ba}
and for the Holst action in \cite{Engle:2009ba}. The result is given by\footnote{Besides this local factor,
the measure of the gravity path integral also includes gauge-f\/ixing conditions and the usual Faddeev--Popov determinant.}
\begin{gather}
\mu_{\rm Pl}=\CV^9 V,
\qquad
\mu_{\rm Holst}=\CV^3 V,
\label{measure-fac}
\end{gather}
where $V$ and $\CV$ are three- and four-dimensional volumes.
Note that, due to the presence of the spatial volume, the measure is not covariant.
This is in fact a familiar phenomenon which can be traced back to the fact that the dif\/feomorphisms
coincide with the gauge transformations generated by the constraints of the canonical formulation only on-shell.
The full path integral is by construction invariant under the gauge symmetries and in \cite{Han:2009bb}
it has been verif\/ied that the measure factors~\eqref{measure-fac} are indeed consistent with this invariance.

The implications of such additional factors on spin foam models have been considered in~\cite{Bojowald:2009im}.
There are two possible approaches to take into account their contributions.
The f\/irst one is to insert them in the initial path integral, as required by the canonical approach,
and to repeat all the steps leading to the spin foam representation.
However, since these are metric dependent contributions, their insertion
would spoil the very f\/irst step of the spin foam quantization~-- the integration over the~$B$ f\/ield.
An attempt to overcome this problem and to get a spin foam model along these lines has been made in~\cite{Baratin:2008du},
but the resulting discrete model looks very complicated and no concrete results have been obtained in this direction.

An alternative method is to use the interpretation of areas and volumes in terms of representations and intertwiners,
and to replace the original function $\mu$ in the path integral measure by a function $\mu(\lambda,\Int)$, local
in the discretized sense, inserted already in the spin foam sum over representations.
This approach was studied in~\cite{Bojowald:2009im} and it was shown that
such contributions would af\/fect the face and edge amplitudes.
On the basis of this observation, it was argued that the proper choice of these amplitudes,
which must be consistent with the measure favored by the canonical analysis, is crucial
for the correct implementation of all the symmetries. A wrong choice will unavoidably
result in an anomalous quantization.
On the other hand, the spin foam approach does not suggest a unique way to f\/ix the face and edge
amplitudes, and therefore the link with the canonical quantization seems to be
the necessary additional input which might allow to f\/ill the gap.

Let us now return to the $\delta$-function contributions to the measure due to the secondary constraints.
Although the trick of~\cite{Henneaux:1994jf} allows to remove them in the partition function, it does not work
for correlation functions~\cite{Alexandrov:2008da}. This implies that we cannot disregard them and they should be included
into the measure also in the discretized path integral used in the spin foam quantization.
It turns out to be possible to derive several important conclusions just from this simple fact.

From the canonical analysis of the Plebanski formulation presented in Section~\ref{subsec_Pleb}, we know
that the secondary second class constraints impose restrictions on the connection variable. Thus, their
discretized version should give rise to certain conditions on the holonomies. We assume that they can be described
by a function $\psi_{\rm discr}\(\gl_{\sigma\tau},x_\tau\)$ where group elements $\gl_{\sigma\tau}$, as in \eqref{cohstates-BFpf},
describe holonomies from the center of a four-simplex $\sigma$ to the center of one of its tetrahedra $\tau$,
and we also included the dependence on the normal $x_\tau$ to the tetrahedron.
An example of such a discretized constraint is given by the one used in \eqref{wl}, which in the Riemannian case
can be formulated as
\begin{gather}
\psi_{\rm discr}\(\gl_{\sigma\tau},x_\tau\)=
(\gl^+_{\sigma \tau})^{-1} \gl^-_{\sigma\tau}u_{x_\tau}^{-1}=\unit,
\label{exampl-dc}
\end{gather}
where we have used the decomposition $g=(g^+,g^-)$ for $\Sp(4)$ group elements and
$u_{x}$ was def\/ined in \eqref{cond_nn} in terms of $g_x$.

Relying on these observations, we conclude that the secondary second class constraints should be incorporated
into the discretized path integral by modifying the integration measure over holonomies.
We suggest that it takes the following form
\begin{gather}
\CD^{(x_\tau)} [\gl_{\sigma\tau}]\sim \delta(\psi_{\rm discr}\(\gl_{\sigma\tau},x_\tau)\)\de g_{\sigma\tau}
\label{measur}
\end{gather}
and satisf\/ies the following covariance property
\begin{gather}
\CD^{(x_\tau)} \[\gl_{\sigma\tau}\, \gb\]=\CD^{(\gb\cdot\, x_\tau)} [\gl_{\sigma\tau}], \qquad \gb\in G,
\label{trmeas}
\end{gather}
which is necessary in order to ensure the Lorentz invariance of the path integral.

To understand the implications of this measure, let us consider the discretized path integral for a single four-simplex
with f\/ixed $B_f$ on the boundary. We then drop the unnecessary index $\sigma$ and denote
$u(f)$ and $d(f)$ two tetrahedra sharing the $f^\text{th}$ triangle
(as the triangle is oriented, one of the tetrahedra is considered as ``up'' and the other as ``down'').
Since there are no internal faces, the only integration involved in
the def\/inition of the corresponding quantum amplitude is the one over holonomies $\gl_{\tau}$
associated with each of the f\/ive tetrahedra. The integration is done with the measure~\eqref{measur},
so the amplitude becomes
\begin{gather}
A[B_f]=\int \prod_\tau \CD^{(x_\tau)} [\gl_\tau]
\prod_f
\exp\big(\I \Tr\big[B_f \gl_{u(f)}^{-1}\gl_{d(f)}^{\hphantom{1}}\big]\big),
\label{ampB}
\end{gather}
where in the exponential one can recognize the discretize BF action.
The amplitude~\eqref{ampB} can be viewed as a simplex boundary state
in the $B_f$ or the so-called f\/lux representation~\mbox{\cite{Baratin:2010nn,Baratin:2010wi}}.
In the total partition function, these contributions should be multiplied by using the non-commutative
star-product\footnote{This product is def\/ined on the plane waves $\eb_g(\xb)=e^{\I\Tr(g\cdot \xb)}$, with $\xb\in\gmat$
and momentum given by the group element $g$, as  $\eb_{g_1}\star \eb_{g_2}=\eb_{g_1 g_2}$ \cite{Baratin:2010nn}.}
and integrated over the bivectors:
\begin{gather}
\CZ=\int\prod_f  \CD[B_f] \big( \mathop{\bigstar}\limits_\sigma  A_\sigma[B_f]\big),
\label{fullpf-noncom}
\end{gather}
where the order in the non-commutative product is dictated by the structure of the simplicial complex $\Delta$.
The measure $\CD[B_f]$, similarly to the measure for holonomies, should contain
the $\delta$-function of the discretized simplicity constraints $\phi_f$
as well as various volume factors of the type~\eqref{measure-fac} which can be usually constructed from
the bivectors. Note that this representation allows to disentangle the issues related to this $B_f$-dependent
factors and to the secondary se\-cond class constraints. It shows that the former
become relevant only when one glues dif\/ferent simplices together, whereas the latter
will af\/fect the contribution of each individual simplex.

To see how it works, we rewrite $A[B_f]$ as the Fourier transform of another amplitude,
\begin{gather}
A[B_f]=\int \prod_f \big[ d g_f\, e^{\I\Tr(B_f g_f)}\big]A[g_f,x_\tau],
\label{amplitgen}
\end{gather}
with
\begin{gather}
A[g_f,x_\tau]=\int\prod_\tau \CD^{(x_\tau)} [\gl_\tau]
\prod_f\delta\big( \gl_{u(f)}^{\hphantom{1}}g_f\gl_{d(f)}^{-1}\big),
\label{simbs_init}
\end{gather}
which thus appears as the simplex boundary state in the ``connection" representation since
it depends on the group elements $g_f$.\footnote{Note that in general the secondary constraints depend on the $B$ f\/ield. As
a result, the measure~\eqref{measur} may depend on the bivectors $B_f$ as well. Here we ignore this possible complication.}
The representation~\eqref{simbs_init} seems to be very similar to the one for the partition function of BF theory
\eqref{ZBF4delta} where the $\delta$-function implements the f\/latness condition. However, as we will see,
the non-trivial measure makes this similarity very superf\/icial and the f\/latness condition does not hold in our case.
In~\cite{Alexandrov:2008da} it was shown that, using the covariance property~\eqref{trmeas}
and inserting integrals over the $\SU(2)$ subgroup attached to each tetrahedron, the amplitude~\eqref{simbs_init}
can be represented as a linear combination of projected spin networks~\eqref{defPojSpinNet},
\begin{gather}
A[g_f,x_\tau]=\sum_{\lambda_f,j_{\tau f},\inter_\tau}\(\prod_f d_{\lambda_f}\)\(\prod_{(\tau,f)} d_{j_{\tau f}}\)
A(\lambda_f,j_{\tau f},\inter_\tau)
\CS_{(\Gamma_\sigma,\vec{\lambda},\vec\jmath^{\,s},\vec\jmath^{\,t},\vec{\imath})}[g_f,x_\tau],
\label{reprampl}
\end{gather}
where $\Gamma_\sigma$ is the graph dual to the boundary of the four-simplex.
The coef\/f\/icients $A(\lambda_f,j_{\tau f},\inter_\tau)$, given below in \eqref{amplit_simplex},
are naturally interpreted as vertex amplitude of the spin foam quantization.
The result \eqref{reprampl} provides a general proof of the observation made in the previous subsections
that in all spin foam models of four-dimensional gravity, the boundary states coincide with
a~certain class of projected spin networks.
Furthermore, below we explain the important implications of this result on the implementation of the simplicity constraints
in spin foam models and on the dynamics of the corresponding quantum theory.

\subsubsection{Quantization of constraints}
\label{subsubsec-quantconstr}

All four-dimensional spin foam models are characterized by the presence of simplicity constraints
which should be implemented in one way or another.
In principle, this problem is one of the central topics of the canonical quantization,
and therefore we should not expect a drastic dif\/ference in the way the constraints are implemented
in the loop and spin foam approaches.
Moreover, the Hilbert spaces before and after the imposition of constraints are identical in the two approaches.
Indeed, on the one hand, the initial state space of the topological BF theory,
once the closure constraint is relaxed as in~\eqref{invN},
coincides with the space of projected spin networks.
On the other hand, the space of the boundary states in a spin foam model of gravity,
resulting from the imposition of the constraints,
must be identical to the kinematical Hilbert space of the loop quantization.
This shows that it is natural to expect that the constraints should be realized in spin foam models
in the same way as the one responsible for the reduction from all projected spin networks
to the kinematical Hilbert space of LQG in its covariant formulation, as was presented in Section~\ref{subsubsec-covLQG}.

Let us recall that this reduction is done due to the fact that the holonomy arguments $g_\ell$ of the projected spin networks
do not span the whole Lorentz group, which in turn can be traced back to the constraints satisf\/ied by
the connection def\/ining these holonomies. Thus, these are the {\it secondary second class constraints}
that are responsible for the reduction of the Hilbert space in the canonical approach, whereas the primary
simplicity constraints are automatically satisf\/ied due to Dirac's quantization rules.
This is in drastic contrast with the usual spin foam strategy where the primary simplicity constraints
are used to reduce the state space of BF theory, whereas the secondary constraints are completely ignored.
Of course, the origin of this discrepancy comes from the fact that the two approaches quantize dif\/ferent symplectic
structures, as was discussed in Section~\ref{subsubsec-problems}:
one is given by Poisson brackets and describes the symplectic structure of BF theory,
and the other is given by Dirac brackets ensuring that we deal with general relativity.

{\sloppy Remarkably, the inclusion of the secondary constraints into the path integral measure leading to \eqref{reprampl}
implies that the constraints are imposed in complete agreement with the canonical approach.
To understand why this is so, let us return to the initial formula for the simplex boundary state in
the connection representation \eqref{simbs_init}. The $\delta$-function expresses the arguments~$g_f$
of the boundary state wave function in terms of the holonomies~$\gl_\tau$. But the latter are
integrated out with the measure \eqref{measur}, which is supposed to contain an additional $\delta$-function.
This $\delta$-function induces certain restrictions on $g_f$, similar to those which hold for~$\gl_\tau$.
For instance, in the simplest case~\eqref{exampl-dc}, the constraint induced on~$g_f$ reads
\begin{gather*}
g_f^-=u_{x_{u(f)}}^{-1}\, g^+_{f}\, u_{x_{d(f)}}.
\end{gather*}
Being used in \eqref{reprampl}, these restrictions precisely correspond to the imposition of
the se\-con\-dary second class constraints on the enlarged Hilbert space of the canonical loop quantization~\mbox{\cite{Alexandrov:2007pq, Alexandrov:2005ar}}. In particular, these constraints should be ref\/lected
in the form of the vertex amplitude $A(\lambda_f,j_{\tau f},\inter_\tau)$,
which might, for example, vanish on a certain set of representation labels.
We consider this issue in more detail in the next subsection.

}

Before that, we however discuss another controversial issue of the spin foam approach,
which is the quantization of the closure constraint~\eqref{closure}. Usually, it is imposed
at the quantum level by requiring the invariance of intertwiners associated to the nodes of the boundary states:
\begin{gather}
\sum_{f\subset\, \tau}\, T_f\, \Int_\tau=0.
\label{clcon}
\end{gather}
On the other hand, the canonical quantization leads to intertwiners which carry an additional
dependence on the normals $x_\tau$ and satisfy instead
the relaxed closure constraint~\eqref{invN}. This constraint can be formulated in the inf\/initesimal form,
analogous to~\eqref{clcon}, as~\cite{Alexandrov:2008da, Alexandrov:2007pq}
\begin{gather}
\sum\limits_{{f\subset\, \tau}}\, T_f\, \Int_\tau(x_\tau)=\hat T\cdot \Int_\tau(x_t),
\label{clconrel}
\end{gather}
where $\hat T$ acts on functions of $x^I\in X$ as a generator of $G$:
\begin{gather*}
\hat T_{IJ}\cdot f(x)=\eta_{IK} x^K \p_J f-\eta_{JK} x^K \p_I f.
\end{gather*}
Although \eqref{clconrel} can be reduced to \eqref{clcon} by integrating the intertwiner over $x_\tau$,
the canonical quantization clearly shows that this should not be done.

A moment of ref\/lection shows that the path integral quantization leads to the exact same conclusion.
Indeed, note that the bivectors $B_f$ appearing as arguments in the quantum amplitude \eqref{ampB}
should satisfy the linear simplicity constraint~\eqref{linearsim} with a f\/ixed normal $x_\tau$.
As a~result, this normal should also be kept f\/ixed in the path integral representation for $A[B_f]$
and, as a consequence, in all the boundary states.
Moreover, it should not be integrated even when two simplex contributions are glued together.
Assume that two simplices $\sigma_1$ and $\sigma_2$ share a~tetrahedron. Then the gluing \eqref{fullpf-noncom}
can be rewritten in the connection representation, and
the total boundary state can be obtained from two individual simplex amplitudes as \cite{Alexandrov:2008da}
\begin{gather*}
A_{12}\[g_f,x_\tau\]
=\int \de \gb_{f_{12}}\,
A_1\[\gl_{f_1},\gl_{f_{12}}\gb_{f_{12}},x_{\tau_1},x_{\tau_{12}}\]
A_2\big[\gl_{f_2},\gb_{f_{12}}^{-1},x_{\tau_2},x_{\tau_{12}}\big],
\end{gather*}
where $f=(f_1,f_2,f_{12})$, $\tau=(\tau_1,\tau_2,\tau_{12})$, the labels~1,~2 refer to non-shared faces and tetrahedra of
the corresponding glued simplices, and the label 12 marks the shared faces and tetrahedron.
Using~\eqref{trmeas}, it is easy to see that the dependence on~$x_{\tau_{12}}$ drops out.
Therefore, an integral over the normals would be completely artif\/icial and moreover would lead
to a divergence in the Lorentzian case. In fact, keeping $x_\tau$ f\/ixed is nothing else
but the usual gauge-f\/ixing of the boost gauge freedom in the path integral.

Thus, both the path integral and the canonical approaches require that the closure constraint be imposed only
in its relaxed form~\eqref{clconrel}.
This is not in contradiction with the classical closure constraint, since the latter
arises only as a classical equation of motion and does not need to be preserved by quantum f\/luctuations.
The necessity to give up the invariance of intertwiners in favor of their covariance
has been recently recognized also in \cite{Rovelli:2010ed}.

\subsubsection{Vertex amplitude}
\label{subsubsec-vertex}

Finally, we consider the implications of our prescription to include secondary second class constraints
in the path integral measure for the vertex amplitude, which is the most important quantity
of the spin foam quantization since it encodes the dynamics of the quantum theory.
This amplitude is given by the coef\/f\/icients $A(\lambda_f,j_{\tau f},\inter_\tau)$ in the decomposition \eqref{reprampl}
and is found to be \cite{Alexandrov:2008da}
\begin{gather}
A(\lambda_f,j_{\tau f},\inter_\tau)=\int \prod_\tau\CD^{(x_\tau)} [\gl_\tau]\,
\CS_{(\Gamma_\sigma,\lambda_f,j_{\tau f},\inter_\tau)}\big[\gl_{u(f)}^{-1}\gl_{d(f)}^{\mathstrut},x_\tau\big].
\label{amplit_simplex}
\end{gather}
It is easy to check that the covariance of the measure \eqref{trmeas} ensures the independence
on the normals $x_\tau$.
Thus, the vertex amplitude is given by an integral over gauge transformations at the nodes of
the projected spin network associated with the boundary of a four-simplex.

This formula is a straightforward generalization of the usual prescription for the evaluation
of vertex amplitudes in the spin foam models considered in the literature. The only dif\/ference
is that the measure in~\eqref{amplit_simplex} is assumed to be non-trivial and, in particular,
to include the secondary constraints. Taking instead the measure to be the usual Haar measure on the group,
$\CD^{(x_\tau)} [\gl_\tau]=\de \gl_\tau$, one reproduces an analogue of~\eqref{EPRLamplit_simplex}.
Moreover, in this case the integrals over the gauge group are equivalent to the integrals over the normals~$x_\tau$.
Therefore, as was discussed in the previous subsection,
they can be viewed as a part of the def\/inition of the boundary states
ensuring the closure constraint~\eqref{clcon}. Then one gets back the simple recipe
that the vertex amplitude is given by the simplex boundary state evaluated on a f\/lat connection,
which can be traced back to the f\/latness condition of BF theory.

However, if one insists on the presence of the secondary constraints in the measure $\CD^{(x_\tau)} [\gl_\tau]$,
this simple prescription is not valid anymore. In particular, f\/lat connections do not play any preferred role
and the dynamics appears to be more complicated. The concrete form of the vertex amplitude depends
on a particular choice of the measure which should be dictated by agreement with the canonical quantization.

\subsubsection{Choice of connection}
\label{subsubsec-choice}

So far our consideration was generic as we did not specify the precise form of the secondary constraints.
These constraints depend on the choice of the connection which is used to def\/ine holonomy variables.
Thus, we encounter the same issue which already arose in Section~\ref{subsubsec-covLQG} when we discussed
the Lorentz-covariant formulation of LQG. Here we should answer the same set of questions:
Which classical variable behaving as a connection should be chosen to def\/ine holonomies?
Which constraints are satisf\/ied by this variable?
And how can these constraints be discretized?

Usually, in the spin foam approach, no connections other than the standard spin connection~$\omega^{IJ}$ appear.
Thus, one may expect that this is the connection to be considered. But then
one runs into problems similar to those which did not allow to use it for covariant formulation of LQG.
The second class constraints satisf\/ied by the spin connection are given by~$\Psi^{ab}$ from~\eqref{seccl-cosntr}
and do not allow a simple geometric interpretation in terms of holonomies constructed from~$\omega^{IJ}$.
Given the troubles with the loop quantization in terms of this connection, it is natural
to ask for something else.

The next evident candidate is the Lorentz generalization of the AB connection $\SSA^{IJ}$ \eqref{conSU2}.
Since it was essential in the covariant reformulation of LQG, one might expect that it provides
a~direct link between LQG and spin foam models.
This connection satisf\/ies the constraints~\eqref{su2con} which allow a very simple discretization as in~\eqref{exampl-dc}.
However, if one plugs these constraints into the formula for the vertex amplitude~\eqref{amplit_simplex},
one f\/inds as a bizarre result that it is given by the vertex amplitude of $\SU(2)$ BF theory.
This puzzle is easily resolved by noting that the above results, and the vertex amplitude~\eqref{amplit_simplex}
in particular, are not applicable in this case because of two reasons. First, the connection $\SSA^{IJ}$
is not a pull-back of a spacetime connection~\cite{Alexandrov:2001wt, Samuel:2000ue}.
In other words, it is not possible to extend it to a spacetime connection which transforms properly under all
gauge transformations including time dif\/feomorphisms. In the absence of such a spacetime interpretation,
it becomes impossible to use this connection in the covariant path integral quantization.
The second reason is even more clear. The connection $\SSA^{IJ}$ dif\/fers from the spin connection~$\omega^{IJ}$
by a complicated term {\it non-vanishing} on the constraint surface. As a result, the Plebanski or even the BF action,
which is the starting point of our analysis, takes a completely dif\/ferent form when it is written in terms of~$\SSA^{IJ}$.
Altogether, these issues make the use of~$\SSA^{IJ}$ in the spin foam quantization very problematic.

There is another possible choice of the connection which avoids the problems of the AB connection $\SSA^{IJ}$
and partially solves the problems of the spin connection $\omega^{IJ}$.
This is the so-called ``shifted connection" introduced in
\cite{Alexandrov:2001pa}. Its name comes from its def\/inition as a shift of the spin connection
by a term proportional to the Gauss constraint,
\begin{gather*}
\SA_a^{IJ} = \omega_a^{IJ}+\frac{I_\gamma^{IJ,\,KL}}{2\(1+\im^{-2}\)}
\,  f^{ST}_{KL,\,PQ}\Pt_a^{PQ}G_{ST}.
\end{gather*}
Here $f^{MN}_{IJ,\,KL}$ are $\so(3,1)$ structure constants, $\Pt_a^{IJ}=g^{-1} g_{ab}\tP^{b,IJ}$
where $\tP^a$ is the canonical f\/ield of the Plebanski formulation def\/ined under \eqref{canactPleb},
$g_{ab}=\eta_{IJ}e^I_a e^J_b$ is the spatial metric and $g$ is its determinant,
and we def\/ined $I_\gamma=\(1-\im^{-1}\,\star\)I  \(1-\im^{-1}\,\star\)$ with
$I^{IJ}_{KL}$ the projector introduced under \eqref{comSSA}.
As it was shown in~\cite{Alexandrov:2001wt}, $\SA^{IJ}$ represents the unique spacetime Lorentz connection
whose holonomies diagonalize the action of the area operator.
Due to this last property, this connection is suitable for the loop quantization.
The corresponding quantization has been called covariant loop quantum gravity (CLQG) \cite{Alexandrov:2002br}
and is inequivalent to LQG.
Its nice feature is that it leads to an area spectrum which is independent of the Immirzi parameter.
This result seems to be much more natural than the standard result of LQG,
where $\im$ appears as a proportionality factor,
since classically $\im$ is a free parameter removable by a canonical transformation.
However, in contrast to the AB connection, the shifted connection is non-commutative.
This fact leads to serious dif\/f\/iculties in quantizing the corresponding symplectic structure
and its proper representation in terms of an operator algebra on the Hilbert space spanned by projected
spin networks has not been found yet.
We refer to~\cite{Alexandrov:2010un} for further details on this approach.

On the other hand, the fact that the shifted connection has a spacetime interpretation and dif\/fers
from the spin connection by only a term vanishing on the surface of the Gauss constraint allows to use
the shifted connection in the spin foam approach.
But we still need to f\/ind a~discretization of the secondary second class constraints,
which in terms of $\SA^{IJ}$ take the following form:
\begin{gather}
I_{KL}^{IJ}\SA_a^{KL}=\Gamma_a^{IJ}(\tP),
\label{contA}
\end{gather}
where $\Gamma_a^{IJ}$ is the ${\rm SL}(2,\Cmat)$ connection compatible with the f\/ield $\tP$ constructed from
the $B$ f\/ield of the Plebanski formulation.
The constraints~\eqref{contA} are somewhat easier than the corresponding constraints~$\Psi^{ab}$~\eqref{seccl-cosntr} on the spin connection, since they involve a projector on the subgroup.
However, the dependence of the right-hand side on the $B$ f\/ield, which is in fact responsible for
the non-commutativity of~$\SA^{IJ}$, introduces additional complications.
As a result, a proper discretization of~\eqref{contA} remains an open question.

Thus, in both the canonical and spin foam approaches, the use of the shifted connection
did not lead to a well-def\/ined model for quantum gravity yet.
However, the dif\/f\/iculties arising on this way seem to be of similar nature and
the shifted connection appears to be a unique object which is suitable for both quantization schemes.
Therefore, from the point of view of the agreement between the canonical and spin foam quantizations,
it can still be viewed as a promising candidate which should be used in the def\/inition of holonomy operators
in both approaches.

\section{Back to three dimensions -- a toy model}
\label{sec-new}

From our discussion in the previous section, it should be clear that all problems and ambiguities arising
in four-dimensional spin foam models come from the dif\/f\/iculties in imposing the simplicity constraints,
which are supposed to turn the trivial dynamics of the topological state sum model into that
of quantum general relativity. Although we have argued that agreement with the canonical approach could play
the role of the guiding principle, it would be nice to have a simple model allowing to test various
ways of implementing the simplicity constraints. In order to do so, such a model should possess the following features:
\begin{itemize}\itemsep=0pt
\item
It should have the form of a Plebanski theory, i.e.\ be written as
\begin{gather}
S_{\rm Th2}=S_{\rm Th1}+{\rm constraints}.
\label{formact}
\end{gather}
\item
Both theories, given by the actions $S_{\rm Th1}$ and $S_{\rm Th2}$, should have known spin foam representations.
\end{itemize}
In this case, we could verify which of the methods to impose the constraints reduces the spin foam quantization of
$S_{\rm Th1}$ to that of $S_{\rm Th2}$. Those methods which do not satisfy
this requirement most likely give wrong quantizations also for the original Plebanski theory~(\ref{plebanski action})
and therefore have few chances to describe quantum gravity.

A model of this kind has been proposed in \cite{Alexandrov:2008da}, where the role of
the theories Th1 and Th2 is played by four-dimensional BF theories with $\Sp(4)$ and $\SU(2)$ gauge groups, respectively.
The constraints that reduce Th1 to Th2 are given by the analogue of the linear simplicity~\eqref{linearsim},
whose discretization reads
\begin{gather}
B_f^{IJ}(x_\tau)_J=0,
\label{linearsim-new}
\end{gather}
and by the secondary constraints \eqref{su2con} relevant for the Lorentz generalization of the AB connection,
which after discretization take the form \eqref{exampl-dc}.
It was shown that one recovers the known spin foam representation of $\SU(2)$ BF theory from the known
spin foam quantization of $\Sp(4)$ BF theory only if one incorporates the secondary constraints
in the way suggested in Section~\ref{subsec_lessons}.
In particular, the vertex amplitude should be modif\/ied according to~\eqref{amplit_simplex}.

However, the model of \cite{Alexandrov:2008da} has one important drawback which makes
the above conclusions uncertain: it has never been shown whether there is an action of the form~\eqref{formact}
with primary constraints~\eqref{linearsim-new} and secondary constraints~\eqref{exampl-dc}.
In other words, the constraint structure was imposed by hand and did not follow from the Hamiltonian
analysis of any action.

On the other hand, in \cite{geiller-noui} it was understood
that one f\/inds a well-def\/ined model with all the above characteristics if one goes down to three dimensions.
In other words, the authors proposed a three-dimensional $\Sp(4)$ Plebanski-like
action in which an analogue of the simplicity constraints is used to recover Riemannian gravity,
i.e.\ $\SU(2)$ BF theory in three dimensions.
Then one can take advantage of the well-established results for
the spin foam quantization of three-dimensional  BF theories to test various proposals,
such as BC, EPRL\footnote{Strictly speaking, we are going
to use the prescription of the EPR model \cite{Engle:2007qf},
since we do not consider the inclusion of the Immirzi parameter.}
and the one from Section~\ref{subsec_lessons},
for imposing the simplicity constraints in four dimensions.

The key observation is that the action
\begin{gather}
\label{plebanski 3d}
S[\omega,B,\lmul]=\int_\CM\de^3x\,\big(\eps^{\mu\nu\rho}\Tr(B_\mu F_{\nu\rho})+\lmul^{\mu\nu}\Tr(B_\mu\star B_\nu)\big),
\end{gather}
in which $B^{IJ}_\mu$ is an $\so(4)$-valued one-form f\/ield, generates the simplicity constraints
\begin{gather}
\label{3d simplicity}
\Phi_{\mu\nu}\equiv
\f{1}{2}\,\eps_{IJKL}B^{IJ}_\mu B^{KL}_\nu = 0,
\end{gather}
which behave like in the four-dimensional Plebanski theory.
In particular, it can be shown that these constraints admit three sectors of solutions.
Apart from a degenerate sector in which the three-volume $\mathcal{V}$ is vanishing,
one has the following solutions:
\begin{itemize}\itemsep=0pt
\item topological sector: $B^{IJ}_\mu=x^Ie^J_\mu-x^Je^I_\mu$,
\item gravitational sector: $B^{IJ}_\mu=\eps^{IJ}_{~~KL}x^Ke^L_\mu$,
\end{itemize}
where $x^I\in\mathbb{R}^4$ is a vector analogous to the normal introduced in Section~\ref{sec_4d}.
It is easy to see that in the topological sector the action~(\ref{plebanski 3d}) is vanishing,
while in the gravitational sector it reduces to the usual $\SU(2)$ Riemannian action for three-dimensional gravity.
The Hamiltonian analysis of this sector shows \cite{geiller-noui} that
the simplicity constraints~$\Phi_{ab}$~(\ref{3d simplicity}) generate secondary constraints
\begin{gather}\label{3d secondary}
\Psi_{ab}\equiv\Tr(D_{(a}B_0\star\!B_{b)})=0,
\end{gather}
which do not commute with the primary ones and together form a second class system.
As their four-dimensional counterpart~$\Psi^{ab}$ given in~\eqref{seccl-cosntr}, they can be seen as constraints
on the components of the spacetime connection~$\omega^{IJ}$.

Now we can quantize the theory \eqref{plebanski 3d} in two ways. First, we impose the constraints (\ref{3d simplicity})
classically and restrict ourselves to the gravitational sector.
Its spin foam quantization is well-known
and provided by the Ponzano--Regge model introduced in Section~\ref{sec_3dSF}.
It is def\/ined by the partition function
\begin{gather}
\mathcal{Z}_{\rm PR}(\Delta^*)
=\sum_{j\rightarrow f}\prod_{f\in\Delta^*}(-1)^{2j_f}(2j_f+1)\prod_{v\in\Delta^*}\{6j\}
\label{pfPR}
\end{gather}
with the vertex amplitude given by the $\SU(2)$ $\{6j\}$ symbol.
On the other hand, we can start from the unconstrained $\Sp(4)$ BF theory and
try to impose the simplicity constraints at the quantum level in order to mimic the situation in four dimensions.
It is clear that the state sum model for the unconstrained part is given by
\begin{gather}
\mathcal{Z}_{\Sp(4)}(\Delta^*)
=\sum_{(j^+,j^-)\rightarrow f}\prod_{f\in\Delta^*}(-1)^{2j_f^+ + 2j_f^-}(2j^+_f+1)(2j^-_f+1)\prod_{v\in\Delta^*}\{6j^+\}\{6j^-\}.
\label{pfSpin4}
\end{gather}
Thus, we come to the following simple question: is there a way to incorporate the constraints
into the partition function~\eqref{pfSpin4} which would reduce it to the Ponzano--Regge model~\eqref{pfPR}?

\subsubsection*{The usual strategy}

\looseness=-1
First we would like to test the ideas used to derive the BC and EPRL spin foam models.
Accor\-ding to them, the simplicity constraints are implemented as restrictions
on the representations and intertwiners labeling the faces and the edges of $\Delta^*$, respectively.
These restrictions are in turn obtained by imposing the primary simplicity constraints~\eqref{3d simplicity}
on the boundary states of the unconstrained theory.
As in four dimensions, at the discrete level one can again distinguish the diagonal and
cross simplicity constraints. The former is always imposed strongly and requires
that the $\Sp(4)$ representations assigned to the faces be always simple, $j^+_f=j^-_f\equiv j_f$.
On the other hand, the imposition of the cross simplicity depends on the chosen approach
and is expected to f\/ix the intertwiners labeling the edges of $\Delta^*$.
In this way one obtains the following results \cite{geiller-noui}:
\begin{itemize}\itemsep=0pt
\item
BC prescription: After the strong imposition of both the diagonal and cross simplicity quadratic
constraints, the partition function becomes
\begin{gather}
\mathcal{Z}_\text{3dBC}(\Delta^*)
=\sum_{j\rightarrow f}\prod_{f\in\Delta^*}\(d_{j_f}\)^2\prod_{v\in\Delta^*}\{6j\}^2.
\label{pf3dBC}
\end{gather}
\item EPR prescription:
One replaces the quadratic cross simplicity by its linear version \eqref{linearsim-new} which
is then imposed in the weak sense.
The resulting partition function reads
\begin{gather}
\mathcal{Z}_\text{3dEPR}(\Delta^*)
=\sum_{j\rightarrow f}\prod_{f\in\Delta^*}\(d_{(j_f/2)}\)^2\prod_{v\in\Delta^*}
\(\{6j\}^2\prod_{\alpha=1}^4f^{i_\alpha}_{i^+_\alpha i^-_\alpha}(j_{\alpha\beta})\),
\label{pf3dEPR}
\end{gather}
with $\alpha,\beta=1,\dots,4$ labeling the edges of the tetrahedron dual to the vertex $v$
and the pair $\alpha\beta$ labeling the corresponding faces.
The fusion coef\/f\/icients $f^i_{i^+i^-}$ attached to edges are given by the evaluation of
a spin network and produce the $\{9j\}$ symbol
\begin{gather*}
f^i_{i^+i^-} =\quad
\vcenter{\hbox{\includegraphics[scale=0.45]{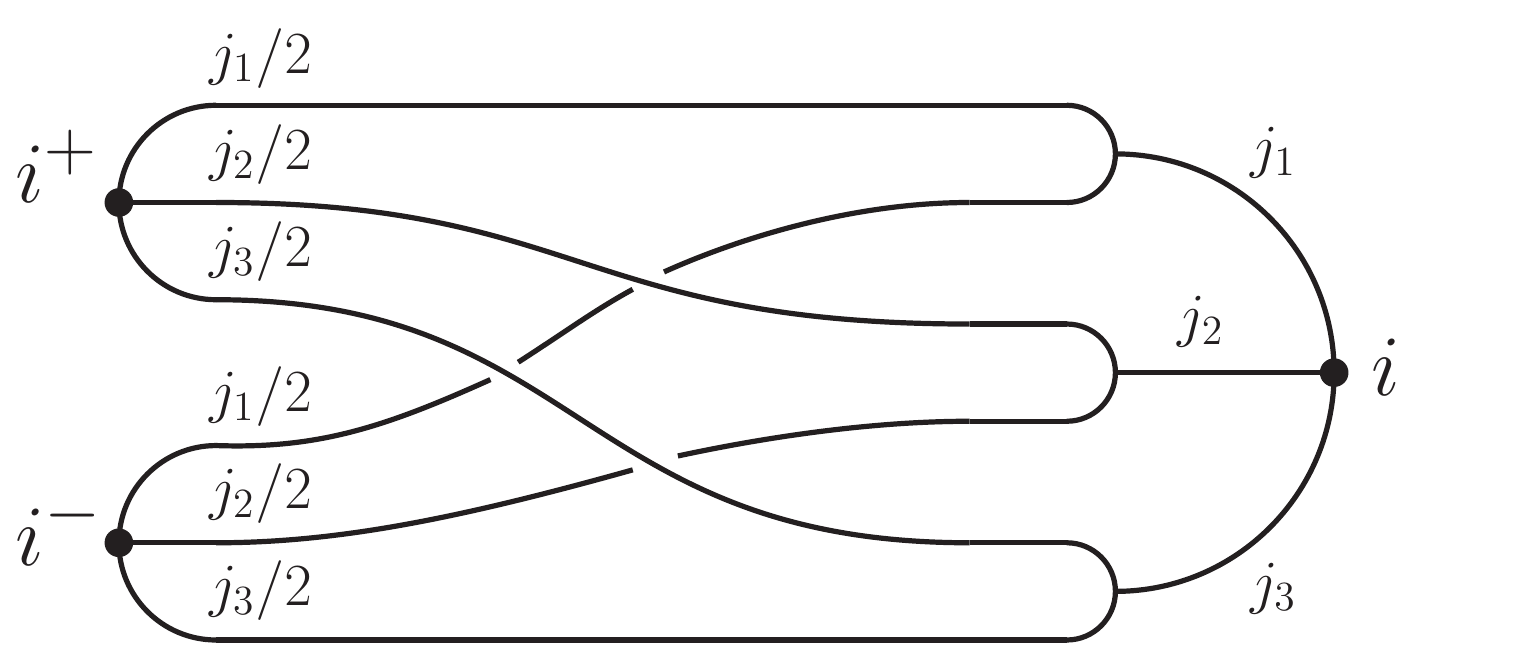}}}
=
\left\{\begin{array}{ccc}
         j_1 & j_2 & j_3\\
         j_1 & j_2 & j_3 \\
         2j_1 & 2j_2 & 2j_3
\end{array}\right\}.
\end{gather*}
Actually, they depend only on three spins $j_{f\supset e}$ since all intertwiners appearing
in their def\/inition are unique.
\end{itemize}
As a result, the EPR-like partition function \eqref{pf3dEPR} dif\/fers from
the one given by the BC prescription \eqref{pf3dBC} only by a normalization factor assigned to the edges.
In fact, this factor appears because the intertwiners obtained by imposing the cross simplicity {\it \`a la} EPR
have not been properly normalized and the operator mapping \eqref{pfSpin4} to \eqref{pf3dEPR}
is not a projector \cite{Kaminski:2009cc}. Once the normalization is corrected, one f\/inds that
\begin{gather*}
\mathcal{Z}_\text{3dEPR-norm}=\mathcal{Z}_\text{3dBC}.
\end{gather*}
This fact has a simple explanation.
Note that since we are in three dimensions, for any simplicial discretization each edge bounds three faces.
Therefore, from the very beginning, up to norma\-li\-za\-tion, there is a unique gauge invariant intertwiner
which can be assigned to the edges of $\Delta^*$.
As a result, in this case the ef\/fect of the cross simplicity is trivial and, after proper normalization,
the BC and EPR models coincide. However, they do become dif\/ferent on a generic discretization.

\looseness=-1
What is important however for our discussion is that neither the BC nor the EPR prescription leads to the Ponzano--Regge
partition function~\eqref{pfPR}. The dif\/ference is twofold.
First of all, the face amplitudes do not reproduce the expected result $d_{j_f}$.
This is in fact not surprising because, as was discussed in Section~\ref{subsubsec-measure},
the lower dimensional amplitudes depend crucially on the path integral measure,
which we did not attempt to f\/ix here properly.
The second dif\/ference is more important: both prescriptions lead to a wrong vertex amplitude.
The resulting amplitude is nothing else but the original vertex amplitude of $\Sp(4)$ BF theory
restricted to the allowed set of representations, exactly as in four-dimensional spin foam models.
However, the correct vertex in our example is the $\SU(2)$ $\{6j\}$ symbol,
which is clearly not equal to its square. Thus, we have to conclude that
the standard spin foam strategy ``f\/irst quantize, then constrain" leads to the wrong quantum dynamics,
as was suspected in Section~\ref{subsubsec-problems} on the basis of the canonical analysis.

\subsubsection*{Improved quantization}

Since the usual methods to incorporate the simplicity constraints into the spin foam state sum failed,
we need to try something else. Our basic observation is that the model \eqref{plebanski 3d},
just like the four-dimensional Plebanski formulation,
possesses the secondary second class constraints~$\Psi_{ab}$~\eqref{3d secondary},
which impose certain restrictions on the components of the connection $\omega$.
As has been argued in Section~\ref{subsec_lessons}, the presence of such constraints
requires a modif\/ication of the integration measure for holonomies.
Here we will show that the results presented in that section allow to recover
all the ingredients of the Ponzano--Regge model and, combining them together,
one f\/inds precisely the state sum \eqref{pfPR}.
It should be noted that the derivation we are going to give here works also in the same way
for the four-dimensional model proposed in \cite{Alexandrov:2008da} with the only dif\/ference
that the three-dimensional simplicial complex should be replaced by its four-dimensional analogue.

Let us f\/ix the normal f\/ield $x^I$, which allows to rewrite the simplicity constraints
in the gravitational sector in the linear form
\begin{gather*}
B^{IJ}_\mu x_J = 0.
\end{gather*}
This normal breaks the $\Sp(4)$ gauge group to an $\SU(2)$ subgroup stabilizing $x^I$.
As a result, the boost components of the Gauss constraint become second class and together with~\eqref{3d secondary}
can be written in the same way as the constraints for the Lorentz extension of the AB connection~\eqref{su2con},~i.e.
\begin{gather}
\(x^{[J}_{\mathstrut}\delta^{I]}_{[K} x_{L]}^{\mathstrut}\) \omega_a^{KL}= x^{[J}\p_a x^{I]}.
\label{su2con-model}
\end{gather}
In particular, in the time gauge $x^I=\delta^I_0$, the meaning of \eqref{su2con-model}
is simply that the left and right chiral parts of the spin connection are the same, $\omega_a^+=\omega_a^-$.
These constraints are easy to discretize and their
discrete version is given by~\eqref{exampl-dc}. It is clear that, when inserted in the measure,
they simply reduce the integration to the $\SU(2)$ subgroup.

Now we are ready to use the prescription of Section~\ref{subsec_lessons} \cite{Alexandrov:2008da}.
First, the vertex amplitude, which is given by \eqref{amplit_simplex} now becomes\footnote{Recall that
we are in three dimensions so that the edges of $\Delta^*$ are dual to triangles of $\Delta$. This is why
the index $\tau$ was replaced by $t$.}
\begin{gather}
A(\lambda_f,j_{t f}) =\int_{G^4} \prod_t\[\de\gl_t\, \delta\((\gl^+_{t})^{-1} \gl^-_{t}u_{x_t}^{-1}\)\]
\CS_{(\Gamma_\tau,\lambda_f,j_{t f})}\big[\gl_{u(f)}^{-1}\gl_{d(f)}^{\mathstrut},x_t\big]
\nonumber\\
\hphantom{A(\lambda_f,j_{t f})}{}
= \{6j\}\prod_f d_{j_f}^{-1} \delta_{j_{u(f)f}j_{d(f)f}},
\label{amplit_simplex-model}
\end{gather}
where the $\{6j\}$ symbol is constructed out of six representations $j_f=j_{u(f)f}=j_{d(f)f}$
and we have used the property\footnote{The normalization of the Clebsch--Gordan coef\/f\/icients
$C^{\,{j_1}\,{j_2}\,{j_3}}_{{m_1}{m_2}{m_3}}$ is chosen as in \cite{Freidel:2007py}.}
\begin{gather*}
\mathop{\sum}\limits_{m,m',n,n'}\ClG{j^+}{j^-}{j_1^{\mathstrut}}{m}{m'}{\ell_1}\,
\Db^{(j^+)}_{mn}(h)\Db^{(j^-)}_{m'n'}(h) \,
\overline{\ClG{j^+}{j^-}{j_2}{n}{n'}{\ell_2}}
=\delta_{j_1j_2} d_{j_1}^{-1}\, \Db^{(j_1)}_{\ell_1 \ell_2}(h).
\end{gather*}
Thus, up to the normalization factor, we have reproduced the vertex amplitude of the Ponzano--Regge model!

However, we should still understand the following issue. The prescription that we are using here
is based on the use of an enlarged state space spanned by all the projected spin networks.
This is ref\/lected, for instance, in the formula~\eqref{reprampl} for the simplex boundary state
in the connection representation where it is represented as a sum over all these spin networks.
On the other hand, the result~\eqref{amplit_simplex-model} implies that the label $\lambda_f$
is auxiliary since the vertex amplitude does not depend on it. How is this degeneracy removed?
Or in other words, how do we reduce to the boundary state space of the Ponzano--Regge model
which is spanned by the usual $\SU(2)$ spin networks?

\label{page-dislabels}
One could expect that this is done by imposing the primary simplicity constraints which
relate the labels $\lambda_f$ to $j_f$, as it happens for example in the EPRL model.
However, this would still leave us with the problem, already discussed in Section~\ref{subsubsec-problems},
that such restrictions on the labels are not enough to reduce
the projected spin networks to the usual ones. In fact, as we will see now, the mechanism
of the reduction is dif\/ferent and is analogous to the one in the canonical quantization.

The idea is to consider the simplex boundary state~\eqref{reprampl} where the projected spin networks
are weighted by the vertex amplitude. Since the latter does not depend on $\lambda_f$, the sum over
this label can be done explicitly. Indeed, it is easy to show the following identity:
\begin{gather*}
\sum_{j^+,j^-}d_{j^+} d_{j^-}
\mathop{\sum}\limits_{m,m',n,n'}\ClG{j^+}{j^-}{j^{\mathstrut}}{m}{m'}{\ell_1}\,
\Db^{(j^+)}_{mn}(g^+)\Db^{(j^-)}_{m'n'}(g^-) \,
\overline{\ClG{j^+}{j^-}{j}{n}{n'}{\ell_2}}
=\delta\( g^- (g^+)^{-1}\) \Db^{(j)}_{\ell_1 \ell_2}(g^+),
\end{gather*}
due to which the simplex amplitude can be evaluated to
\begin{gather}
A[g_f,x_t]=\sum_{j_f} \{6j\}\[\prod_f d_{j_f}\,\delta\big(g_f^-(x) (g^+_{f}(x))^{-1}\big)\]
\CS_{(\Gamma_\tau,\vec \jmath)}[g_f^+(x)],
\label{modelampl}
\end{gather}
where we have introduced $g_f(x)=g_{x_{u(f)}}^{-1} g_f g_{x_{d(f)}}^{\mathstrut}$
and $\CS_{(\Gamma_\tau,\vec \jmath) }$ is the usual $\SU(2)$ spin network associated with the boundary graph
of a tetrahedron $\tau$. This result shows that the sum over the auxiliary representation labels restores
the secondary second class constraints on holonomies, which in turn provide the reduction from the enlarged
state space to the kinematical Hilbert space of the Ponzano--Regge model.
It is worth to note that the primary simplicity constraints have not been used so far.
It is the secondary constraints that allowed to reproduce both the kinematical Hilbert space and
the vertex amplitude implementing the quantum dynamics.

The last step is to glue the amplitudes \eqref{modelampl} associated to dif\/ferent tetrahedra together to
get the full partition function. This can be done by using \eqref{fullpf-noncom} with \eqref{amplitgen},
and this is the only place where the primary simplicity constraints appear explicitly.
Due to these constraints and the $\delta$-function in \eqref{modelampl},
the integral over the bivectors $B_f$ generates the $\delta$-function on the $\SU(2)$ subgroup only.
It imposes the f\/latness condition on the product of the holonomies $h_{\tau f}\equiv g_{\tau f}^+(x)=g_{\tau f}^-(x)$
around each face. Expanding the $\delta$-function in the sum over representations, it is easy to see that
the full partition function is given by
\begin{gather*}
\CZ=\sum_{j_f}\sum_{j_{\tau f}}\prod_{(\tau,f)} \[d_{j_{\tau f}}\int_{\SU(2)}\de h_{\tau f}\]
\prod_f \[d_{j_f}\chi_{j_f}\(\prod_{\tau\supset f}h_{\tau f} \)\]
\prod_\tau \[ \{6j\}\,\CS_{(\Gamma_\tau,\vec \jmath_\tau)}[h_{\tau f}]\].
\end{gather*}
It is immediate to check that, doing the remaining integration and contracting all indices,
one reproduces the Ponzano--Regge state sum \eqref{pfPR} with the same face, edge and vertex amplitudes.

This derivation explicitly demonstrates how the second class constraints should be incorporated into
the discretized path integral and how they allow to recover the right quantization for the constrained theory.
In fact, it is not really surprising that our procedure led to the correct partition function
because the formula \eqref{fullpf-noncom} we have started with, after integrating out the of\/f-diagonal degrees
of freedom f\/ixed by the second class constraint, is trivially equivalent to the initial expression \eqref{initPRpf}
for the Ponzano--Regge partition function. What we have done is just comparing the spin foam
representations of \eqref{fullpf-noncom} and \eqref{initPRpf} obtained after certain manipulations,
which of course cannot spoil the initial equality.
In more complicated cases, it is impossible to solve the second class constraints explicitly
and the strategy based on the partition function \eqref{fullpf-noncom}
and the vertex amplitude \eqref{amplit_simplex} is the only one which remains at our disposal.

\section{Conclusions}

In this work, we have reviewed the loop and spin foam approaches to quantum gravity in three and four spacetime dimensions,
focusing on the issue of their mutual consistency.
Let us summarize the main lessons to be learnt from this study.

In the three-dimensional Riemannian case, we have a beautiful consistency between the canonical quantization
and the covariant path integral represented by the Ponzano--Regge (for $\Lambda=0)$ and the Turaev--Viro (for $\Lambda>0$)
spin foam models. Moreover, both models and their canonical counterparts
are easily extended to include point-like particles.
Nevertheless, even this nice picture contains important gaps.
The most serious one is that, whereas the canonical quantization dual to the Ponzano--Regge model is obtained by
using the standard loop quantization technique, the Turaev--Viro model has been reproduced only from
the so-called combinatorial quantization presented in Section~\ref{subsubsec-combin}.
In this case the loop approach has not been realized yet to a full extent and its agreement with other approaches,
which have proven already to be quite successful, remains an open issue.
In particular, the crucial question is whether it gives rise to the quantum group structure
provided in this case by $\Uq(\so(4))$? We have presented some steps towards this result
in Section~\ref{subsec-positive}.

It is worth stressing that three-dimensional gravity in the Lorentzian signature
represents so far almost an unexplored territory.
The physical states are in general unknown, and no spin foam
quantization (at least mathematically well-def\/ined) has been found so far.
Only the case where $\Lambda$ is positive has been studied in great details because it
is dual to three-dimensional Riemannian gravity with a negative $\Lambda$.
On the other hand, gravity with $\Lambda<0$ is much more interesting because this is the only
case where black hole solutions exist in three dimensions.
One should probably pay more attention to these cases
since they are still much simpler than the physically interesting Lorentzian four-dimensional case.
They can provide new important lessons and generate new interesting ideas.

In four dimensions, the situation is not so clear. On the one hand, the spin foam models that have recently been
introduced have the advantage of solving some long standing problems in the spin foam quantization, and they provide
structures reminiscent of those in loop quantum gravity. For example, the construction of the EPRL model
seems to give rise to the reduction from the full gauge symmetry group to its $\SU(2)$ subgroup, so that
the boundary states of this model are claimed to coincide with the kinematical states of LQG.
On the other hand, a detailed comparison with the canonical approach reveals crucial dif\/ferences
and furthermore raises some questions about the validity of
the whole quantization procedure used to get to these models.
Thus, here one cannot claim that a complete agreement has been achieved, even at the kinematical level.

\looseness=-1
Why is the four-dimensional case so dif\/ferent from the three-dimensional one?
The main dif\/ference between them can be traced back to the existence, in four dimensions,
of certain second class constraints called simplicity. This becomes especially clear
in the Plebanski formulation presented in Section~\ref{subsec_Pleb}.
These constraints are responsible for breaking the topological invariance and generating the non-trivial dynamics
of general relativity. At the same time, they also represent the deviation point between the two quantization
approaches: while in the canonical approach the simplicity constraints are incorporated at the classical level,
either by means of the Dirac bracket or by solving them explicitly, in the spin foam approach
they are supposed to be imposed only at the quantum level.
This dif\/ference in the quantization strategy is at the origin of all the other dif\/ferences
in the results and f\/inal constructions of LQG and spin foam models.

Moreover, as we brief\/ly reviewed in Section~\ref{subsubsec-problems}, it is this unusual
strategy to deal with second class constraints used in the spin foam approach that is the reason
for various inconsistencies expected to appear at the dynamical level.
In fact, this strategy dif\/fers from the usual one implied by Dirac's quantization
in two aspects. First, as we already said, the simplicity constraints are imposed {\it after} quantization,
so that one quantizes really a {\it non-physical symplectic structure}. Second, it completely ignores
the presence of the {\it secondary constraints} conjugated to the primary simplicity.
A nice demonstration that such a strategy is indeed erroneous is provided by the model
introduced in \cite{geiller-noui} and presented in Section~\ref{sec-new}.

But is there an alternative way to get spin foam models in four dimensions and how should
a~model consistent with the canonical quantization look like?
We have presented some tentative answers to these questions in Section~\ref{subsec_lessons}.
In our opinion, it is crucial to start with the path integral with the right measure,
which must agree with the measure implied by the canonical quantization.
In particular, besides various volume-dependent factors indispensable for the correct implementation
of the dif\/feomorphism symmetry, the measure should contain $\delta$-functions of the secondary second
class constraints mentioned above. It turns out that the inclusion of these constraints
has far reaching consequences and strongly af\/fects the dynamics of the theory.

First of all, the secondary second class constraints ensure the reduction from an enlarged
Hilbert space consisting of all projected spin networks
to the kinematical Hilbert space in exactly the same way as the Lorentz covariant formulation of LQG
reduces to the usual formulation based on the $\SU(2)$ gauge group, as recalled in Section~\ref{subsubsec-covLQG}.
Thus, it becomes possible to achieve agreement between the canonical and spin foam quantizations
at least at the kinematical level. But most importantly, the inclusion of the secondary constraints
changes the standard prescription for the vertex amplitude: it is still given by an integral
of the simplex boundary state, but now with a modif\/ied measure (see~\eqref{amplit_simplex}).
In fact, it is quite natural to expect such a modif\/ication, since it introduces a new input
into the quantity determining the dynamics at the quantum level. Without it, the dynamics
turns out to be completely determined by the kinematics via kinematical boundary states.
In contrast, now it becomes dependent on the particular form of the Hamiltonian, since
the secondary constraints do depend on its form.

Again all these conclusions f\/ind a nice conf\/irmation in the model of Section~\ref{sec-new}.
Of course, whether they can be realized in the full four-dimensional theory of quantum gravity
remains to be seen. Nevertheless, we hope that the ideas and techniques presented in this review
will be helpful to go along this direction.

\subsection*{Acknowledgements}

This research is supported by contract ANR-09-BLAN-0041.

\LastPageEnding

\end{document}